\documentclass[12pt]{article}
\usepackage{float}
\usepackage{amsmath}
\usepackage{amssymb}
\usepackage{graphicx}
\usepackage{color}
\usepackage{neil}
\usepackage{tikz}
\usepackage[authoryear]{natbib}

\newtheorem{example}{Example}
\newtheorem{remark}{Remark}
\newtheorem{assumption}{Assumption}
\newtheorem{definition}{Definition}

\newtheorem{theorem}{Theorem}
\allowdisplaybreaks
\input{tcilatex}
\begin{document}

\title{An estimator for predictive regression: \\
reliable inference for financial economics\thanks{%
Thanks to Isaiah Andrews, Joseph Blitzstein, Andrew Patton, Ashesh
Rambachan, Julia Shephard and particularly Jihyun Kim and Nour Meddahi. The
code which produces all the simulation results is in \texttt{sign1.r}. The
code which produces all the empirical results (including downloading the
data) is in \texttt{RegFinance1.r} and \texttt{Recursive1.r}.}}
\author{Neil Shephard \\
\textit{Department of Economics and }\\
\textit{Department of Statistics,}\\
\textit{Harvard University}\\
\texttt{shephard@fas.harvard.edu}}
\maketitle

\begin{abstract}
Estimating linear regression using least squares and reporting robust
standard errors is very common in financial economics, and indeed, much of
the social sciences and elsewhere. \ For thick tailed predictors under
heteroskedasticity this recipe for inference performs poorly, sometimes
dramatically so. Here, we develop an alternative approach which delivers an
unbiased, consistent and asymptotically normal estimator so long as the
means of the outcome and predictors are finite. \ The new method has
standard errors under heteroskedasticity which are easy to reliably estimate
and tests which are close to their nominal size. \ The procedure works well
in simulations and in an empirical exercise. \ An extension is given to
quantile regression. \ 
\end{abstract}


\section{Introduction}

Think about an outcome variable $Y_{1}$ and $p$ predictors $\mathbf{Z}%
_{1}=(Z_{1},...,Z_{p})^{\mathtt{T}}$. \ Throughout assume $\mathrm{E}[Y_{1}]$
and $\mathrm{E}[\mathbf{Z}_{1}]$ exist (meaning $\mathrm{E}\left\vert
Y_{1}\right\vert <\infty $ and $\mathrm{E}\left\vert \mathbf{Z}%
_{1}\right\vert <\infty $). Write $\mathbf{X}_{1}^{\mathtt{T}}=\left\{
1,\left( \mathbf{Z}_{1}\mathbf{-}\mathrm{E}[\mathbf{Z}_{1}]\right) ^{\mathtt{%
T}}\right\} ^{\mathtt{T}}$, where $^{\mathtt{T}}$ denotes a transpose, then $%
\mathrm{E}\left\vert \mathbf{X}_{1}\right\vert <\infty $. I will work with a
linear in parameters \textquotedblleft predictive
regression,\textquotedblright 
\begin{equation}
\mathrm{E}[Y_{1}|\mathbf{X}_{1}=\mathbf{x}_{1}]=\mathbf{x}_{1}^{\mathtt{T}}%
\mathbf{\beta ,\quad }\text{where\quad }\mathbf{x}_{1}^{\mathtt{T}}=\left\{
1,\left( \mathbf{z}_{1}\mathbf{-}\mathrm{E}[\mathbf{Z}_{1}]\right) ^{\mathtt{%
T}}\right\} ^{\mathtt{T}}\mathbf{.}  \label{predict4}
\end{equation}%
$\mathbf{\beta =(}\beta _{0},\beta _{1},...,\beta _{p}\mathbf{)}^{\mathtt{T}%
} $ is my estimand and inference about $\mathbf{\beta }$ is my goal. \ 

The motivation for this paper is that thick tailed predictors with
heteroskedastic outcomes are very common in financial economics. Finance
researchers nearly always assume that the mean of asset returns exists,
while the vast bulk believe the variance exists. Due to the empirical
evidence of their sample instability and the results from applying extreme
value theory to estimate the data's tail index, many are skeptical that
third or fourth moments exist (e.g. the accessible review of \cite{Cont(01)}%
). This challenges traditional least squares based \textquotedblleft robust
standard errors\textquotedblright\ type inference methods nearly universally
used in financial economics, which rely on these higher order moments for
their justification. This credibility gap rarely impacts the way applied
researchers behave, perhaps understandably so because it is less than clear
what action to take without potentially employing quite complicated methods.
This paper provides a simple solution to this problem.

More broadly, the use of traditional least squares based robust standard
errors is very common in many areas of applied statistics (e.g. see the
beginning of \cite{KingRoberts(15)} for a discussion of the use in political
science and \cite{MacKinnon(12)} for a discussion of the econometrics
literature). \ The methods developed here could prove useful in other
applied fields, for thick tailed data is very common, although often less
apparent than in the data rich environment of financial economics. \ 

The core of this paper focuses on the sample $\left( \mathbf{Z}%
_{1},Y_{1}\right) ,...,\left( \mathbf{Z}_{n},Y_{n}\right) $, a sequence of
pairs of i.i.d. random variables which each obeys (\ref{predict4}),
highlighting 
\begin{eqnarray*}
\widehat{\mathbf{\psi }} &=&\frac{1}{n}\sum_{j=1}^{n}\mathbf{Z}_{j},\quad 
\mathbf{X}_{j}=(1,(\mathbf{Z}_{j}-\widehat{\mathbf{\psi }})^{\mathtt{T}})^{%
\mathtt{T}}, \\
\widehat{\mathbf{\beta }} &=&\underset{\mathbf{b}}{\arg }\underset{}{\min }%
\sum_{j=1}^{n}S_{j}(\mathbf{b}),\quad S_{j}(\mathbf{b})=\frac{1}{2}%
\left\Vert \mathbf{X}_{j}\right\Vert _{2}^{-1}\left( Y_{j}-\mathbf{X}_{j}^{%
\mathtt{T}}\mathbf{b}\right) ^{2},\quad \left\Vert \mathbf{X}_{j}\right\Vert
_{2}=\sqrt{\sum_{i=1}^{p+1}X_{j,i}^{2}}.
\end{eqnarray*}

The presence of $\left\Vert \mathbf{X}_{j}\right\Vert _{2}^{-1}$ reduces the
influence of thick tailed predictors, making valid inference possible for
problems in financial economics. \ Downweighting extreme predictors is at
the heart of the \textquotedblleft bounded-influence
function\textquotedblright\ part of the robustness literature. \ A classic
reference to that work is \cite{KraskerWelsch(82)}. My focus is on the
contribution $\left\Vert \mathbf{X}_{j}\right\Vert _{2}^{-1}$ can make to
allowing valid inference about $\mathbf{\beta }$ under heteroskedasticity. \
\ \ \ 

Then%
\begin{equation*}
\frac{\partial S_{j}(\mathbf{\beta })}{\partial \mathbf{\beta }}=-\mathbf{G}%
_{j}\left( Y_{j}-\mathbf{X}_{j}^{\mathtt{T}}\mathbf{\beta }\right) ,\quad 
\mathbf{G}_{j}=\left\Vert \mathbf{X}_{j}\right\Vert _{2}^{-1}\mathbf{X}_{j},
\end{equation*}%
so, if the symmetric $\sum_{j=1}^{n}\mathbf{G}_{j}\mathbf{X}_{j}^{\mathtt{T}%
} $ is invertible, then 
\begin{equation*}
\widehat{\mathbf{\beta }}=S_{\mathbf{G,X}}^{-1}S_{\mathbf{G},Y},\quad S_{%
\mathbf{G,X}}=\frac{1}{n}\sum_{j=1}^{n}\mathbf{G}_{j}\mathbf{X}_{j}^{\mathtt{%
T}},\quad S_{\mathbf{G},Y}=\frac{1}{n}\sum_{j=1}^{n}\mathbf{G}_{j}Y_{j}.
\end{equation*}%
Crucially 
\begin{equation*}
\left\Vert \mathbf{G}_{j}\right\Vert _{\infty }=\underset{i=1,...,p+1}{\max }%
|G_{j,i}|\;\leq 1,
\end{equation*}%
which will drive the robustness of $\widehat{\mathbf{\beta }}$ to thick
tailed predictors. \ 

$\widehat{\mathbf{\beta }}$ will be conditionally (on the predictors)
unbiased, consistent, asymptotically normal with a variance which can be
estimated by 
\begin{equation}
\frac{1}{n}S_{G\mathbf{,X}}^{-1}S_{\widehat{U}^{2}\mathbf{G,G}}S_{G\mathbf{,X%
}}^{-1},\quad S_{\widehat{U}^{2}\mathbf{G,G}}=\frac{1}{n}\sum_{j=1}^{n}1_{%
\left\Vert |\mathbf{X}_{1}|/\mathrm{E}|\mathbf{X}_{1}|\right\Vert _{\infty
}\leq cn^{1/5}}\widehat{U}_{j}^{2}\mathbf{G}_{j}\mathbf{G}_{j}^{\mathtt{T}}
\label{SE computation}
\end{equation}%
where $U_{j}=Y_{j}-\mathbf{X}_{j}^{\mathtt{T}}\widehat{\mathbf{\beta }}$, so
long as $\mathrm{Var}(U_{1})<\infty $ and $\mathrm{E}|\mathbf{X}_{1}|<\infty 
$. \ In practice I take $c=10$, so the truncation $1_{\left\Vert |\mathbf{X}%
_{1}|/\mathrm{E}|\mathbf{X}_{1}|\right\Vert _{\infty }\leq cn^{1/5}}$ is
irrelevant except for the most extraordinary predictors. \ Without the
truncation we need the additional condition that $\mathrm{Var}(\mathbf{X}%
_{1})<\infty $. \ 

In comparison, Assumption 4 of \cite{White(80)} spells out that \cite%
{Eicker(67)}, \cite{Huber(67)} and \cite{White(80)} robust standard errors
needs $\mathrm{Var}(\mathbf{X}_{1}^{2})<\infty $ for inference on $\mathbf{%
\beta }$ based on least squares 
\begin{equation*}
\widehat{\mathbf{\beta }}_{LS\ }=S_{\mathbf{Z,Z}}^{-1}S_{\mathbf{Z,}Y},\quad
S_{\mathbf{X,X}}=\frac{1}{n}\sum_{j=1}^{n}\mathbf{X}_{j}\mathbf{X}_{j}^{%
\mathtt{T}},\quad S_{\mathbf{X,}Y}=\sum_{j=1}^{n}\mathbf{X}_{j}Y_{j},
\end{equation*}%
to be asymptotically valid. Recall these standard errors are based on 
\begin{equation*}
\frac{1}{n}S_{\mathbf{X,X}}^{-1}\mathbf{S}_{\widehat{U}_{LS}^{2}\mathbf{X,X}%
}S_{\mathbf{X,X}}^{-1},\quad \mathbf{S}_{\widehat{U}_{LS}^{2}\mathbf{X,X}}=%
\frac{1}{n}\sum_{j=1}^{n}\widehat{U}_{LS,j}^{2}\mathbf{X}_{j}\mathbf{X}_{j}^{%
\mathtt{T}},\quad \widehat{U}_{LS,j}=(Y_{j}-\mathbf{X}_{j}^{\mathtt{T}}%
\widehat{\mathbf{\beta }}_{LS}).
\end{equation*}%
Unfortunately, financial economists may well not have those four moments
available to them. \ Monte Carlo and empirical results presented later will
demonstrate this asymptotic worry is important in practice. \ Further, the
results suggest that even if more moments exist than four the finite sample
inference is still very fragile unless $n$ is very substantial. \ Overall,
in my opinion, the evidence suggests \cite{Eicker(67)}, \cite{Huber(67)} and 
\cite{White(80)} type \textquotedblleft robust standard
errors\textquotedblright\ are not credible in financial economics. \ $%
\widehat{\mathbf{\beta }}$ is one potential solution. \ \ 

The same line of argument holds for inference on the $\tau $-quantile
regression: 
\begin{equation*}
Q_{Y_{1}|\mathbf{X}_{1}=\mathbf{x}_{1}}(\tau )=\mathbf{x}_{1}^{\mathtt{T}}%
\mathbf{\beta ,\quad }\tau \in (0,1).
\end{equation*}%
The estimand is, again, $\mathbf{\beta }$. Recall the check-function
notation $\rho _{\tau }(u)=u\left( \tau -I_{u<0}\right) $. I advocate the
estimator 
\begin{equation*}
\widehat{\mathbf{\beta }}=\underset{\mathbf{b}}{\arg }\underset{}{\min }%
\sum_{j=1}^{n}S_{j}(\mathbf{b}),\quad S_{j}(\mathbf{b})=\left\Vert \mathbf{X}%
_{j}\right\Vert _{2}^{-1}\rho _{\tau }(Y_{j}-\mathbf{X}_{j}^{\mathtt{T}}%
\mathbf{b}),
\end{equation*}%
noting $S_{j}(\mathbf{b})$ is convex in $\mathbf{b}$ with bounded
subderivative 
\begin{equation*}
\partial S_{j}(\mathbf{b})=-\left\Vert \mathbf{X}_{j}\right\Vert _{2}^{-1}%
\mathbf{X}_{j}(\tau -1_{Y_{j}<\mathbf{X}_{j}^{\mathtt{T}}\mathbf{b}}).
\end{equation*}%
This is an alternative to the celebrated \cite{KoenkerBassett(78)} estimator 
\begin{equation*}
\widehat{\mathbf{\beta }}_{KB}=\underset{\mathbf{b}}{\arg }\underset{}{\min }%
\sum_{j=1}^{n}S_{j}^{\ast }(\mathbf{b}),\quad S_{j}^{\ast }(\mathbf{b})=\rho
_{\tau }(Y_{j}-\mathbf{X}_{j}^{\mathtt{T}}\mathbf{b}),
\end{equation*}%
which has the unbounded subderivative%
\begin{equation*}
\partial S_{j}^{\ast }(\mathbf{b})=-\mathbf{X}_{j}(\tau -1_{Y_{j}<\mathbf{X}%
_{j}^{\mathtt{T}}\mathbf{b}}).
\end{equation*}%
When $\tau =1/2$ then $\widehat{\mathbf{\beta }}_{KB}$ is, famously, the
least absolute deviation (LAD) estimator of Boscovich from 1805.
Unfortunately, inference on $\widehat{\mathbf{\beta }}_{KB}$ is, again, not
robust to thick tailed predictors and so is not, in my opinion, credible for
financial economics. \ \ The math is more complicated for $\widehat{\mathbf{%
\beta }}_{KB}$ than $\widehat{\mathbf{\beta }}_{LS\ }$, but the source of
weakness is exactly the same. \ $\widehat{\mathbf{\beta }}$ is one potential
consistent and asymptotically normal solution. \ Is $\widehat{\mathbf{\beta }%
}$ easy to compute? \ $\widehat{\mathbf{\beta }}$ is $\widehat{\mathbf{\beta 
}}_{KB}$ applied to the preprocessed data $Y_{j}^{\ast }=$\ $\left\Vert 
\mathbf{X}_{j}\right\Vert _{2}^{-1}Y_{j}$ and $\mathbf{X}_{j}^{\ast }=$ $%
\left\Vert \mathbf{X}_{j}\right\Vert _{2}^{-1}\mathbf{X}_{j}$, noting%
\begin{eqnarray*}
\left\Vert \mathbf{X}_{j}\right\Vert _{2}^{-1}\rho _{\tau }(Y_{j}-\mathbf{X}%
_{j}^{\mathtt{T}}\mathbf{b}) &=&(\left\Vert \mathbf{X}_{j}\right\Vert
_{2}^{-1}Y_{j}-\left\Vert \mathbf{X}_{j}\right\Vert _{2}^{-1}\mathbf{X}_{j}^{%
\mathtt{T}}\mathbf{b})\left( \tau -I_{(\left\Vert \mathbf{X}_{j}\right\Vert
_{2}^{-1}Y_{j}-\left\Vert \mathbf{X}_{j}\right\Vert _{2}^{-1}\mathbf{X}_{j}^{%
\mathtt{T}}\mathbf{b})<0}\right)  \\
&=&\rho _{\tau }(Y_{j}^{\ast }-\mathbf{X}_{j}^{\ast \mathtt{T}}\mathbf{b}).
\end{eqnarray*}%
\ The preprocessing stabilizes statistical inference, while existing
software can be used without any further changes. \ \ \ 

The remaining parts of this paper are as follows. In Section \ref{sect:lit}
I will focus on a scalar predictor and explain where $\widehat{\mathbf{\beta 
}}$ comes from and derive its main inferential properties. \ While doing
this I will review the literature on this topic, linking results across
different intellectual fields. \ 

In Section \ref{sect:identify} I provide conditions for identifying $\beta $
and derive a corresponding method of moments estimator $\widehat{\beta }$. \
Section \ref{sect:conditionProp} holds the main condition properties of $%
\widehat{\beta }$, conditioning on the predictors. \ Section \ref%
{sect:uncondition} contains the corresponding unconditional properties of $%
\widehat{\beta }$. \ In both sections $\widehat{\beta }$ is compared to the
corresponding least squares estimator $\widehat{\beta }_{LS}$. \ Section \ref%
{sect:simulation exper} presents the results from various simulation
experiments to see how effective the asymptotics guidance is. \ 

Section \ref{sect:empirical} contains results from a massive number of
hypothesis tests using $\widehat{\beta }$, where I identify stocks with high
betas or low betas. \ This allows me to form high (or low) beta portfolios,
which is a potentially useful investment vehicle for investors unable to
take on financial leverage (e.g. young savers into pensions). \ I also study
how these procedures work as they are rolled through the time series
database. \ 

Section \ref{sect:quantile} extends the work to quantile based estimation,
focusing on median predictive regression. \ I state my conclusions in
Section \ref{sect:conclusion}. Any lengthy proof of a Theorem stated in the
main text is given in the Appendix.

\section{Why is $\protect\widehat{\mathbf{\protect\beta }}$ interesting and
the literature\label{sect:lit}}

The main virtues of $\widehat{\mathbf{\beta }}$ are seen in the most
stripped down case: the focus of this section. \ \ 

Assume a linear in parameters \textquotedblleft predictive
regression\textquotedblright\ 
\begin{equation}
\mathrm{E}[Y_{1}|X_{1}=x_{1}]=\beta _{1}x_{1},  \label{eqn:prediction}
\end{equation}%
for outcome $Y_{1}$ and scalar predictor $X_{1}=Z_{1}$, where $\mathrm{E}%
[Y_{1}]=\mathrm{E}[Z_{1}]=0$. \ As each item is a scalar, no bolding will be
used here. Upper cases denote random variables, lower cases fixed numbers. \
\ 

Then, 
\begin{equation*}
\left\Vert x_{1}\right\Vert _{2}=|x_{1}|,\quad \quad g_{1}=\left\Vert
x_{1}\right\Vert _{2}^{-1}x_{1}=sign(x_{1}),
\end{equation*}%
so%
\begin{equation*}
\widehat{\beta }_{1}=\underset{b_{1}}{\arg }\underset{}{\min }%
\sum_{j=1}^{n}S_{j}(b_{1}),\quad S_{j}(b_{1})=\frac{1}{2}|x_{j}|^{-1}\left(
y_{j}-x_{j}b_{1}\right) ^{2},\quad \frac{\partial S_{j}(b_{1})}{\partial
b_{1}}=-sign(x_{j})\left( y_{j}-x_{j}b_{1}\right) ,
\end{equation*}%
implying 
\begin{equation*}
\widehat{\beta }_{1}=S_{G,X}^{-1}S_{G,Y}=\frac{\sum_{j=1}^{n}sign(x_{j})y_{j}%
}{\sum_{j=1}^{n}|x_{j}|},
\end{equation*}%
where $S_{G,X}=\frac{1}{n}\sum_{j=1}^{n}g_{j}x_{j}=\frac{1}{n}%
\sum_{j=1}^{n}\left\vert x_{j}\right\vert $ and $S_{G,Y}=\frac{1}{n}%
\sum_{j=1}^{n}g_{j}y_{j}=\frac{1}{n}\sum_{j=1}^{n}sign(x_{j})y_{j}$.

I give nine features of $\widehat{\beta }_{1}$, weaving them together with a
literature review.

\subsection{Ways of deriving $\protect\widehat{\protect\beta }_{1}$\ }

The first three features are different ways of deriving $\widehat{\beta }%
_{1} $. \ 

First, multiply both sides of (\ref{eqn:prediction}) by $g_{1}$, then 
\begin{equation*}
g_{1}\mathrm{E}[Y_{1}|X_{1}=x_{1}]=\beta _{1}sign(x_{1})x_{1}=\beta
_{1}|x_{1}|.
\end{equation*}%
If $\mathrm{E}[G_{1}X_{1}]$ and $\mathrm{E}[G_{1}Y_{1}|$ exist, then
unconditionally 
\begin{equation*}
\mathrm{E}[G_{1}Y_{1}]=\beta _{1}\mathrm{E}[G_{1}X_{1}].
\end{equation*}%
Crucially $|G_{1}|\leq 1$, so a sufficient condition for $\mathrm{E}%
[G_{1}Y_{1}|$ to exist is that $\mathrm{E}|Y_{1}|<\infty $. \ The same
argument implies $\mathrm{E}[G_{1}Z_{1}]$ exists if $\mathrm{E}\mathbf{|}%
X_{1}|<\infty $. \ If, in addition, $\mathrm{E}[G_{1}Z_{1}]=\mathrm{E}%
\mathbf{|}X_{1}|>0$ then \ 
\begin{equation*}
\beta _{1}=\frac{\mathrm{E}[G_{1}Y_{1}]}{\mathrm{E}[G_{1}Z_{1}]}=\frac{%
\mathrm{E}[sign(Z_{1})Y_{1}]}{\mathrm{E}\mathbf{|}Z_{1}|}.
\end{equation*}%
So a sufficient condition for $\beta _{1}$ to be identified is $0<\mathrm{E}%
\mathbf{|}Z_{1}|<\infty $ and $\mathrm{E}|Y_{1}|<\infty $. \ Let $\left(
X_{1},Y_{1}\right) ,...,\left( X_{n},Y_{n}\right) $ be a sequence of pairs
of random variables which each obeys (\ref{eqn:prediction}). Then 
\begin{equation}
\widehat{\beta }_{1}=\frac{\sum_{j=1}^{n}sign(X_{j})Y_{j}}{%
\sum_{j=1}^{n}|X_{j}|},  \label{eqn:weirdEst3}
\end{equation}%
is a method of moments estimator. By the strong law of large numbers, under
just two conditions, $0<\mathrm{E}|X_{1}|<\infty $ and $\mathrm{E}%
|Y_{1}|<\infty $, 
\begin{equation*}
\widehat{\beta }_{1}\overset{p}{\rightarrow }\frac{\mathrm{E}%
[sign(X_{1})Y_{1}]}{\mathrm{E}|X_{1}|}=\beta _{1}.
\end{equation*}%
Hence $\widehat{\beta }_{1}$ is consistent if the data is a tad less thick
tailed than, for example, Cauchy random variables.\ 

Second, $\widehat{\beta }_{1}$ is an Instrumental Variable (IV) estimator,
where the \textquotedblleft instruments\textquotedblright\ are $sign(X_{j})$%
. Often, IV estimators behave notoriously poorly in many of their
applications as the \textquotedblleft relevance\textquotedblright\ condition
of instrumental variables is \textquotedblleft weak\textquotedblright\ (e.g.
the reviews in \cite{AndrewsStockSun(19)}). This is not the case here, as
the relevance condition $\mathrm{E}[sign(X_{1})X_{1}]=\mathrm{E}|X_{1}|>0$
should hold strongly.

Third, $\widehat{\beta }_{1}$ is the maximum quasi-likelihood (ML) estimator
from the contrived model $Y_{j}|X_{j}=x_{j}\overset{indep}{\sim }N(\beta
_{1}x_{j},|x_{j}|\sigma ^{2})$. This implies the existence of a
quasi-likelihood $\log L(\beta _{1})=\frac{\beta _{1}}{\sigma ^{2}}%
\sum_{j=1}y_{j}sign(x_{j})-\frac{1}{2\sigma ^{2}}\beta
_{1}^{2}\sum_{j=1}|x_{j}|$, which downweights predictors with very large $%
|x_{j}|$ compared to the tradition homoskedastic quasi-likelihood case. $%
\log L(\beta _{1})$ invites a Gaussian prior for $\beta _{1}$ given the
predictors, delivering a Gaussian quasi-posterior for $\beta _{1}$ given the
outcomes and predictors.

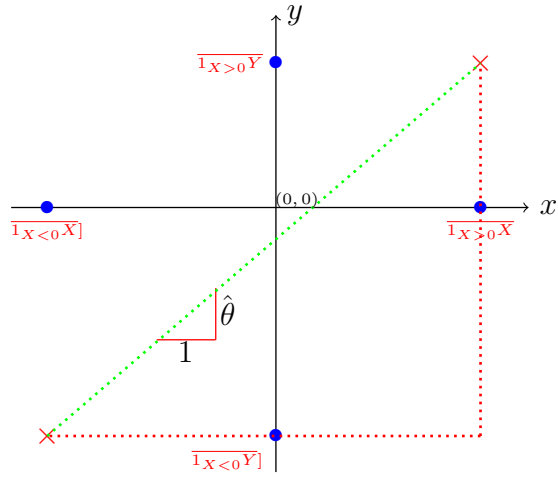
\begin{figure}[tbh]
\center
\begin{tikzpicture}[xscale=1.6,yscale=1.6]

\draw[->,color=black,line width=0.5pt] (-2.2,0) -- (2.1,0) node[right] {$x$};
\draw[->,color=black,line width=0.5pt] (0,-2.2) -- (0,1.6) node[right] {$y$};

\draw[color=red,line width=0.5pt] (-0.99,-1.1) -- (-0.5,-1.1); \draw[color=red,line width=0.5pt] (-0.5,-1.1) -- (-0.5,-0.67);
\node[line width=2pt] at (-0.75,-1.2) {1};  \node[line width=2pt] at (-0.4,-0.85) {$\hat{\theta}$};
%

\node[color=red,left,line width=2pt] at (0.0,1.2) {\tiny $\overline{1_{X>0}Y}$};  \node[color=blue,inner sep=0.5pt] at (0,1.2) {$\bullet$};
\node[color=red,left,line width=2pt] at (0.0,-2.1) {\tiny $\overline{1_{X<0}Y}]$};  \node[color=blue,inner sep=0.5pt] at (0,-1.9) {$\bullet$};

\node[color=red,below,line width=2pt] at (1.7,0.0) {\tiny $\overline{1_{X>0}X}$};  \node[color=blue,inner sep=0.5pt] at (1.7,0.0) {$\bullet$};
\node[color=red,below,line width=2pt] at (-1.9,0.0) {\tiny $\overline{1_{X<0}X}]$};  \node[color=blue,inner sep=0.5pt] at (-1.9,0.0) {$\bullet$};

\draw[dotted,color=red,line width=1pt] (1.7,-1.9) -- (1.7,1.2) node {$\times$};

\draw[dotted,color=red,line width=1pt] (1.7,-1.9) -- (-1.9,-1.9) node {$\times$};

\draw[dotted,color=green,line width=1pt] (-1.9,-1.9) -- (1.7,1.2) node {};

\node[right] at (-0.1,0.06) {\tiny $(0,0)$};

\end{tikzpicture}
\caption{Slope of the green line is $\hat{\protect\theta}$. Length of
horizontal red line is $\overline{1_{X>0}X}-\overline{1_{X<0}X}>0$, the
vertical red line moves down from $\overline{1_{X>0}Y}$ to $\overline{%
1_{X<0}Y}$. }
\label{fig:weirdest}
\end{figure}

My fourth point is different. \ Divide the top and bottom of (\ref%
{eqn:weirdEst3}) by $n$ and write 
\begin{equation*}
\widehat{\beta }_{1}=\frac{\overline{1_{X>0}Y}-\overline{1_{X<0}Y}}{%
\overline{1_{X>0}X}-\overline{1_{X<0}X}},\quad \text{where, e.g.,}\quad 
\overline{1_{X>0}Y}=\frac{1}{n}\sum_{j=1}^{n}1_{X_{j}>0}Y_{j},
\end{equation*}%
then the geometry of the estimator is shown in Figure \ref{fig:weirdest},
where the slope of the green line is $\widehat{\beta }_{1}$. The length of
horizontal red line is $\overline{1_{X>0}X}-\overline{1_{X<0}X}>0$, while
the vertical red line moves down from $\overline{1_{X>0}Y}$ to $\overline{%
1_{X<0}Y}$. 

\subsection{Major properties of $\protect\widehat{\protect\beta }_{1}$}

The next two features are the main inferential properties of $\widehat{\beta 
}_{1}$.

Fifth, in terms of conditional inference, if the pairs $(X_{j},Y_{j})$ are
independent and (\ref{eqn:prediction}) holds for each $j$, then $\mathrm{E}[%
\widehat{\beta }_{1}|(\mathbf{X}=\mathbf{x})]=\beta _{1}$, where $\mathbf{X=}%
(X_{1},...,X_{n})$ and the observed predictors $x=(x_{1},...,x_{n})$.
Further, for $j=1,...,n,$ if $\sigma _{j}^{2}(x_{j})=\mathrm{Var}%
(Y_{j}|X_{j}=x_{j})<\infty $, 
\begin{equation*}
\mathrm{Var}[\widehat{\beta }_{1}|\left( \mathbf{X}=\mathbf{x}\right) ]=%
\frac{\sum_{j=1}^{n}\sigma _{j}^{2}(x_{j})}{\left(
\sum_{j=1}^{n}|x_{j}|\right) ^{2}}=\frac{1}{n}\frac{\frac{1}{n}%
\sum_{j=1}^{n}\sigma _{j}^{2}(x_{j})}{\left( \frac{1}{n}%
\sum_{j=1}^{n}|x_{j}|\right) ^{2}}.
\end{equation*}%
Then $\frac{1}{n}\sum_{j=1}^{n}\sigma _{j}^{2}(x_{j})$ can be estimated by $%
\frac{1}{n}\sum_{j=1}^{n}(Y_{j}-\widehat{\beta }_{1}x_{j})^{2}$ (this will
be discussed in more detail shortly). This makes inference based on the
approximate pivot%
\begin{equation*}
\widehat{T}_{\widehat{\beta }}=\frac{\widehat{\beta }_{1}-\beta _{1}}{\sqrt{%
\frac{\sum_{j=1}^{n}(Y_{j}-\widehat{\beta }_{1}x_{j})^{2}}{\left(
\sum_{j=1}^{n}|x_{j}|\right) ^{2}}}}
\end{equation*}%
effective for thick tailed heteroskedastic data. \ Whether the researcher
assumes homoskedasticity or not does not change the form of $\widehat{T}_{%
\widehat{\beta }}$. \ It was this property which initially made me
interested in $\widehat{\beta }_{1}$.

Sixth, in terms of unconditional inference, if the pairs $(X_{j},Y_{j})$ are
independent and identically distributed (i.i.d.), then the strong law of
large numbers implies that, as $n\rightarrow \infty $, 
\begin{equation*}
\widehat{\beta }_{1}\xrightarrow{p}\frac{\mathrm{E}[sign(X_{1})Y_{1}]}{%
\mathrm{E}|X_{1}|}=\frac{\mathrm{E}[1_{X_{1}>0}Y_{1}]-\mathrm{E}%
[1_{X_{1}<0}Y_{1}]}{\mathrm{E}[1_{X_{1}>0}X_{1}]-\mathrm{E}[1_{X_{1}<0}X_{1}]%
}=\beta ^{\ast },
\end{equation*}%
so long as $\mathrm{E}|Y_{1}|<\infty $ and $0<\mathrm{E}|X_{1}|<\infty $.
Here, $\beta _{1}^{\ast }$ is a \textquotedblleft
pseudo-true\textquotedblright\ value of $\beta $. If (\ref{eqn:prediction})
holds, then Adam's Law implies that $\mathrm{E}[1_{X_{1}>0}Y_{1}]=\beta _{1}%
\mathrm{E}[1_{X_{1}>0}X_{1}]$ and $\mathrm{E}[1_{X_{1}<0}Y_{1}]=\beta _{1}%
\mathrm{E}[1_{X_{1}<0}X_{1}]$, so $\beta _{1}^{\ast }=\beta _{1}$, forcing $%
\widehat{\beta }_{1}\xrightarrow{p}\beta _{1}$. \ Further, defining $%
U_{1}=Y_{1}-X_{1}\beta _{1}$ and additionally assuming $\mathrm{Var}%
(U_{1})<\infty $, then unconditionally 
\begin{equation*}
\sqrt{n}(\widehat{\beta }_{1}-\beta _{1})\xrightarrow{d}N\left( 0,\frac{%
\mathrm{Var}(U_{1})}{\left\{ \mathrm{E}|X_{1}|\right\} ^{2}}\right) ,
\end{equation*}%
hence heteroskedasticity has no impact on the limit distribution of $%
\widehat{\beta }_{1}$. Further, $\mathrm{E}|X_{1}|$ can be estimated by $%
\frac{1}{n}\sum_{j=1}^{n}|X_{j}|$. \ Finally, define $\widehat{U}%
_{j}=Y_{j}-X_{j}\widehat{\beta }_{1}=U_{j}-X_{j}(\widehat{\beta }_{1}-\beta
_{1})$ where $U_{j}=Y_{j}-X_{j}\beta _{1}$, so 
\begin{equation*}
\widehat{\mathrm{Var}(U_{1})}=\frac{1}{n}\sum_{j=1}^{n}\widehat{U}_{j}^{2}=%
\frac{1}{n}\sum_{j=1}^{n}U_{j}^{2}+(\widehat{\beta }_{1}-\beta _{1})^{2}%
\frac{1}{n}\sum_{j=1}^{n}X_{j}^{2}-2(\widehat{\beta }_{1}-\beta _{1})\frac{1%
}{n}\sum_{j=1}^{n}U_{j}X_{j}.
\end{equation*}%
If $\mathrm{Var}(X_{1})<\infty $ exists then $\mathrm{Var}(U_{1})$ can be
consistently estimated by $\widehat{\mathrm{Var}(U_{1})}$. More broadly, if $%
\mathrm{Var}(X_{1})$ does not exist then $\widehat{\mathrm{Var}(U_{1})}$
performs poorly in theory and in simulations (this will be reported in
Sections \ref{sect:ses} and \ref{sect:simulation exper}). \ Instead, a
weighted version, which clips the estimator for large absolute predictors, $%
\frac{1}{n}\sum_{j=1}^{n}\widehat{U}_{j}^{2}w(X_{j})$ where the weight $%
w(x)=1_{\frac{|x|}{\mathrm{E}|X_{1}|}<dn^{1/5}}$, is consistent for $\mathrm{%
Var}(U_{1})$, requiring, again just requiring $\mathrm{E}|X_{1}|<\infty $. \
In simulations, I take $d=10$ (so if $n=10$ then $dn^{1/5}\simeq 16$), so
the clipping will have literally no impact on nearly all applied work. \
However, the evidence suggests the weight is a worthwhile guardrail for very
thick tailed data.

\subsection{Relating $\protect\widehat{\protect\beta }$ to other estimators\
\ }

Seventh, in the context of linear regressions for stable random variables, 
\cite{BlattbergSargent(71)} derived 
\begin{equation*}
\widetilde{\beta }_{1}=\sum_{j=1}^{n}|x_{j}|^{c}sign(x_{j})Y_{j}/%
\sum_{j=1}^{n}|x_{j}|^{1+c},\quad c>0,
\end{equation*}%
regarding the predictors as non-stochastic, minimizing the $\alpha $-stable
scale in the class of linear unbiased estimators. When $c=0$, $\widetilde{%
\beta }_{1}$ would be $\widehat{\beta }_{1}$, but they did not cover that
case, nor its analytic properties. When $c>0$ standard errors are not robust
to heteroskedasticity. \cite{SamorodnitskyRachevKurzStoyanov(07)} studied
the distributional properties of $\widetilde{\beta }_{1}$ under very heavy
tails in $X_{1}$, but assuming independence between the predictors and the
regression errors. This independence assumption takes them outside our
interests. \cite{GorjiAminghafari(19)} builds on \cite{BlattbergSargent(71)}
and \cite{SamorodnitskyRachevKurzStoyanov(07)} towards non-parametric
regression.

Eighth, \cite{SoShin(99)} studied estimating autoregressions with the
instrument $sign(Y_{j-1})$, yielding an estimator 
\begin{equation*}
\overline{\beta }_{1}=\sum_{j=2}^{n}sign(Y_{j-1})Y_{j}/%
\sum_{j=2}^{n}|Y_{j-1}|.
\end{equation*}%
To \cite{SoShin(99)}, $\overline{\beta }_{1}$ had attractive properties in
heavy tailed time series. They call this a \textquotedblleft Cauchy
estimator\textquotedblright , after \cite{Cauchy(1836)} (who followed up his
first paper with 6 others around this topic), noting that $\overline{\beta }%
_{1}$ can be thought of as an instrumental variable estimator and a
generalized least squares estimator. The historians of least squares and
regression in statistics usually associate Cauchy's work with numerical
interpolation (e.g. \cite{Seal(67)}, Ch. 13 of \cite{Farebrother(98)} and
Ch. 4 of \cite{HeydeSeneta(77)}). \ It matches $\widehat{\beta }_{1}$ only
in the scalar case with no intercept. \cite{HeydeSeneta(77)} detail Cauchy's
work on regression from a modern perspective. Ch. 13.5 of \cite{Linnik(61)}
discusses the multivariate Cauchy's method, proving it is unbiased and
derives the variance of $\widehat{\beta }_{1}$ in the scalar case for
non-stochastic predictors under homoskedasticity.

\cite{PhillipsParkChang(04)} generalized the \cite{SoShin(99)} use of $%
sign(Y_{j-1})$ in an autoregression to an \textquotedblleft instrument
generating function\textquotedblright\ $F(Y_{j-1})$, where $F$ is an
asymptotically homogenous function.  \cite{KimMeddahi(20)} mention the \cite%
{SoShin(99)} approach to fitting autoregressions in the context of time
series of realized volatility type objects (which tend to be thick tailed).
They also link to \cite{SamorodnitskyRachevKurzStoyanov(07)}. Their interest
was in consistent estimation for time series with heavy tailed regression
errors. \ \cite{IbragimovKimShrobotov(20)} look at time series estimators of 
$\overline{\beta }_{1}$ type under volatility clustering. \ 

\cite{MikoschdeVries(13)} studied the properties of least squares under
heavy tailed predictors, providing interesting results and references. \cite%
{HillRenault(10)} devised trimming methods for the GMM which allow for both
heavy tailed variables and Gaussian limit theory. \cite%
{HallinSwanVerdeboutVeredas(10)} looked at using rank based methods in
regression context with heavy tailed data, while \cite%
{ButlerMacDonaldNelsonWhite(90)} use adaptive statistical models for
regressions, assuming predictors and prediction errors are independent.

Ninth, more broadly, \cite{BalkemaEmbrechts(18)} provides a review of a
substantial literature on robust estimation and heavy tailed data, as well
as comparing procedures using Monte Carlo methods. Related work is \cite%
{KurzKimLoretan(14)}. \cite{NolanOjeda(13)} looks at linear regressions with
heavy tailed errors. \ Of course, these last two papers interface with the
influential robustness literature, reviewed by \cite%
{HampelRonchettiRousseeuwStahel(05)}. \cite{SunZhouFan(20)} is an
interesting recent paper which is close to our setup. \ It uses a Huber loss
for a linear predictive regression where the threshold is selected
adaptively so that asymptotically they still recover the estimand, the
parameter indexing the predictive regression. The \cite{SunZhouFan(20)}
procedure is likely to be asymptotically more efficient than $\widehat{\beta 
}_{1}$. 

The use of%
\begin{equation*}
\widehat{\mathbf{\beta }}=\underset{\mathbf{b}}{\arg }\underset{}{\min }%
\sum_{j=1}^{n}\left\Vert \mathbf{X}_{j}\right\Vert _{2}^{-1}\left( Y_{j}-%
\mathbf{X}_{j}^{\mathtt{T}}\mathbf{b}\right) ^{2},
\end{equation*}%
is a special case of the important wide ranging work on the bounded
influence function literature, which goes back to \cite%
{Mallows(73),Mallows(75)} and \cite{KraskerWelsch(82)}. \ Most of their
focus is on efficient robust estimation of $\mathbf{\beta }$ (or, in
particular, M-estimator generalizations). \ My interest is in being able to
estimate the standard errors under heteroskedasticity so that inference in
financial economics is reliable. \ 

\subsection{Comparing $\protect\widehat{\protect\beta }$ to least squares}

I now compare $\widehat{\beta }_{1}$ to the ML estimator from the
conditionally Gaussian linear regression $Y_{j}|X_{j}=x_{j}\overset{indep}{%
\sim }N(\beta _{1}x_{j},\sigma ^{2})$, 
\begin{equation*}
\widehat{\beta }_{LS,1}=\frac{\sum_{j=1}^{n}X_{j}Y_{j}}{%
\sum_{j=1}^{n}X_{j}^{2}},
\end{equation*}%
the celebrated \textquotedblleft least squares\textquotedblright\ estimator.
Of course, under homoskedasticity, $\widehat{\beta }_{LS,1}$ will be more
efficient than $\widehat{\beta }_{1}$. For i.i.d. pairs, famously, if enough
moments exist, then unconditionally 
\begin{equation*}
\sqrt{n}(\widehat{\beta }_{LS,1}-\beta _{1})\xrightarrow{d}N\left( 0,\frac{%
\mathrm{Var}(X_{1}U_{1})}{\left\{ \mathrm{E}[X_{1}^{2}]\right\} ^{2}}\right)
.
\end{equation*}%
Then $\frac{1}{n}\sum_{j=1}^{n}X_{j}^{2}$ estimates $\mathrm{E}[X_{1}^{2}]$.
Define $\widehat{U}_{LS,j}=Y_{j}-X_{j}\widehat{\beta }_{LS,1}=U_{j}-X_{j}(%
\widehat{\beta }_{LS,1}-\beta _{1})$ where, again, $U_{j}=Y_{j}-X_{j}\beta
_{1}$. Then 
\begin{equation*}
\widehat{\mathrm{Var}(X_{1}U_{1})}=\frac{1}{n}\sum_{j=1}^{n}X_{j}^{2}%
\widehat{U}_{LS,j}^{2}=\frac{1}{n}\sum_{j=1}^{n}X_{j}^{2}U_{j}^{2}+(\widehat{%
\beta }_{LS,1}-\beta _{1})^{2}\frac{1}{n}\sum_{j=1}^{n}X_{j}^{4}-2(\widehat{%
\beta }_{1}-\beta _{1})\frac{1}{n}\sum_{j=1}^{n}U_{j}X_{j}^{3},
\end{equation*}%
which needs $\mathrm{E}(X_{1}^{4})<\infty $ to behave well as an estimator
of $\mathrm{Var}(X_{1}U_{1})$, in theory and in simulations. \ If enough
moments exist, then, taken together, this motivates robust standard errors
based on $\frac{1}{n}\sum_{j=1}^{n}X_{j}^{2}\widehat{U}_{LS,j}^{2}$,
following \cite{Eicker(67)}, \cite{Huber(67)} and \cite{White(80)} (where
Assumption 4 spells out the need for $\mathrm{E}(X_{1}^{4})<\infty $). \
Robust standard errors are used in vast numbers of applied papers.
Unfortunately $\widehat{\mathrm{Var}(X_{1}U_{1})}$ is a poor estimator
unless (i) $n$ is very large, (ii) $U_{1}$ is known to be independent of $%
X_{1}$ or (iii) the predictors are thin tailed. \ This makes valid inference
based on the asymptotic pivot%
\begin{equation*}
\widehat{T}_{LS}=\frac{\widehat{\beta }_{LS,1}-\beta _{1}}{\sqrt{\frac{%
\sum_{j=1}^{n}X_{j}^{2}(Y_{j}-\widehat{\beta }X_{j})^{2}}{\left(
\sum_{j=1}^{n}X_{j}^{2}\right) ^{2}}}},
\end{equation*}%
challenging for data in finance. \ It is well known that $\widehat{T}_{LS}$
often has poor finite sample properties, although the infeasible version of
this $T_{LS}=(\widehat{\beta }_{LS,1}-\beta _{1})/\sqrt{\mathrm{Var}(%
\widehat{\beta }_{LS,1}|(\mathbf{X}=\mathbf{x}))}$ does not. Some try to
mend this problem using a bootstrap of the approximate pivot $\widehat{T}%
_{LS}$ or an Edgeworth expansion, e.g. \cite{MacKinnonWhite(85)}, \cite%
{Hall(92)}, \cite{MacKinnon(12)} and \cite{HausmanPalmer(12)}.

As I said, under homoskedasticity $\widehat{\beta }_{LS,1}$ will be more
efficient than $\widehat{\beta }_{1}$ (e.g. the Gauss-Markov Theorem or, in
the Gaussian outcomes case, Cram\'{e}r-Rao inequality), with 
\begin{equation*}
\frac{\mathrm{Var}(\widehat{\beta }_{1})}{\mathrm{Var}(\widehat{\beta }%
_{LS,1})}\simeq \frac{\left\{ \mathrm{E}[X_{1}^{2}]\right\} }{\left\{ 
\mathrm{E}|X_{1}|\right\} ^{2}}\geq 1,
\end{equation*}%
by Jensen's inequality. This point goes back at least to Ch. 13.5 of \cite%
{Linnik(61)}. \ If $X_{1}\in \{-1,1\}$, with equal probability, then $%
\mathrm{E}[X_{1}^{2}]/\left\{ \mathrm{E}|X_{1}|\right\} ^{2}=1$, so $%
\widehat{\beta }$ is fully efficient. Of course it is, $sign(X_{j})=X_{j}$
in that case, so $\widehat{\beta }_{1}=\widehat{\beta }_{LS,1}$. \ Under $%
X_{1}\sim N(0,\lambda ^{2})$, then $\mathrm{E}[X_{1}^{2}]/\left\{ \mathrm{E}%
|X_{1}|\right\} ^{2}=\pi /2\simeq 1.57$ hence the $\widehat{\beta }_{LS,1}$
is substantially more efficient than $\widehat{\beta }_{1}$. Under the
thicker tailed $X_{1}\sim Laplace(0,\lambda )$, so $\mathrm{E}%
[X_{1}^{2}]/\left\{ \mathrm{E}|X_{1}|\right\} ^{2}=2$. If $X_{1}$ is very
thick tailed, then $\mathrm{E}[X_{1}^{2}]$ can go to infinity in cases where 
$\mathrm{E}|X_{1}|$ is finite. \ Then $\widehat{\beta }_{LS,1}$ is a much
more precise estimator, on average, but the Gaussian CLT no longer holds for 
$\widehat{\beta }_{LS,1}$ in cases where the CLT for $\widehat{\beta }_{1}$
is still useful. This suggests CLT for $\widehat{\beta }_{1}$ may be a more
practical guide in the kind of thick tailed data often seen in finance, for
example.\ 


\section{Identification and estimation of $\mathbf{\protect\beta }$\label%
{sect:identify}}

Again, think about an outcome variable $Y_{1}$ and $p$ predictors $\mathbf{Z}%
_{1}=(Z_{1},...,Z_{p})^{\mathtt{T}}$, \ where $\mathrm{E}\left\vert
Y_{1}\right\vert <\infty $ and $\mathrm{E}\left\vert \mathbf{Z}%
_{1}\right\vert <\infty $. The following is enough to establish
identification of $\mathbf{\beta }$.

\begin{assumption}[Joint law of $(\mathbf{Z}_{1},Y_{1})$]

\begin{enumerate}
\item[A1.] $\mathrm{E}\left\vert Y_{1}\right\vert <\infty $ and $\mathrm{E}%
\left\vert \mathbf{Z}_{1}\right\vert <\infty $. Write $\mathbf{\mathbf{\psi =%
}}\mathrm{E}[\mathbf{Z}_{1}]$, 
\begin{equation*}
\mathbf{X}_{1}(\mathbf{\mathbf{\psi }})^{\mathtt{T}}=\left\{ 1,\left( 
\mathbf{Z}_{1}\mathbf{-\mathbf{\psi }}\right) ^{\mathtt{T}}\right\} ^{%
\mathtt{T}},\quad \text{and\quad }\mathbf{G}_{1}(\mathbf{\mathbf{\psi }}%
)=\left\Vert \mathbf{X}_{1}(\mathbf{\mathbf{\psi }})\right\Vert _{2}^{-1}%
\mathbf{X}_{1}(\mathbf{\mathbf{\psi }}).
\end{equation*}

\item[A2.] There exists a single $\mathbf{\beta }$ such that 
\begin{equation}
\mathrm{E}[Y_{1}|\mathbf{Z}_{1}=\mathbf{z}_{1}]=\mathbf{x}_{1}^{\mathtt{T}}%
\mathbf{\beta ,\quad x}_{1}^{\mathtt{T}}=\left\{ 1,\left( \mathbf{z}_{1}%
\mathbf{-\mathbf{\psi }}\right) ^{\mathtt{T}}\right\} ^{\mathtt{T}}
\end{equation}%
so all $\mathbf{z}_{1}$.

\item[A3.] 
\begin{equation*}
\mathrm{E}\left[ \mathbf{G}_{1}(\mathbf{\mathbf{\psi }})\mathbf{X}_{1}(%
\mathbf{\mathbf{\psi }})^{\mathtt{T}}\right] =\mathrm{E}\left[ \left\Vert 
\mathbf{X}_{1}(\mathbf{\mathbf{\psi }})\right\Vert _{2}^{-1}\mathbf{X}_{1}(%
\mathbf{\mathbf{\psi }})\mathbf{X}_{1}(\mathbf{\mathbf{\psi }})^{\mathtt{T}}%
\right]
\end{equation*}%
is positive definite.
\end{enumerate}
\end{assumption}

As A1 includes an intercept, note that $\left\Vert \mathbf{X}_{1}(\mathbf{%
\mathbf{\psi }})\right\Vert _{2}\geq 1$, while $\left\Vert \mathbf{G}_{1}(%
\mathbf{\mathbf{\psi }})\right\Vert _{\infty }\leq 1$.

\begin{theorem}
\label{thm:basic} Under A1, 
\begin{equation*}
\mathrm{E}\left[ \mathbf{G}_{1}(\mathbf{\mathbf{\psi }})\mathbf{X}_{1}(%
\mathbf{\mathbf{\psi }})^{\mathtt{T}}\right] 
\end{equation*}%
exists and is symmetric, positive semidefinite. Under A1+A3, it is also
positive definite. \ 
\end{theorem}

Proof. \ $\left\Vert \mathbf{G}_{1}(\mathbf{\mathbf{\psi }})\right\Vert
_{\infty }\leq 1$ so A1 implies $\mathrm{E}\left[ \mathbf{G}_{1}(\mathbf{%
\mathbf{\psi }})\mathbf{X}_{1}(\mathbf{\mathbf{\psi }})^{\mathtt{T}}\right] $
exists. \ By construction, it is symmetric, positive semi-definite. \
Assumption A3 pushes this to positive definite. \ QED. \ 

The estimand will be $\mathbf{\beta }$, while $\mathbf{\mathbf{\psi =}}%
\mathrm{E}[\mathbf{Z}_{1}]$ will be a \textquotedblleft
nuisance\textquotedblright . \ Using Theorem \ref{thm:basic}, the following
is straightforward.

\begin{theorem}[Identification]
\label{thm:identification}Assume A1-A3, then 
\begin{eqnarray}
\mathbf{\mathbf{\psi }} &\mathbf{=}&\mathrm{E}[\mathbf{Z}_{1}]
\label{identifiedtheta} \\
\mathbf{\beta } &\mathbf{=}&\left\{ \mathrm{E}\left[ \mathbf{G}_{1}(\mathbf{%
\mathbf{\psi }})\mathbf{X}_{1}(\mathbf{\mathbf{\psi }})^{\mathtt{T}}\right]
\right\} ^{-1}\mathrm{E}[\mathbf{G}_{1}(\mathbf{\mathbf{\psi }})Y_{1}]. 
\notag
\end{eqnarray}
\end{theorem}

This Theorem says that $\mathbf{\beta }$ and $\mathbf{\mathbf{\psi }}$ can
be uniquely determined, that is, identified, from $\mathrm{E}[\mathbf{Z}%
_{1}] $, $\mathrm{E}[\mathbf{G}_{1}(\mathbf{\mathbf{\psi }})Y_{1}]$ and $%
\mathrm{E}\left[ \mathbf{G}_{1}(\mathbf{\mathbf{\psi }})\mathbf{X}_{1}(%
\mathbf{\mathbf{\psi }})^{\mathtt{T}}\right] $. \ Further, and crucially,
all three of these terms are guaranteed to exist if both $\mathrm{E}[\mathbf{%
Z}_{1}]$ and $\mathrm{E}[Y_{1}]$ exist. Adding Assumption A3 is the only
substantial assumption made beyond the core model A1 and A2. \ \ \ \ 

Now turn to estimation. \ 

Let $\left( \mathbf{Z}_{1},Y_{1}\right) ,...,\left( \mathbf{Z}%
_{n},Y_{n}\right) $ be a sequence of pairs of random variables which each
obeys A1-A3.

Define 
\begin{eqnarray*}
\overline{\mathbf{Z}} &=&\frac{1}{n}\sum_{j=1}^{n}\mathbf{Z}_{j},\quad 
\mathbf{X}_{j}=(1,(\mathbf{Z}_{j}-\overline{\mathbf{Z}})^{\mathtt{T}})^{%
\mathtt{T}},\quad \mathbf{G}_{j}=\left\Vert \mathbf{X}_{j}\right\Vert
_{2}^{-1}\mathbf{X}_{j}, \\
S_{\mathbf{G,X}} &=&\frac{1}{n}\sum_{j=1}^{n}\mathbf{G}_{j}\mathbf{X}_{j}^{%
\mathtt{T}},\quad S_{\mathbf{G},Y}=\frac{1}{n}\sum_{j=1}^{n}\mathbf{G}%
_{j}Y_{j}.
\end{eqnarray*}%
Importantly, $\left\Vert \mathbf{G}_{j}\right\Vert _{\infty }\leq 1$ and $S_{%
\mathbf{G,X}}$ is symmetric and positive semi-definite. \ \ 

We now introduce a method of moment estimator of $(\mathbf{\mathbf{\psi ,}%
\beta })$.

\begin{definition}
Assume $S_{\mathbf{G,X}}$ is positive definite. \ Using the moment condition
(\ref{identifiedtheta}), define a method of moment estimator%
\begin{equation*}
\widehat{\mathbf{\mathbf{\psi }}}=\overline{\mathbf{Z}},\quad \text{and\quad 
}\widehat{\mathbf{\beta }}=S_{\mathbf{G,X}}^{-1}S_{\mathbf{G},Y}.
\end{equation*}
\end{definition}

$\widehat{\mathbf{\beta }}$ is an instrumental variable (IV) estimator, that
uses the $\mathbf{G}_{j}$ as instruments (although note that $\widehat{%
\mathbf{\mathbf{\psi }}}$ is buried within $\mathbf{G}_{j}$).

\begin{example}
If $p=1$ and no intercept, so $X_{j}=Z_{j}-\overline{Z}$, then $\mathbf{G}%
_{j}=\left\Vert \mathbf{X}_{j}\right\Vert _{2}^{-1}\mathbf{X}_{j}=sign(Z_{j}-%
\overline{Z})$, 
\begin{equation*}
\widehat{\beta }=\frac{\sum_{j=1}^{n}sign(Z_{j}-\overline{Z})Y_{j}}{%
\sum_{j=1}^{n}|Z_{j}-\overline{Z}|},
\end{equation*}%
which is non-centered (prediction) version of the estimator discussed in
Section \ref{sect:lit}. \ \ 
\end{example}

It is sometimes helpful to unpack $\mathbf{\beta }$ into its elements $%
\mathbf{\beta }=(\beta _{0},\mathbf{\beta }_{1:p}^{\mathtt{T}})^{\mathtt{T}}$%
.

\begin{theorem}
\label{thm:unpack}Assume $S_{\mathbf{G,X}}>0$. \ Define the weights and
weighted averages 
\begin{equation*}
w_{j}=\frac{\left\Vert \mathbf{X}_{j}\right\Vert _{2}^{-1}}{%
\sum_{i=1}^{n}\left\Vert \mathbf{X}_{i}\right\Vert _{2}^{-1}},\quad 
\widetilde{Y}=\sum_{j=1}^{n}w_{j}Y_{j},\quad \widetilde{\mathbf{Z}}%
=\sum_{j=1}^{n}w_{j}\mathbf{Z}_{j},
\end{equation*}%
and the weighted sums of outer products%
\begin{equation*}
\widetilde{S}_{\mathbf{Z}-\widetilde{\mathbf{Z}},\mathbf{Z}-\widetilde{%
\mathbf{Z}}}=\sum_{j=1}^{n}w_{j}\left( \mathbf{Z}_{j}-\widetilde{\mathbf{Z}}%
\right) \left( \mathbf{Z}_{j}-\widetilde{\mathbf{Z}}\right) ^{\mathtt{T}%
},\quad \widetilde{S}_{\mathbf{Z}-\widetilde{\mathbf{Z}},Y-\widetilde{Y}%
}=\sum_{j=1}^{n}w_{j}\left( \mathbf{Z}_{j}-\widetilde{\mathbf{Z}}\right)
\left( Y_{j}-\widetilde{Y}\right) .
\end{equation*}%
Then 
\begin{equation*}
\widehat{\beta }_{1:p}=\widetilde{S}_{\mathbf{Z}-\widetilde{\mathbf{Z}},%
\mathbf{Z}-\widetilde{\mathbf{Z}}}^{-1}\widetilde{S}_{\mathbf{Z}-\widetilde{%
\mathbf{Z}},Y-\widetilde{Y}},\quad \widehat{\beta }_{0}=\widetilde{Y}-\left( 
\widetilde{\mathbf{Z}}-\overline{\mathbf{Z}}\right) ^{\mathtt{T}}\widehat{%
\gamma },
\end{equation*}%
delivering the $j$-th residual $\widehat{U}_{j}=\left( Y_{j}-\widehat{\beta }%
_{0}\right) -(Z_{j}-\overline{\mathbf{Z}})^{\mathtt{T}}\widehat{\beta }_{1:p}
$, $j=1,...,n$.
\end{theorem}

Proof. Given in the Appendix.

\begin{example}
\label{Ex:p is one}If $p=1$ then%
\begin{equation*}
\widetilde{Y}=\frac{\sum_{j=1}^{n}\frac{Y_{j}}{\sqrt{1+\left( Z_{j}-%
\overline{Z}\right) ^{2}}}}{\sum_{j=1}^{n}\frac{1}{\sqrt{1+\left( Z_{j}-%
\overline{Z}\right) ^{2}}}},\quad \widetilde{Z}=\frac{\sum_{j=1}^{n}\frac{%
Z_{j}}{\sqrt{1+\left( Z_{j}-\overline{Z}\right) ^{2}}}}{\sum_{j=1}^{n}\frac{1%
}{\sqrt{1+\left( Z_{j}-\overline{Z}\right) ^{2}}}},\quad \widehat{\beta }%
_{0}=\widetilde{Y}-\left( \widetilde{Z}-\overline{Z}\right) \widehat{\beta }%
_{1}
\end{equation*}%
$\widehat{\beta }_{0,LS}=\overline{Y}$ and 
\begin{equation*}
\widehat{\beta }_{1}=\frac{\sum_{j=1}^{n}\frac{\left( Z_{j}-\widetilde{Z}%
\right) \left( Y_{j}-\widetilde{Y}\right) }{\sqrt{1+\left( Z_{j}-\overline{Z}%
\right) ^{2}}}}{\sum_{j=1}^{n}\frac{\left( Z_{j}-\widetilde{Z}\right) ^{2}}{%
\sqrt{1+\left( Z_{j}-\overline{Z}\right) ^{2}}}},\quad \widehat{\beta }%
_{1,LS}=\frac{\sum_{j=1}^{n}\left( Z_{j}-\overline{Z}\right) Y_{j}}{%
\sum_{j=1}^{n}\left( Z_{j}-\overline{Z}\right) ^{2}}.
\end{equation*}%
\ \ 
\end{example}

\section{Properties of $\protect\widehat{\mathbf{\protect\beta }}$}

\subsection{Conditional properties of $\protect\widehat{\mathbf{\protect%
\beta }}$\label{sect:conditionProp}}

There are two broad ways of performing inference on the parameters that
index a predictive regression: conditionally and unconditionally. \ First,
focus on the conditional case. \ 

\begin{assumption}[Conditional assumptions]

\begin{enumerate}
\item[B1.] The matrix%
\begin{equation*}
S_{\mathbf{G,X}}=\frac{1}{n}\sum_{j=1}^{n}\mathbf{G}_{j}\mathbf{X}_{j}^{%
\mathtt{T}}
\end{equation*}
is positive definite.

\item[B2.] The pairs $(\mathbf{Z}_{1},Y_{1}),...,(\mathbf{Z}_{n},Y_{n})$ are
independent.

\item[B3.] $\mathrm{Var}(Y_{j}|\mathbf{Z}_{j}=\mathbf{z}_{j})=\sigma
_{j}^{2}<\infty $, for $j=1,2...,n$.

\item[B4.] The matrix%
\begin{equation*}
S_{\sigma ^{2}\mathbf{G,G}}=\frac{1}{n}\sum_{j=1}^{n}\sigma _{j}^{2}\mathbf{G%
}_{j}\mathbf{G}_{j}^{\mathtt{T}}
\end{equation*}
is positive definite. \ 

\item[B5.] $\mathrm{E}[|U_{j}|^{3}|(\mathbf{Z}=\mathbf{z})]<\infty $, for $%
j=1,2...,n$, where $U_{j}=Y_{j}-\mathbf{X}_{j}^{\mathtt{T}}\mathbf{\beta }$.
\end{enumerate}
\end{assumption}

Write all the random predictors as $\mathbf{Z}=(\mathbf{Z}_{1},...,\mathbf{Z}%
_{n})$ and their observed version in the sample as $\mathbf{z=(z}_{1},...,%
\mathbf{z}_{n}\mathbf{)}$\textbf{.} \ 

\begin{theorem}
\label{thm:conditional}If B1-B2 and A1-A2 hold for each $j=1,...,n$, then $%
\mathrm{E}[\widehat{\mathbf{\beta }}|(\mathbf{Z}=\mathbf{z})]=\mathbf{\beta }
$\textbf{. }If B1-B3 and A1-A2 hold, then 
\begin{equation*}
\mathrm{Var}(\widehat{\mathbf{\beta }}|(\mathbf{Z}=\mathbf{z}))=\frac{1}{n}%
\Psi _{n},\quad \Psi _{n}=S_{\mathbf{G,X}}^{-1}S_{\sigma ^{2}\mathbf{G,G}}S_{%
\mathbf{G,X}}^{-1}.
\end{equation*}%
Further, under B1-B5, there exists a constant $c>0$, such that 
\begin{eqnarray*}
&&\sup_{A\in \mathcal{C}_{p+1}}\left\vert \Pr (\sqrt{n}\left( \widehat{%
\mathbf{\beta }}-\mathbf{\beta }\right) \in A|(\mathbf{Z}=\mathbf{z}))-\Pr
(N(0,\Psi _{n})\in A|(\mathbf{Z}=\mathbf{z}))\right\vert \\
&\leq &c\frac{(p+1)^{1/4}}{n^{1/2}}\left( n^{-1}\sum_{j=1}^{n}\varsigma
_{j}\right) ,
\end{eqnarray*}%
where $\mathcal{C}_{p+1}$ denotes the set of all convex subsets of $R^{p+1}$
and 
\begin{equation*}
\varsigma _{j}=\mathrm{E}[|U_{j}|^{3}|(\mathbf{Z}=\mathbf{z})]\left\Vert
S_{\sigma ^{2}\mathbf{G,G}}^{-1/2}\mathbf{G}_{j}\right\Vert _{2}^{3}.
\end{equation*}
\end{theorem}

\textbf{Proof. }Given in the Appendix.

The first two results are of a familiar form. \ The last result is a type of
Berry-Esseen bound. Although the Berry-Esseen bound looks at first sight
asymptotic, it is not, it is exact. Assumption B5 is a Lyapunov-type
condition. \ It has the asymptotic implications that under B1-B5, so as $n$
increases%
\begin{equation*}
\sqrt{n}\Psi _{n}^{-1/2}\left( \widehat{\mathbf{\beta }}-\mathbf{\beta }%
\right) |(\mathbf{Z}=\mathbf{z})\overset{d}{\rightarrow }N(0,I_{p+1}).
\end{equation*}%
The Berry-Esseen bound provides guidance if $p$ increases with $n$ (so long
as $S_{\mathbf{G,X}}$ and $S_{\sigma ^{2}\mathbf{G,G}}^{-1}$ are well
behaved as $p$ increases). \ \ 

\begin{remark}
Assume $S_{\mathbf{X,X}}=\frac{1}{n}\sum_{j=1}^{n}\mathbf{X}_{j}\mathbf{X}%
_{j}^{\mathtt{T}}$ is non-singular. Then Theorem \ref{thm:conditional}
corresponds to the classical $\mathrm{E}[\widehat{\mathbf{\beta }}_{LS}|(%
\mathbf{Z}=\mathbf{z})]=\mathbf{\beta }$ and\textbf{\ } 
\begin{equation*}
\mathrm{Var}(\widehat{\mathbf{\beta }}_{LS}|(\mathbf{Z}=\mathbf{z}))=\frac{1%
}{n}\Xi _{n},\quad \Xi _{n}=S_{\mathbf{X,X}}^{-1}S_{\sigma ^{2}\mathbf{X,X}%
}S_{\mathbf{X,X}}^{-1},
\end{equation*}%
where $S_{\sigma ^{2}\mathbf{X,X}}=\frac{1}{n}\sum_{j=1}^{n}\sigma _{j}^{2}%
\mathbf{X}_{j}\mathbf{X}_{j}^{\mathtt{T}}$. Assume $S_{\sigma ^{2}\mathbf{X,X%
}}$\ is positive definite. \ The corresponding Berry-Esseen bound is 
\begin{eqnarray*}
&&\sup_{A\in \mathcal{C}_{p+1}}\left\vert \Pr (\sqrt{n}\left( \widehat{%
\mathbf{\beta }}_{LS}-\mathbf{\beta }\right) \in A|(\mathbf{Z}=\mathbf{z}%
))-\Pr (N(0,\Xi _{n})\in A|(\mathbf{Z}=\mathbf{z}))\right\vert \\
&\leq &c_{LS}\frac{(p+1)^{1/4}}{n^{1/2}}\left( n^{-1}\sum_{j=1}^{n}\varsigma
_{LS,j}\right) ,
\end{eqnarray*}%
where $\mathcal{C}_{p+1}$ denotes the set of all convex subsets of $R^{p+1}$
and 
\begin{equation*}
\varsigma _{LS,j}=\mathrm{E}[|U_{j}|^{3}|(\mathbf{Z}=\mathbf{z})]\left\Vert
S_{\sigma ^{2}\mathbf{X,X}}^{-1/2}\mathbf{X}_{j}\right\Vert _{2}^{3}.
\end{equation*}%
Finally, 
\begin{equation*}
\sqrt{n}\Xi _{n}^{-1/2}\left( \widehat{\mathbf{\beta }}_{LS}-\mathbf{\beta }%
\right) |(\mathbf{Z}=\mathbf{z})\overset{d}{\rightarrow }N(0,I_{p+1}).
\end{equation*}
\end{remark}

\begin{example}
(Continuing Example \ref{Ex:p is one}) Recall $p=1$ then%
\begin{equation*}
\mathrm{E}[\widehat{\beta }_{1}]=\beta _{1},\quad \mathrm{Var}(\widehat{%
\beta }_{1})=\frac{\sum_{j=1}^{n}\frac{\left( Z_{j}-\widetilde{Z}\right)
^{2}\sigma _{j}^{2}}{1+\left( Z_{j}-\overline{Z}\right) ^{2}}}{\left[
\sum_{j=1}^{n}\frac{\left( Z_{j}-\widetilde{Z}\right) ^{2}}{\sqrt{1+\left(
Z_{j}-\overline{Z}\right) ^{2}}}\right] ^{2}},\quad \mathrm{E}[\widehat{%
\beta }_{1,LS}]=\gamma ,\quad \mathrm{Var}(\widehat{\beta }_{1,LS})=\frac{%
\sum_{j=1}^{n}\left( Z_{j}-\overline{Z}\right) ^{2}\sigma _{j}^{2}}{\left[
\sum_{j=1}^{n}\left( Z_{j}-\overline{Z}\right) ^{2}\right] ^{2}},
\end{equation*}%
while 
\begin{equation*}
\mathrm{E}[\widehat{\beta }_{0}]=\beta _{0},\quad \mathrm{Var}(\widehat{%
\beta }_{0})=\sum_{j=1}^{n}\sigma _{j}^{2}\lambda _{j}^{2},\quad \mathrm{E}[%
\widehat{\beta }_{0,LS}]=\beta _{0},\quad \mathrm{Var}(\widehat{\beta }%
_{0,LS})=\sum_{j=1}^{n}\sigma _{j}^{2}n^{-2},
\end{equation*}%
where 
\begin{equation*}
\lambda _{j}=w_{j}-\left( \widetilde{Z}-\overline{Z}\right) w_{j}^{\prime
},\quad w_{j}=\frac{\frac{1}{\sqrt{1+\left( Z_{j}-\overline{Z}\right) ^{2}}}%
}{\sum_{j=1}^{n}\frac{1}{\sqrt{1+\left( Z_{j}-\overline{Z}\right) ^{2}}}}%
,\quad w_{j}^{\prime }=\frac{\frac{\left( Z_{j}-\widetilde{Z}\right) }{\sqrt{%
1+\left( Z_{j}-\overline{Z}\right) ^{2}}}}{\sum_{j=1}^{n}\frac{\left( Z_{j}-%
\widetilde{Z}\right) ^{2}}{\sqrt{1+\left( Z_{j}-\overline{Z}\right) ^{2}}}}.
\end{equation*}
\end{example}

\subsection{Unconditional inference\label{sect:uncondition}}

The corresponding result for unconditional inference is stated in Theorem %
\ref{thm:uncond}. \ The proof is a straightforward application of the usual
limit theory for the method of moments, thinking of the problem as a type of
two step estimation problem (e.g. \cite{NeweyMcFadden(94)}). \ 

\begin{theorem}
\label{thm:uncond}Assume $\mathbf{\theta =(\psi }^{\mathtt{T}}\mathbf{,\beta 
}^{\mathtt{T}}\mathbf{)}^{\mathtt{T}}\in \Theta $, a compact parameters
space. \ Assume that the pairs $(\mathbf{Z}_{1},Y_{1}),...,(\mathbf{Z}%
_{n},Y_{n})$ are i.i.d. obeying A1-A3 and write $\mathbf{\theta }_{0}$ as
the true values under this sampling. Writing $U_{1}=Y_{1}-\mathbf{X}_{1}^{%
\mathtt{T}}\mathbf{\beta }_{0}$, $\mathbf{X}_{1}=\mathbf{Z}_{1}-\mathbf{\psi 
}_{0}$ and $\mathbf{G}_{1}=\left\Vert \mathbf{X}_{1}\right\Vert _{2}^{-1}%
\mathbf{X}_{1}$, then as $n\rightarrow \infty $, so%
\begin{equation*}
\sqrt{n}\left( \widehat{\mathbf{\beta }}-\mathbf{\beta }_{0}\right) \overset{%
d}{\rightarrow }N(0,\left\{ \mathrm{E}[\mathbf{G}_{1}\mathbf{X}_{1}^{\mathtt{%
T}}\mathbf{]}\right\} ^{-1}\mathrm{E}(\sigma _{1}^{2}\mathbf{G}_{1}\mathbf{G}%
_{1}^{\mathtt{T}})\left\{ \mathrm{E}[\mathbf{G}_{1}\mathbf{X}_{1}^{\mathtt{T}%
}\mathbf{]}\right\} ^{-1}),
\end{equation*}%
assuming $\mathrm{E}[\sigma _{1}^{2}]<\infty $, where $\sigma _{1}^{2}=%
\mathrm{Var}(Y_{1}|\mathbf{Z}_{1})$.\ 
\end{theorem}

\textbf{Proof.} Given in the Appendix.

Section \ref{sect:identify} showed that the existence of $\mathrm{E}[\sigma
_{1}^{2}]<\infty $ and $\mathrm{E}[\mathbf{X}_{1}]$ is a sufficient
condition for $\mathrm{E}[\mathbf{G}_{1}\mathbf{X}_{1}^{\mathtt{T}}\mathbf{]}
$ and $\mathrm{E}(\sigma _{1}^{2}\mathbf{G}_{1}\mathbf{G}_{1}^{\mathtt{T}})$
to both exist. \ Hence the central limit theory for $\widehat{\mathbf{\beta }%
}$ can hold in cases where the variance of $Y_{1}$ and the variance of $%
\mathbf{X}_{1}$ do not exist. What is needed is the existence of the
conditional variance of the outcomes given the predictors. \ To remove the
conditional variance assumption a switch in estimand is needed, e.g. to one
based on quantiles. \ This will be discussed in Section \ref{sect:quantile}.

\begin{remark}
The result above compares to the classical 
\begin{equation*}
\sqrt{n}\left( \widehat{\mathbf{\beta }}_{LS}-\mathbf{\beta }\right) \overset%
{d}{\rightarrow }N\left( 0,\left\{ \mathrm{E}[\mathbf{X}_{1}\mathbf{X}_{1}^{%
\mathtt{T}}]\right\} ^{-1}\mathrm{E}[\sigma _{1}^{2}\mathbf{X}_{1}\mathbf{X}%
_{1}^{\mathtt{T}}]\left\{ \mathrm{E}[\mathbf{X}_{1}\mathbf{X}_{1}^{\mathtt{T}%
}]\right\} ^{-1}\right) ,
\end{equation*}%
assuming $\mathrm{E}[\mathbf{X}_{1}\mathbf{X}_{1}^{\mathtt{T}}]$ and $%
\mathrm{E}[\sigma _{1}^{2}\mathbf{X}_{1}\mathbf{X}_{1}^{\mathtt{T}}]$ exist
and $\mathrm{E}[\mathbf{X}_{1}\mathbf{X}_{1}^{\mathtt{T}}]$ is invertible.
\end{remark}

\begin{example}
(Continuing Example \ref{Ex:p is one}) When $p=1$, then write $\widetilde{%
\mathrm{E}}[Z_{1}]=\frac{\mathrm{E}\left[ \left\Vert \mathbf{X}%
_{1}\right\Vert _{2}^{-1}Z_{1}\right] }{\mathrm{E}\left[ \left\Vert \mathbf{X%
}_{1}\right\Vert _{2}^{-1}\right] }$, recalling $\left\Vert \mathbf{X}%
_{1}\right\Vert _{2}=\sqrt{1+(Z_{1}-\mathrm{E}[Z_{1}])^{2}}$, so 
\begin{equation*}
Avar(\widehat{\beta }_{1})=\frac{1}{n}\frac{\mathrm{E}\left[ \sigma _{1}^{2}%
\frac{(Z_{1}-\widetilde{\mathrm{E}}[Z_{1}])^{2}}{1+(Z_{1}-\mathrm{E}%
[Z_{1}])^{2}}\right] }{\mathrm{E}\left[ \frac{(Z_{1}-\widetilde{\mathrm{E}}%
[Z_{1}])^{2}}{\sqrt{1+(Z_{1}-\mathrm{E}[Z_{1}])^{2}}}\right] ^{2}},\quad
Avar(\widehat{\beta }_{1,LS})=\frac{1}{n}\frac{\mathrm{E}\left[ \sigma
_{1}^{2}(Z_{1}-\mathrm{E}[Z_{1}])^{2}\right] }{\mathrm{E}\left[ (Z_{1}-%
\mathrm{E}[Z_{1}])^{2}\right] ^{2}}.
\end{equation*}%
\ \ 
\end{example}

\begin{remark}
The i.i.d. assumption on the sequence of pairs $(\mathbf{Z}_{1},Y_{1}),(%
\mathbf{Z}_{2},Y_{2}),...,(\mathbf{Z}_{n},Y_{n})$ in Theorem \ref{thm:uncond}
is not what drives the result. \ That assumption can be replaced by assuming
the sequence is a martingale difference with respect to the sequence's
natural filtration. \ \ \ 
\end{remark}

\subsection{Estimating the standard errors\label{sect:ses}}

In practice estimating $\mathrm{E}[\sigma _{1}^{2}\mathbf{G}_{1}\mathbf{G}%
_{1}^{\mathtt{T}}]$ or $\mathrm{E}[\sigma _{1}^{2}\mathbf{X}_{1}\mathbf{X}%
_{1}^{\mathtt{T}}]$ is delicate. \ The estimation challenge for thick tailed
predictors has been understated in the applied finance literature. \ 

Focus on the scalar predictor case with no intercept, so $\mathbf{\beta }%
=\beta _{1}$, to concentrate on the important ideas. Then $\mathbf{G}_{j}=1$%
.\ The extension to the general case is immediate. \ Then 
\begin{equation*}
\widehat{U}_{j}=Y_{j}-X_{j}\widehat{\beta }=\left( Y-X_{j}\beta _{1}\right) 
\mathbf{-}X_{j}\left( \widehat{\beta }_{1}-\beta _{1}\right)
=U_{j}-X_{j}\left( \widehat{\beta }-\beta \right) ,
\end{equation*}%
so define 
\begin{eqnarray*}
\widehat{\mathrm{Var}(U_{1})} &=&\frac{1}{n}\sum_{j=1}^{n}\widehat{U}_{j}^{2}
\\
&=&\frac{1}{n}\sum_{j=1}^{n}U_{j}^{2}+\left( \widehat{\beta }_{1}-\beta
_{1}\right) ^{2}\frac{1}{n}\sum_{j=1}^{n}X_{j}^{2}-2\left( \widehat{\beta }%
_{1}-\beta _{1}\right) \frac{1}{n}\sum_{j=1}^{n}U_{j}X_{j},
\end{eqnarray*}%
as an estimator of $\mathrm{Var}(U_{1})$. \ Then, $\frac{1}{n}%
\sum_{j=1}^{n}U_{j}^{2}$ converges to $\mathrm{Var}(U_{1})$ using the strong
law of large numbers as $\mathrm{E}[U_{1}]=0$. What happens to the two other
terms in the above expression? \ 

Now, if $\mathrm{E}[X_{1}^{2}]<\infty $, under the conditions of Theorem \ref%
{thm:uncond}, 
\begin{equation*}
\left\{ \sqrt{n}\left( \widehat{\beta }_{1}-\beta _{1}\right) \right\} ^{2}%
\frac{1}{n}\sum_{j=1}^{n}X_{j}^{2}\overset{d}{\rightarrow }\mathrm{E}%
X_{1}^{2}\frac{\mathrm{Var}(U_{1})}{\left\{ \mathrm{E}|X_{1}|\right\} ^{2}}%
\chi _{1}^{2},
\end{equation*}%
while 
\begin{equation*}
\sqrt{n}\left( \widehat{\beta }_{1}-\beta _{1}\right) \left( \frac{1}{n}%
\sum_{j=1}^{n}U_{j}X_{j}\right) \overset{p}{\rightarrow }0,
\end{equation*}%
as long as $\mathrm{E}[U_{1}X_{1}]$ exists (in which case $\mathrm{E}%
[U_{1}X_{1}]=0$). Then, under the conditions of Theorem \ref{thm:uncond}, 
\begin{equation*}
\widehat{\mathrm{Var}(U_{1})}\overset{p}{\rightarrow }\mathrm{E}[\sigma
^{2}(X_{1})]=\mathrm{Var}(U_{1}).
\end{equation*}%
\ \ \ \ 

But what happens if $\mathrm{E}[X_{1}^{2}]$ does not exists? \ The CLT for $%
\sqrt{n}\left( \widehat{\beta }_{1}-\beta _{1}\right) $ does not change and $%
\frac{1}{n}\sum_{j=1}^{n}U_{j}^{2}$ is well behaved. \ However, trouble
brews in the terms $\frac{1}{n}\sum_{j=1}^{n}X_{j}^{2}$ and $\frac{1}{n}%
\sum_{j=1}^{n}U_{j}X_{j}$. \ In our simulations, $\widehat{\mathrm{Var}%
(U_{1})}$ becomes inadequate if $\mathrm{E}[X_{1}^{2}]$ ceases to exist. \ 

To entirely circumvent this problems, I use a weighted estimator, where the
weights will be denoted $w(x)$, 
\begin{eqnarray*}
\widetilde{\mathrm{Var}(U_{1})} &=&\frac{1}{n}\sum_{j=1}^{n}\left\{ \widehat{%
U}_{j}w(X_{j})\right\} ^{2},\quad w(x)=1_{\frac{|x|}{\mathrm{E}|X_{1}|}%
<dn^{1/5}} \\
&=&\frac{1}{n}\sum_{j=1}^{n}U_{j}^{2}w(X_{j})+\left( \widehat{\beta }%
_{1}-\beta _{1}\right) ^{2}\frac{1}{n}\sum_{j=1}^{n}X_{j}^{2}w(X_{j}) \\
&&-2\left( \widehat{\beta }_{1}-\beta _{1}\right) \frac{1}{n}%
\sum_{j=1}^{n}U_{j}X_{j}w(X_{j}).
\end{eqnarray*}%
This clips regression residuals associated with large predictors. \ Then,%
\begin{equation*}
\begin{array}{lll}
\frac{1}{n}\sum_{j=1}^{n}U_{j}^{2}1_{|X_{j}|<dn^{1/5}\mathrm{E}|X_{1}|} & 
\overset{p}{\rightarrow } & \mathrm{E}[\sigma ^{2}(X_{1})] \\ 
\frac{1}{n}\sum_{j=1}^{n}X_{j}^{2}1_{\frac{|x_{j}|}{\mathrm{E}|X_{1}|}%
<dn^{1/5}} & \leq & d^{2}n^{2/5}\left( \mathrm{E}|X_{1}|\right) ^{2}, \\ 
\left\vert \frac{1}{n}\sum_{j=1}^{n}U_{j}X_{j}w(X_{j})\right\vert & \leq & 
dn^{1/5}\mathrm{E}|X_{1}|\frac{1}{n}\sum_{j=1}^{n}\left\vert
U_{j}\right\vert .%
\end{array}%
\end{equation*}%
As $\sqrt{n}\left( \widehat{\beta }_{1}-\beta _{1}\right) =O_{p}(1)$, these
three results imply that 
\begin{equation*}
\widetilde{\mathrm{Var}(U_{1})}\overset{p}{\rightarrow }\mathrm{E}[\sigma
^{2}(X_{1})],
\end{equation*}%
even if $\mathrm{E}[X_{1}^{2}]$ does not exist. \ 

In our simulation and empirical work we take $d=10$. \ If $n=100$, the
scaled threshold is $dn^{1/5}\simeq 16$. \ This implies the weight function
will not clip any regression residual unless the associated predictor is
extraordinarily unusual. \ For thin tailed predictors, clipping is neither
helpful or harmful. \ For thick tailed predictors, it is deeply important. \
As researchers usually do not know the tail behavior of their predictors,
the safe approach is to always include the weighting function. \ 

More broadly, the same line of argument applies to 
\begin{equation*}
\frac{1}{n}\sum_{j=1}^{n}\left\{ \widehat{U}_{j}w(\mathbf{X}_{j})\right\}
^{2}\mathbf{G}_{j}\mathbf{G}_{j}^{\mathtt{T}}\overset{p}{\rightarrow }%
\mathrm{E}(\sigma _{1}^{2}\mathbf{G}_{1}\mathbf{G}_{1}^{\mathtt{T}}),
\end{equation*}%
where now $w(\mathbf{x})=1_{\left\{ \max_{i}\frac{|x_{i}|}{\mathrm{E}%
|X_{1,i}|}\right\} <dn^{1/5}}$, so long as $\mathrm{E}[\sigma
_{1}^{2}]<\infty $. More straightforwardly, by the strong law of large
numbers 
\begin{equation*}
\frac{1}{n}\sum_{j=1}^{n}\mathbf{G}_{j}\mathbf{X}_{j}^{\mathtt{T}}\overset{p}%
{\rightarrow }\mathrm{E}(\mathbf{G}_{1}\mathbf{X}_{1}^{\mathtt{T}}),
\end{equation*}%
so long as $\mathrm{E}(\mathbf{G}_{1}\mathbf{X}_{1}^{\mathtt{T}})$ exists --
which is true so long as $\mathrm{E}|\mathbf{X}_{1}|$ exists. \ 

Thus asymptotically valid estimators of the standard errors can be computed
without needing any more assumptions about higher order moments. \ 

As mentioned in the introduction, for least squares, their robust standard
errors need at least the fourth moments of the predictors to be valid. \
This is spelt out in Assumption 4 of \cite{White(80)}.

\begin{example}
(Continuing from Example \ref{Ex:p is one}) When $p=1$ then 
\begin{equation*}
\widehat{\mathrm{Var}(\widehat{\beta }_{1})}=\frac{\sum_{j=1}^{n}W_{j}\frac{%
\left( Z_{j}-\widetilde{Z}\right) ^{2}\widehat{U}_{j}^{2}}{1+\left( Z_{j}-%
\overline{Z}\right) ^{2}}}{\left[ \sum_{j=1}^{n}\frac{\left( Z_{j}-%
\widetilde{Z}\right) ^{2}}{\sqrt{1+\left( Z_{j}-\overline{Z}\right) ^{2}}}%
\right] ^{2}},\quad \widehat{\mathrm{Var}(\widehat{\beta }_{1,LS})}=\frac{%
\sum_{j=1}^{n}\left( Z_{j}-\overline{Z}\right) ^{2}\widehat{U}_{j,LS}^{2}}{%
\left[ \sum_{j=1}^{n}\left( Z_{j}-\overline{Z}\right) ^{2}\right] ^{2}},
\end{equation*}%
where $\widehat{U}_{j}=\left( Y_{j}-\widetilde{Y}\right) -\widehat{\gamma }%
\left( X_{j}-\widetilde{X}\right) $, $\widehat{U}_{j,LS}=\left( Y_{j}-%
\overline{Y}\right) -\widehat{\gamma }_{LS}\left( X_{j}-\overline{X}\right) $
and $W_{j}=$ $1_{\frac{|X_{i}-\overline{X}|}{\overline{|X-\overline{X}|}}%
<dn^{1/5}}$.
\end{example}

\section{Simulation experiment\label{sect:simulation exper}}

In this section I focus on the performance of the approximate pivots 
\begin{equation*}
\widehat{T}_{\widehat{\beta }}=\frac{\widehat{\beta }_{1}-\beta _{1}}{\sqrt{%
\widehat{\mathrm{Var}(\widehat{\beta }_{1})}}},\quad \text{and}\quad 
\widehat{T}_{LS}=\frac{\widehat{\beta }_{1,LS}-\beta _{1}}{\sqrt{\widehat{%
\mathrm{Var}(\widehat{\beta }_{1,LS})}}}
\end{equation*}%
in the case with an intercept and one predictor, so $\beta =(\beta
_{0},\beta _{1})^{T}$, where the weight function for $\widehat{T}_{\widehat{%
\beta }}$ is $w(x)=1_{\left\vert x\right\vert <\widehat{c}10n^{0.2}}$, $%
\widehat{c}=\frac{1}{n}\sum_{j=1}^{n}|X_{j}-\overline{X}|$. The truncation
in $w(x)$ has no impact except for the very extreme cases we discuss in
Section \ref{sect:no var sim}, which studies $\widehat{T}_{\widehat{\beta }}$
in the case where $\mathrm{Var}(X_{1})=\infty $. \ Recall, theory suggests
that weighting is needed in that case. \ \ 

The simulation design I initially use is based around regressing stock
returns on a broad based index, to estimate a \textquotedblleft
beta\textquotedblright . \ The design mimics the empirical challenge tackled
in the next section. \ That challenge looks at 2 years of weekly percentage
arithmetic returns on a major U.S. company, $Y_{j}$, and $X_{j}$ will be the
S\&P500 index arithmetic returns. \ In the empirical work this will be
implemented for more than 400 individual companies, using hypothesis tests
based on $\widehat{T}_{\widehat{\beta }}$ and $\widehat{T}_{LS}$ to identify
stocks with very high or very low betas. \ \ 

\subsection{Initial experiment}

Assume 
\begin{equation*}
X_{j}\overset{indep}{\sim }\psi +\sigma _{X}\sqrt{\frac{\nu -2}{\nu }}%
V_{j},\quad V_{1}\sim t_{\nu },\quad \nu >2,\quad j=1,...,n,
\end{equation*}%
and 
\begin{equation*}
Y_{j}|\left( X_{j}=x\right) \overset{indep}{\sim }N(\beta _{0}+\beta
_{1}(x-\psi ),\sigma ^{2}).
\end{equation*}%
This implies that, for every value of $\nu $, the $\mathrm{E}[X_{1}]=\psi $, 
$\mathrm{E}[Y_{1}]=\beta _{0}$, and $\mathrm{Var}(X_{1})=\sigma
_{X}^{2}<\infty $. \ The simulations will have homoskedasticity, but the
approximate pivots will be computed without imposing that. Importantly if $%
\nu <4$\ the $\widehat{T}_{LS}$ is not asymptotically N(0,1), while $%
\widehat{T}_{\widehat{\beta }}$ will be. \ Hence $\widehat{T}_{LS}$ is
expected to have poor performance unless $\nu $ is substantially above 4. \
This is what you will see in the simulations. \ 

To calibrate this using universally available data, I look at weekly
arithmetic (total) returns on the SPDR S\&P 500 ETF Trust (SPY) from 1st
August 2018 to 4th August 2020, downloaded using the \texttt{R} package 
\texttt{Quantmod} from Yahoo's database. \ The \texttt{R} code for this
download, delivering a vector of weekly returns \textquotedblleft
X\textquotedblright\ is
\begin{verbatim}
getSymbols("SPY",from='2018-08-01',to='2020-08-04',verbose = TRUE); head(SPY);
XdataM = (to.weekly(SPY))$SPY.Adjusted
X = data.matrix(100*diff(XdataM)/lag(XdataM))[-1];
\end{verbatim}

The sample mean and standard deviation suggest taking $\sigma _{X}=3.24$, $%
\psi =0.21$, and the ML estimator of $\nu $, computed using \texttt{R}'s
function \texttt{fitdistr} for the student-t distribution, is 2.16 with a
standard error of around 0.5. This is not an unusual result --- typical
extreme value theory estimates of the tail index of equity indexes suggest 2
but not 4 moments exist. \ \ \ \ 

To nudge towards safer grounds for least squares, in the simulations we will
use $\nu =2.4$, a little larger than I saw in the data. \ Later, we will
explore many different values of $\nu $.\ \ 

The initial focus is on the case where $n=100$, $\psi =0.21$, $\beta _{0}=0$%
, $\nu =2.4$, $\sigma =2$, $\beta _{1}=1$ and $\sigma _{X}=3.24$. In this
case, the predictors have 2 but not 3 moments. \ 

I will initially summarize results using QQ plots, based on 3,000
replications, comparing the simulated quantiles of $\widehat{T}_{\widehat{%
\beta }}$ and $\widehat{T}_{LS}$ to quantiles of their $N(0,1)$ baseline. \
\ 

\begin{figure}[tb]
\centering
\includegraphics[width=0.24\textwidth]{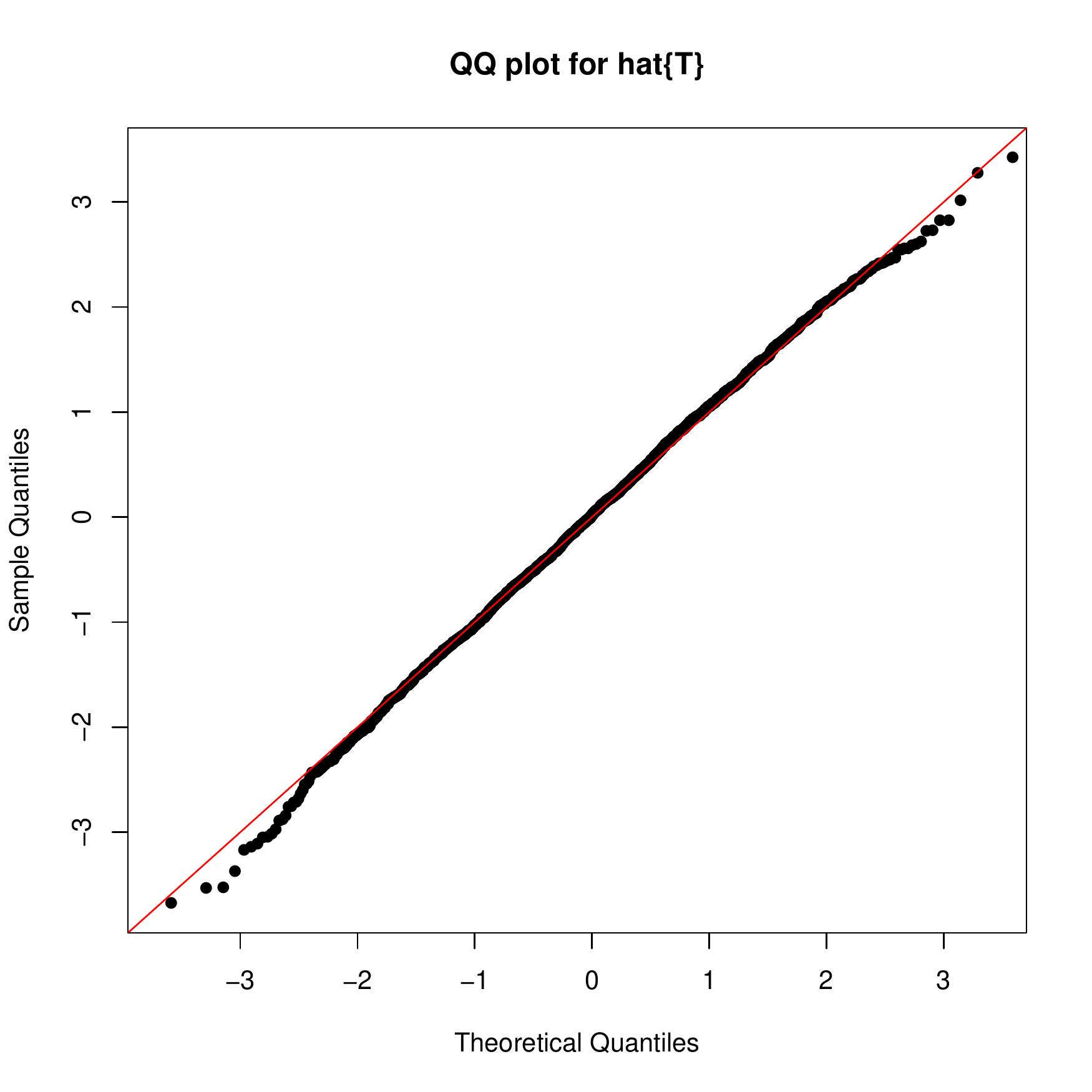} %
\includegraphics[width=0.24\textwidth]{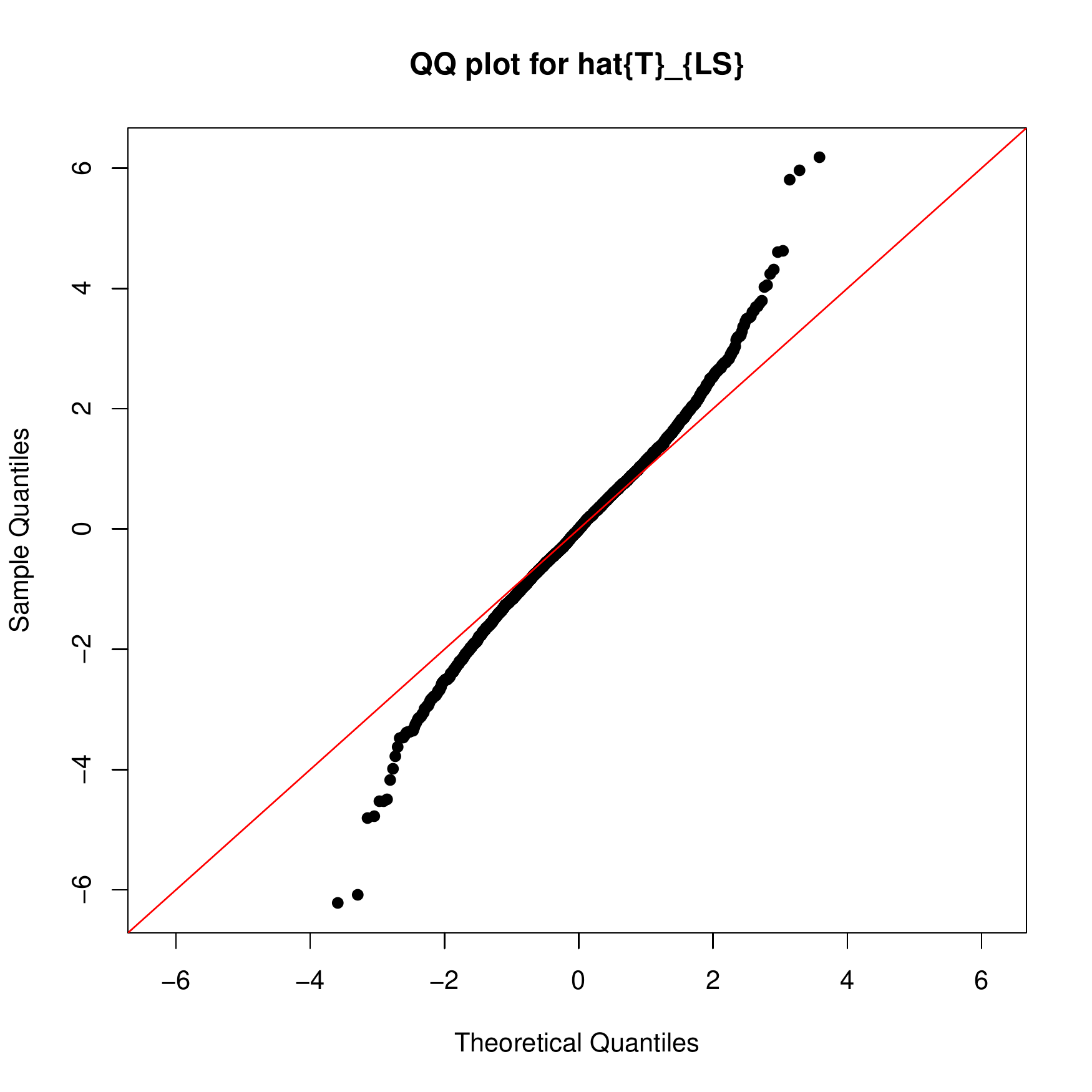} \vrule
\includegraphics[width=0.24\textwidth]{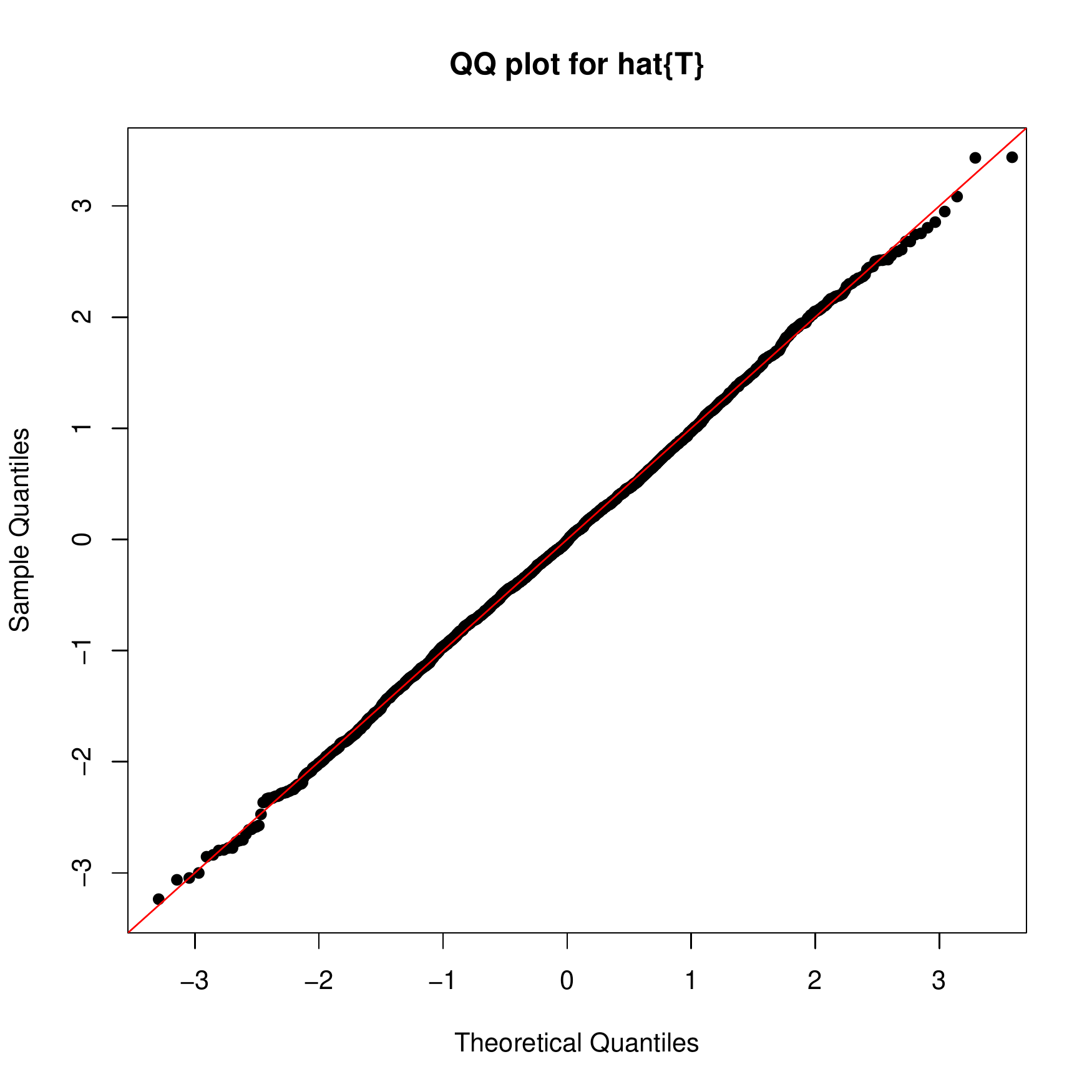} %
\includegraphics[width=0.24\textwidth]{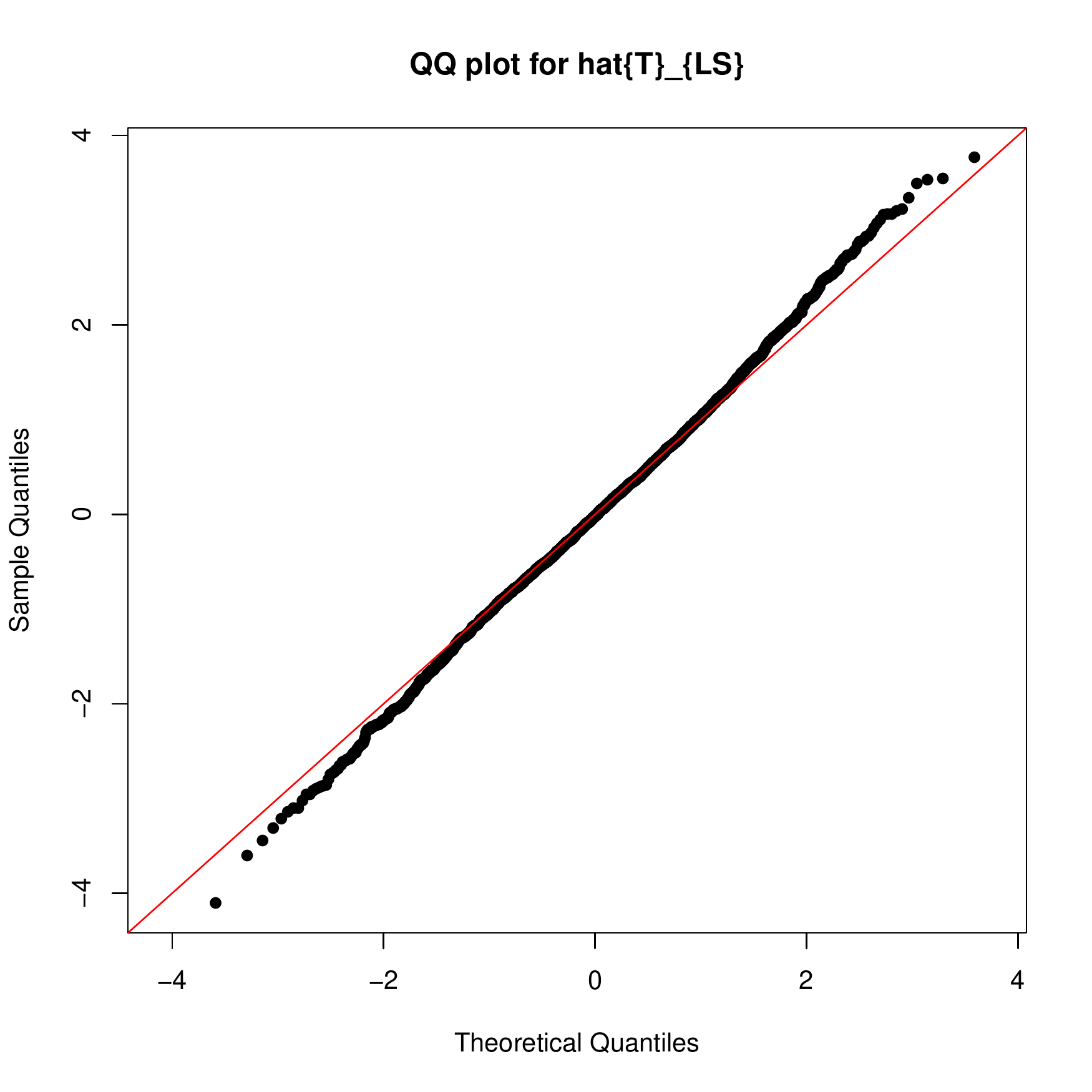}
\caption{QQ plots for $\protect\widehat{T}_{\protect\widehat{\protect\beta }%
} $ and $\protect\widehat{T}_{LS}$. Simulation designed to match empirical
data we see for weekly financial returns on U.S. stocks for major companies,
where the predictor is the main major index. The left hand side corresponds
to $\protect\nu =2.4$, the right hand side has $\protect\nu =4.4$.
Throughout, $n=100$.}
\label{fig:QQplots}
\end{figure}

The resulting QQ plots for $\widehat{T}_{\widehat{\beta }}$ and $\widehat{T}%
_{LS}$ are given in the left hand side of Figure \ref{fig:QQplots} for $\nu
=2.4$. The results for $\widehat{T}_{\widehat{\beta }}$ are strong, matching
the Gaussian limit theorem throughout except perhaps in the very extreme
tails. The results for $\widehat{T}_{LS}$ are terrible. 
Perhaps poor behavior for $\widehat{T}_{LS}$ is to be expected given $\nu $
is low. \ \ 

The right hand side of Figure \ref{fig:QQplots} gives the results for the
corresponding easier case of $\nu =4.4$. The performance of $\widehat{T}_{%
\widehat{\beta }}$ does not change very much, again providing very solid
results. \ The $\widehat{T}_{LS}$ is much better than in the heavier tail
case, no longer terrible, just poor. \ This $\nu =4.4$ case is a situation
where the limit theory for $\widehat{T}_{LS}$ is valid, it is just not very
accurate in practice. \ \ 

\subsection{More extensive experiments}

To compare performance in a wide set of diverse environments, I used the Cram%
\'{e}r-von-Mises statistic to measure how non-Gaussian $500,000$
replications of $\widehat{T}_{\widehat{\beta }}$ and $\widehat{T}_{LS}$
were, for a variety of values of $n$ and $\nu $, as well as $\sigma $. The
Cram\'{e}r-von-Mises test statistic is reviewed in \cite%
{BaringhouseHenze(17)}. \ It is viewed as one of the most powerful
distributional tests. \ To benchmark the values of the Cram\'{e}r-von-Mises
statistic of normality, the green horizontal lines are the 0.5, 0.95 and
0.99 quantiles of the distribution of the Cram\'{e}r-von-Mises test
statistic computed using 500,000 replications under the null of i.i.d. $%
N(0,1)$. \ The test is implemented in \texttt{R} using the function \texttt{%
cvm.test(data, "pnorm",mean=0,sd=1)}.

\begin{figure}[tb]
\centering\includegraphics[width=0.24\textwidth]{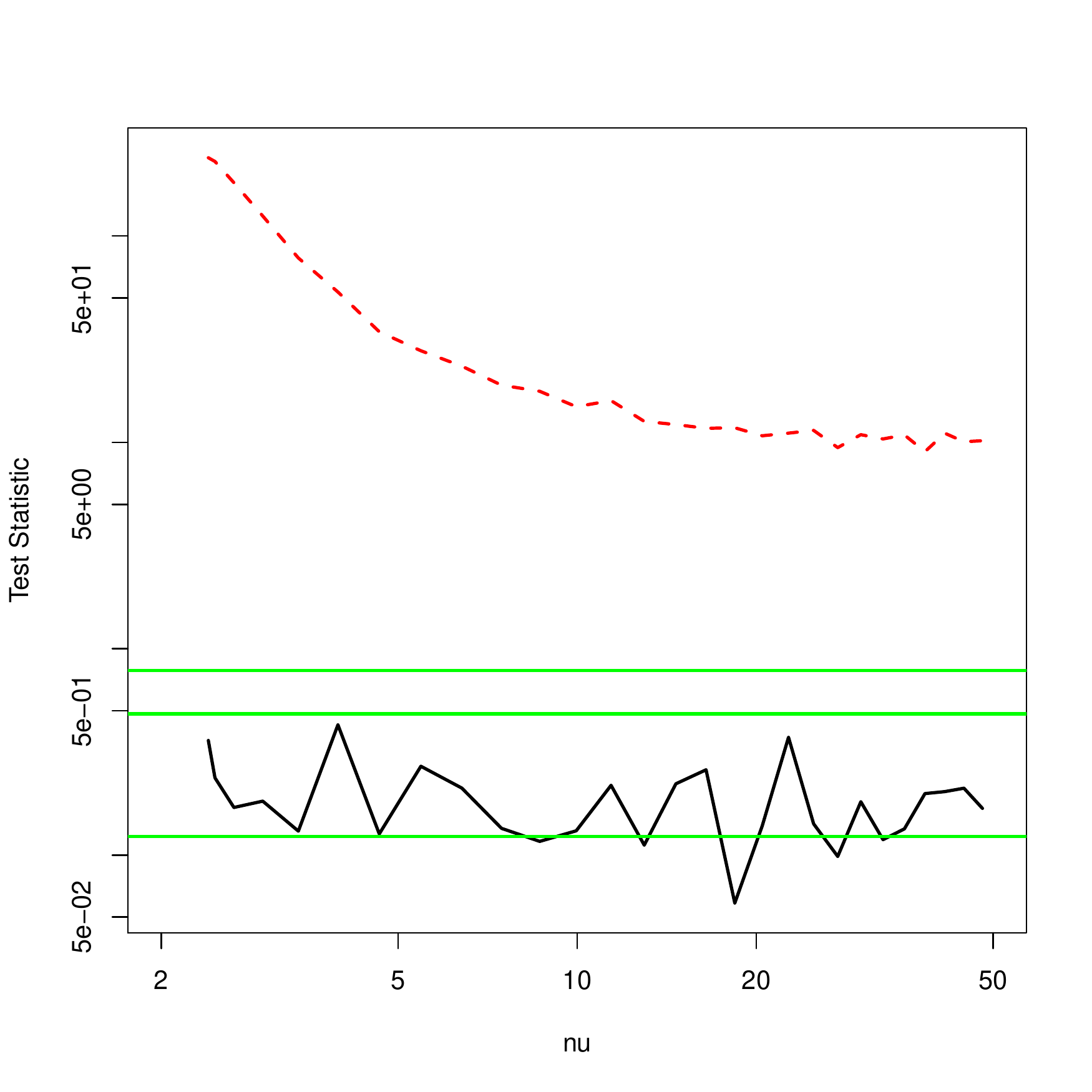} %
\includegraphics[width=0.24\textwidth]{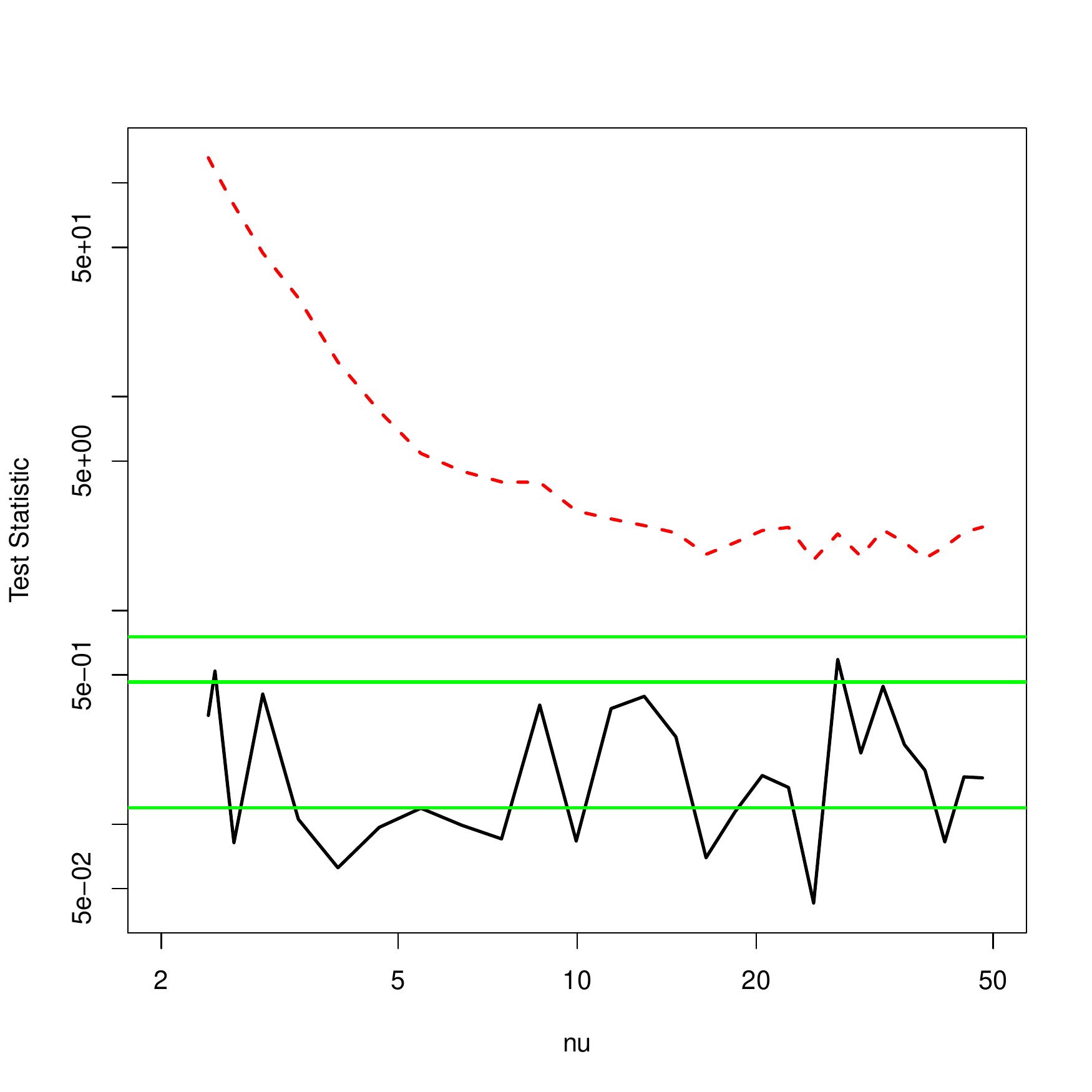} \vrule%
\includegraphics[width=0.24\textwidth]{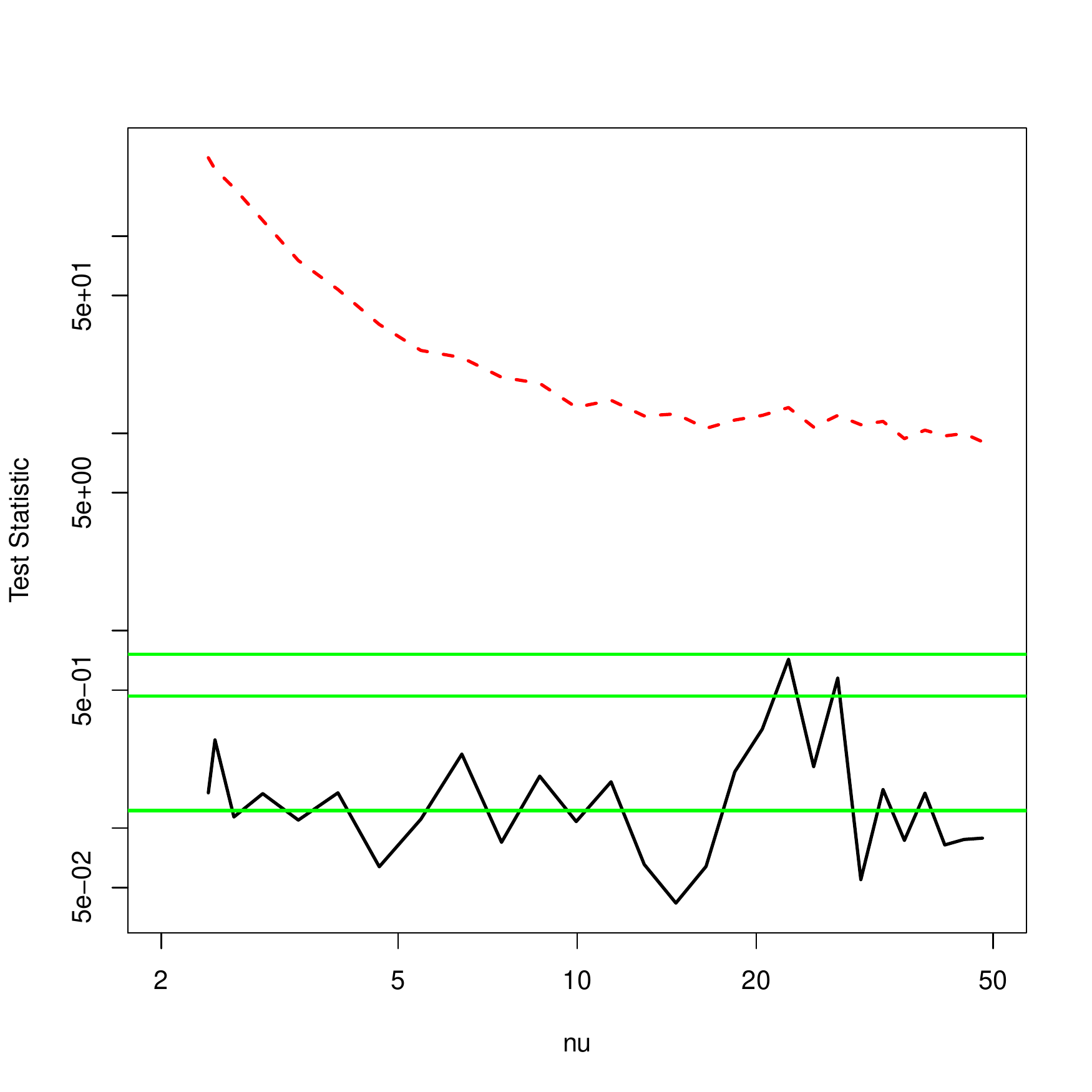} %
\includegraphics[width=0.24\textwidth]{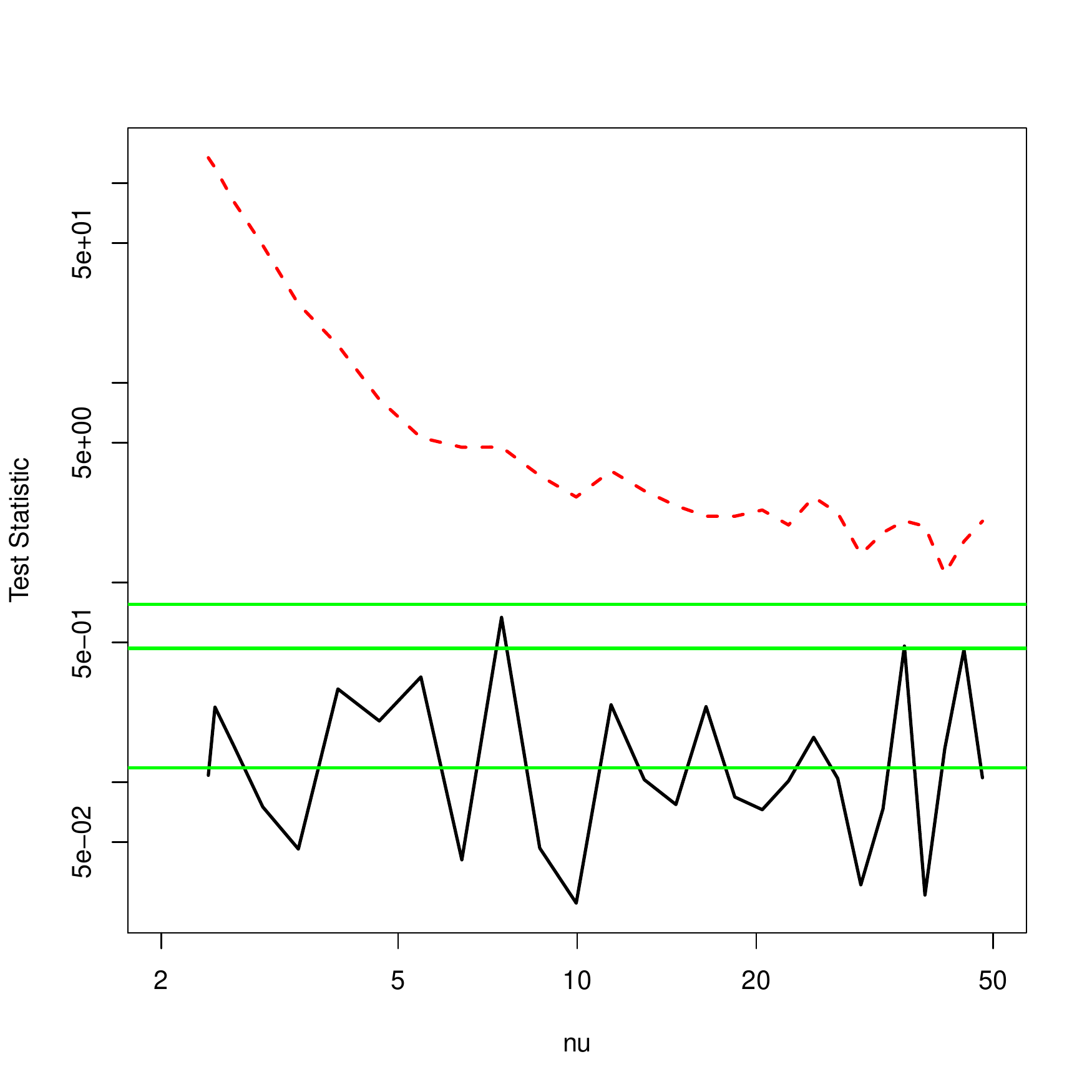}
\caption{Cram\'{e}r-von-Mises statistics test of N(0,1) for $\protect%
\widehat{T}_{\protect\widehat{\protect\beta }}$ (black line) and $\protect%
\widehat{T}_{LS}$ (dotted red line) for various values of $n$, $\protect\nu $
and $\protect\sigma $. \ Throughout $500,000$ replications are used. \ Green
horizontal lines represent the 0.5, 0.95 and 0.99 quantiles of the
Cramer-Von-Mises statistic when the data is i.i.d. $N(0,1)$. Results to the
left hand side have $\protect\sigma =2$, and results to the right have $%
\protect\sigma =4$. 1st and 3rd graph have $n=100$; 2nd and 4th have $n=250$%
. \ Throughout the $x$-axis is $\protect\nu $, on the log-scale. The $y$%
-axis is also drawn on the log-scale. \ }
\label{fig:loglogplot}
\end{figure}

The results are given in Figure \ref{fig:loglogplot}. \ Crucially, notice
that all plots use log-scales on both the $x$-axis and the $y$-axis. \ The
dotted red line is the result for $\widehat{T}_{LS}$, while the black line
is the corresponding results for $\widehat{T}_{\widehat{\beta }}$. \ The $x$%
-axis is the value of $\nu $; the $y$-axis is the Cram\'{e}r-von-Mises test
statistic. The 1st and 3rd graph have $n=100$, the 2nd and 4th have $n=250$.
\ The left hand side corresponds to $\sigma =2$, the right hand side $\sigma
=4$. \ 

The results for $\widehat{T}_{\widehat{\beta }}$ are encouraging. \ Even
with 500,000 replications it is typically not possible to reasonably reject
normality of $\widehat{T}_{\widehat{\beta }}$ even with $n=100$. \ When $\nu 
$ is at the bottom end of the plots, $\nu =2.4$ and $n=100$, there is some
evidence of a tiny amount of non-normality. \ The value of $\sigma $ does
not impact the results materially. \ As $n$ increases to 250 the results
improve, a little. \ 

For $\widehat{T}_{LS}$ the results uniformly reject normality, typically
dramatically. \ As $\nu $ increases the rejections become less significant,
as expected. \ As $n$ increases to 250 the results improve, but only by a
little.

\subsection{Pushing to the case where $\mathrm{Var}(X_{1})=\infty $\label%
{sect:no var sim}}

To assess the $N(0,1)$ approximation for $\widehat{T}_{\widehat{\beta }}$\
when $\mathrm{Var}(X_{1})=\infty $, I ran a separate experiment. \ This
experiment is less relevant to major equity data, although some commodity
market price moves have extraordinarily thick tails and it is interesting
theoretically. I excluded $\widehat{T}_{LS}$ from consideration, as the
previous experiment has shown it would have weak performance and the
asymptotics is far from being valid. \ \ 

It was less than clear to me how to generate predictors which are broadly
comparable in scale across difference values of $\nu $. \ I eventually
settled on 
\begin{equation*}
X_{j}\overset{indep}{\sim }0.21+\sigma _{X}V_{1}/\mathrm{E}\left\vert
V_{1}\right\vert ,\quad V_{1}\sim t_{\nu }
\end{equation*}%
which scales the student-t random variable so $\mathrm{E}|X_{1}-0.21|=\sigma
_{X}$ whatever the value of $\nu $. \ 

\begin{figure}[tb]
\centering\includegraphics[width=7cm]{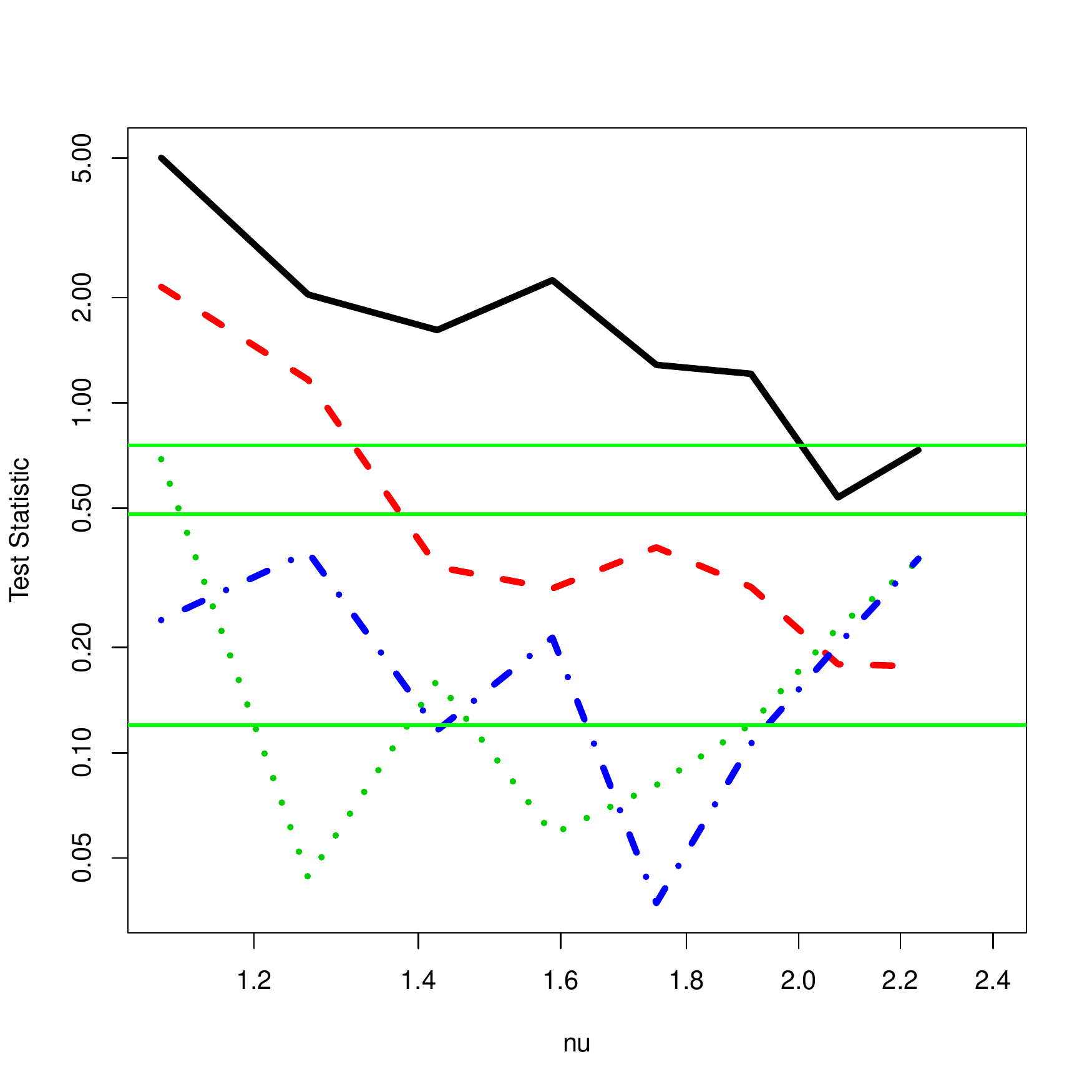}
\caption{Cram\'{e}r-von-Mises statistics test of $N(0,1)$ for $\protect%
\widehat{T}_{\protect\widehat{\protect\beta }}$ and various values of $n$
and $\protect\sigma $. \ 500,000 replications are used. \ Green horizontal
lines represent the 0.5, 0.95 and 0.99 quantiles of the Cramer-Von-Mises
statistic when the data is i.i.d. $N(0,1)$. Throughout the x-axis is $%
\protect\nu $. The y-axis is drawn on the log-scale. Sample size are: $n=100$
(black line), $250$ (red dotted line), $1,000$ (green dots) and $10,000$
(blue line).}
\label{fig:tails}
\end{figure}

Figure \ref{fig:tails} shows the results for $n=100$ (black line), $250$
(red dotted line), $1,000$ (green dots) and $10,000$ (blue line), here with $%
\sigma =2$ throughout. The Figures again plot the Cram\'{e}r-von-Mises test
statistic against $\nu $ based on $500,000$ replications. \ \ 
The results are encouraging, although not entirely positive. \ For small $n$
and small $\nu $, significant distortion is present. Samples of around 1,000
do deliver results for $\widehat{T}_{\widehat{\beta }}$ which are hard to
reject from a null of $N(0,1)$, even in cases where the predictors are just
slightly less thick tailed than Cauchy. \ 

Without the weighting function in $\widehat{T}_{\widehat{\beta }}$, these
simulations fall apart when $\mathrm{Var}(X_{1})=\infty $. \ In that
situation, I found no sign that increasing $n$ improves the behavior of $%
\widehat{T}_{\widehat{\beta }}$. \ This is in line with the suggestions from
the theory than once moments somewhat above the second do to not exist then
the weight function becomes useful and eventually essential if $\mathrm{Var}%
(X_{1})=\infty $. I recommend always including the weight function in
practice. It is harmless for thin tailed data and essential for very thick
tailed data. \ \ \ \ \ \

\section{Empirical work\label{sect:empirical}}

\subsection{Background}

Some investors, such as young workers saving into pensions who do not have a
mortgage, are unable to take on the level of financial leverage they may
rationally desire due to administrative rules or the inability to borrow
against their human capital. \ One viable investment strategy is to
overweight their portfolio with high beta stocks, that is, stocks which move
more strongly with the main market indexes than most stocks. This is
discussed in \cite{Black(72)}, amongst others. \ In finance betas are
usually measured by regressing the returns on the individual stock on a wide
market index. \ Such betas are used directly in vast numbers of empirical
papers and drive other methods such as \textquotedblleft Fama-MacBeth
regressions\textquotedblright . \ All of these empirical results are fragile
due to the thick tailed index returns. \ 

Typically high beta investments will have high risk in order to potentially
capture high expected returns. \ The opposite of this is attractive to a
different type of investor. \ Some investors search out low beta stocks,
hoping to have low risk, positive risk premium although relatively low
expected return. \ This is discussed in \cite{BakerBradleyWurgler(11)}, who
also review the extensive literature on this topic. \ 

\subsection{Selection by hypothesis test\ }

But how to select high beta stocks and low beta stocks? \ Once selected,
these groups of stocks could potentially be placed into portfolios or
packaged as low and high beta ETFs. \ 

In this Section selection will be regarded as a hypothesis testing problem.
The $s$-th stock will be labelled a high beta stock if we can reject the null%
\begin{equation*}
H_{0}:\beta _{s}\leq 1.4,\quad \text{against\quad }H_{1}:\beta _{s}>1.4,
\end{equation*}%
where $\beta _{s}$ is the beta of the $s$-th stock. \ 

I will label the $s$-th stock a low beta stock if the null 
\begin{equation*}
H_{0}:\beta _{s}\geq 0.8,\quad \text{against\quad }H_{1}:\beta _{s}<0.8,
\end{equation*}%
is rejected. \ 

The tests will be based on the approximate pivots 
\begin{equation*}
\widehat{T}_{\widehat{\beta }}\quad \text{and\quad }\widehat{T}_{LS}
\end{equation*}%
rejecting the nulls using a one-sided test with nominal size of 5\%. I will
compare the results for $\widehat{T}_{\widehat{\beta }}$ with the one based
on $\widehat{T}_{LS}$. \ 

\subsection{Data}

I downloaded the data from Yahoo for stock prices from 1 August 2018 to 4th
August 2020. \ The S\&P500 was measured using the SPDR S\&P 500 ETF Trust
(SPY). \ I converted these into weekly percentage arithmetic returns. \
These weekly returns will be compared to the returns on 416 individual
stocks, which are components of the S\&P500. \ The list of the stocks is
available in \texttt{RegFinance1.r}, which produces all the results given in
this Section. \ 

Why weekly returns? \ There are virtues in using higher frequency data than
weekly returns to estimate betas. \ In theory they can produce vastly more
precise estimates. \ High frequency versions of regression include \cite%
{BarndorffNielsenShephard(04multi)}, \cite%
{BarndorffNielsenHansenLundeShephard(11multi)} and \cite%
{BollerslevPattonLiQuaedvlieg(20)}. \ These sophisticated data hungry
methods try to overcome the impact of nonsynchronous trading and
differential rates of price discovery. \ However, they miss the impact of
overnight returns. \ Daily returns capture overnight effects, but have
significant lead-lag correlations. \ The hope with weekly returns is that
most of the impact of these dependencies will be averaged away or dwarfed by
other long-term effects. \ Many practitioners go further than this and use 2
to 5 years of monthly returns, but we do not follow that route. \ Further,
some of the volatility clustering seen in finance is taken out by using
weekly returns rather than high frequency returns.

\subsection{Cross-sectional results\ \ }

\begin{figure}[tb]
\centering\includegraphics[width=0.48\textwidth]{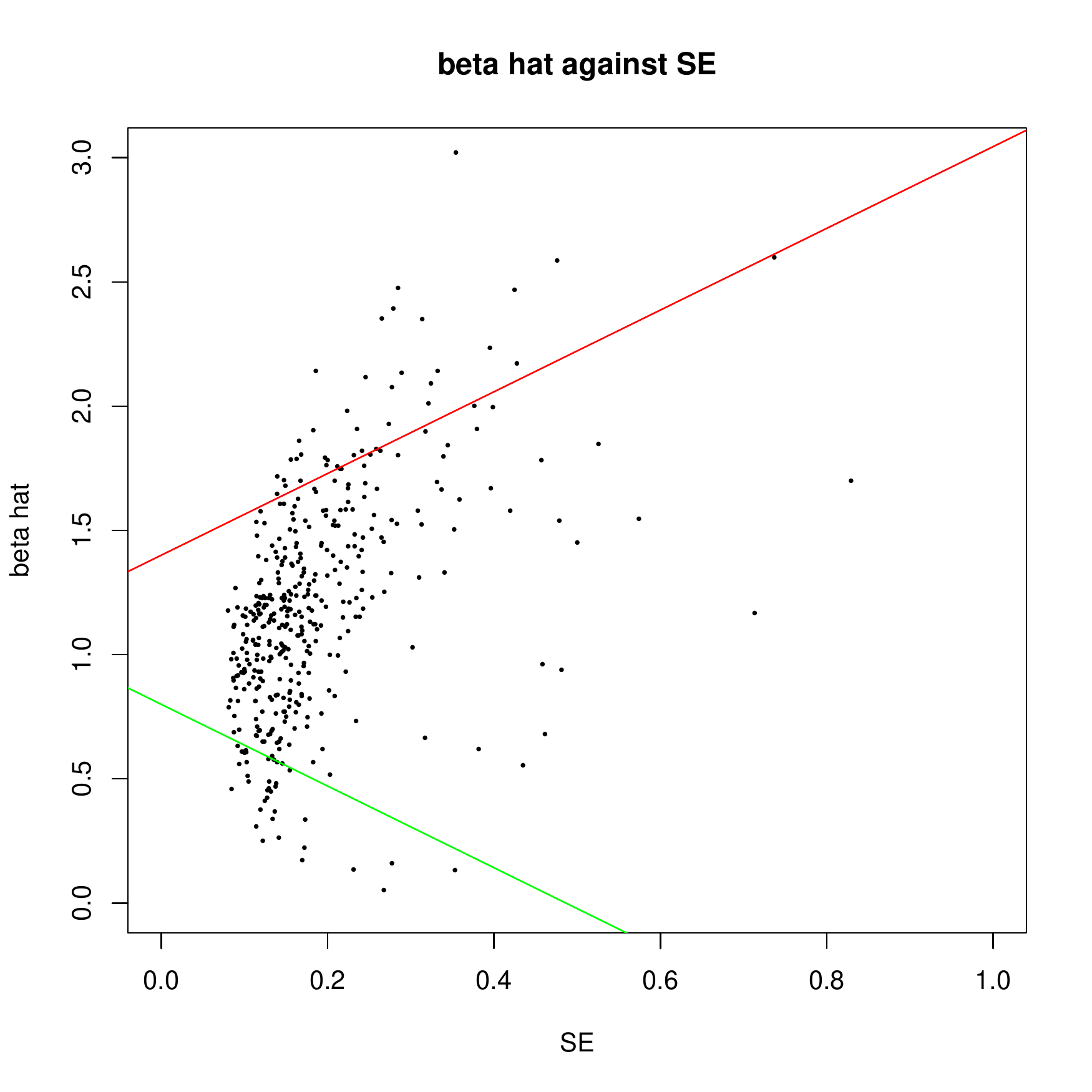}%
\includegraphics[width=0.48\textwidth]{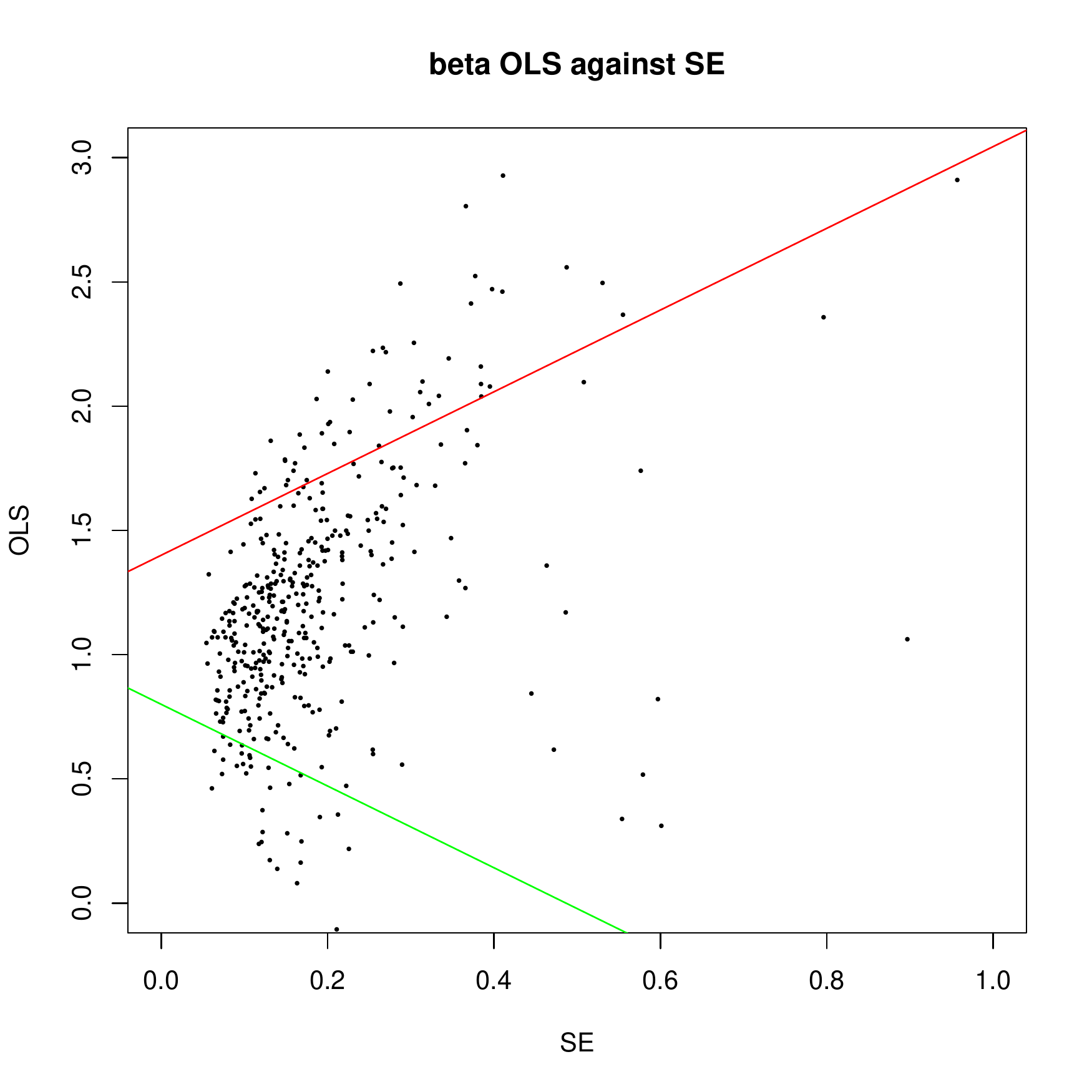}
\caption{The left hand side is the cross-section of $\protect\widehat{%
\protect\beta }$ against its standard error. \ The right hand side shows the
corresponding results for $\protect\widehat{\protect\beta }_{LS}$. The red
line has an intercept of 1.4 and slope of 1.64. \ Values above the red line
are labelled high beta stocks. \ The green line has an intercept of 0.8 and
slope of -1.64. \ Values below the green line are labelled low beta stocks.}
\label{fig:summaryFig}
\end{figure}

The right hand side of Figure \ref{fig:summaryFig} plots the cross-section
of $\hat{\beta}_{LS}$ against $SE(\hat{\beta}_{LS})$. \ The left hand side
gives the corresponding result for $\hat{\beta}$. \ The major impact of
moving from $\hat{\beta}_{LS}$ to $\hat{\beta}$ is that $\hat{\beta}_{LS}$
delivers some estimators with tiny standard errors, which is not the case
with $\hat{\beta}$. \ 
Indeed the smallest standard error for $\hat{\beta}_{LS}$ in the
cross-section is around 0.054, while the minimum for $SE(\hat{\beta})$ is
around 0.080 --- around 50\% higher. \ 

Table \ref{tab:betasummary} provides summary statistics on the cross-section
of $\hat{\beta}$ and $\hat{\beta}_{LS}$. The estimates have roughly the same
level, with roughly the same spread. The $\hat{\beta}$ are slightly lower
than the $\hat{\beta}_{LS}$ in this sample, across the distribution. \ Over
the entire cross-section $SE(\hat{\beta})$ is a little above $SE(\hat{\beta}%
_{LS})$, as we would expect, but the difference is very modest. This
suggests that a potential worry over $\hat{\beta}$ being generally much less
precise than $\hat{\beta}_{LS}$ is not compelling here. \ Table \ref%
{tab:betasummary} also details the cross-section of $\widehat{\alpha }$
against $\widehat{\alpha }_{LS}$. \ These are very similar, which is also
true of the $SE(\widehat{\alpha })$ and $SE(\widehat{\alpha }_{LS})$. 
\begin{table}[tbp]
\center%
\begin{tabular}{c|cc|cc||cc|cc}
& $\hat{\beta}$ & $\hat{\beta}_{LS}$ & $SE(\hat{\beta})$ & $SE(\hat{\beta}%
_{LS})$ & $\hat{\alpha}$ & $\hat{\alpha}_{LS}$ & $SE(\hat{\alpha})$ & $SE(%
\hat{\alpha}_{LS})$ \\ \hline
Mean & 1.17 & 1.21 & 0.184 & 0.184 & 0.067 & 0.100 & 0.402 & 0.420 \\ 
Q(0.1) & 0.619 & 0.644 & 0.103 & 0.087 & -0.548 & -0.427 & 0.240 & 0.247 \\ 
Q(0.5) & 1.16 & 1.17 & 0.155 & 0.152 & 0.164 & 0.148 & 0.347 & 0.359 \\ 
Q(0.9) & 1.76 & 1.83 & 0.287 & 0.304 & 0.546 & 0.547 & 0.603 & 0.663 \\ 
\hline
\end{tabular}%
\caption{Cross-sectional summaries of the betas and alphas, estimated by $%
\hat{\protect\beta}$, $\hat{\protect\beta}_{LS}$ and $\hat{\protect\alpha}$, 
$\hat{\protect\alpha}_{LS}$. The cross-section is over 400 individual stock
returns, individually regressed against the S\&P500 index. }
\label{tab:betasummary}
\end{table}

\begin{figure}[tb]
\centering
\includegraphics[width=0.48\textwidth]{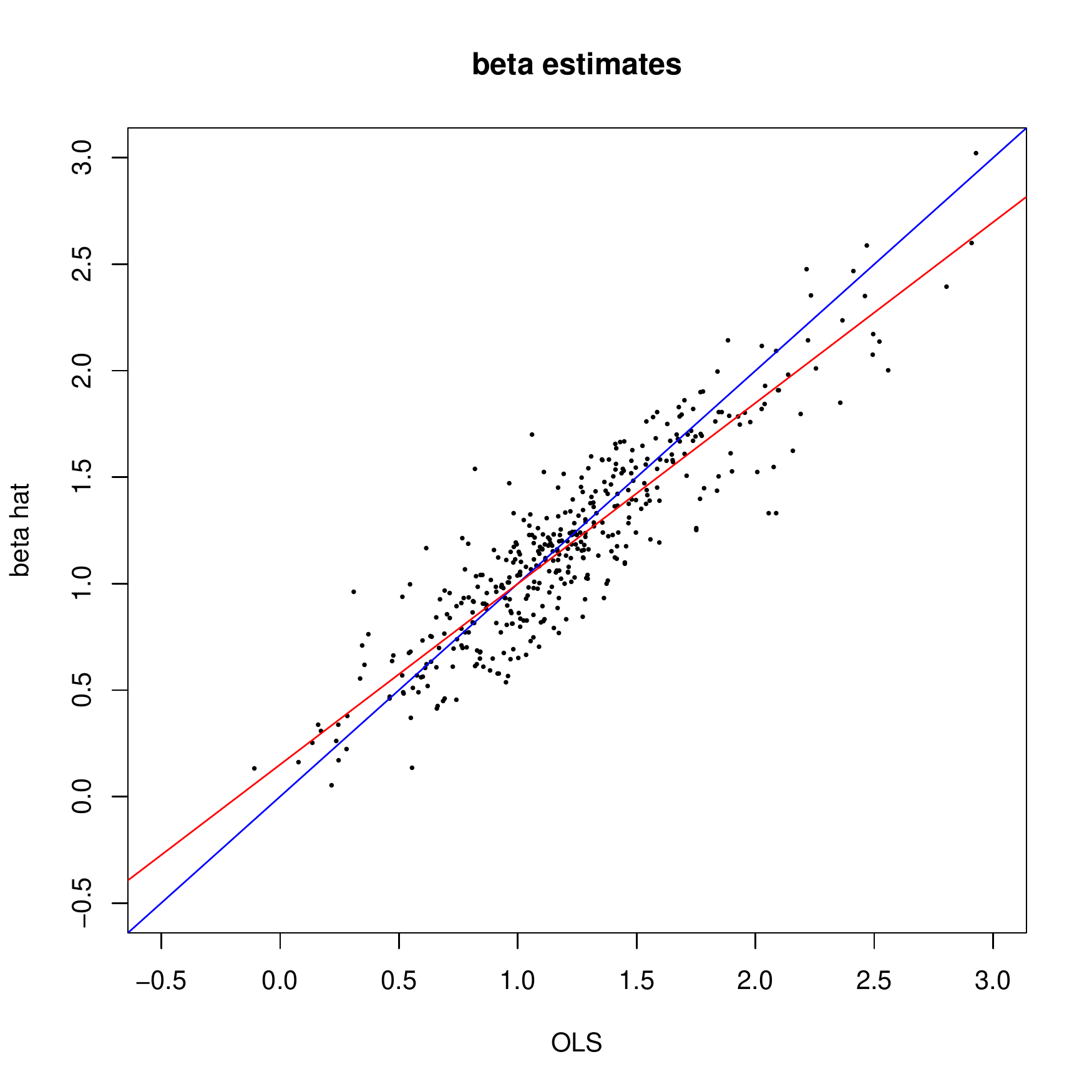} %
\includegraphics[width=0.48\textwidth]{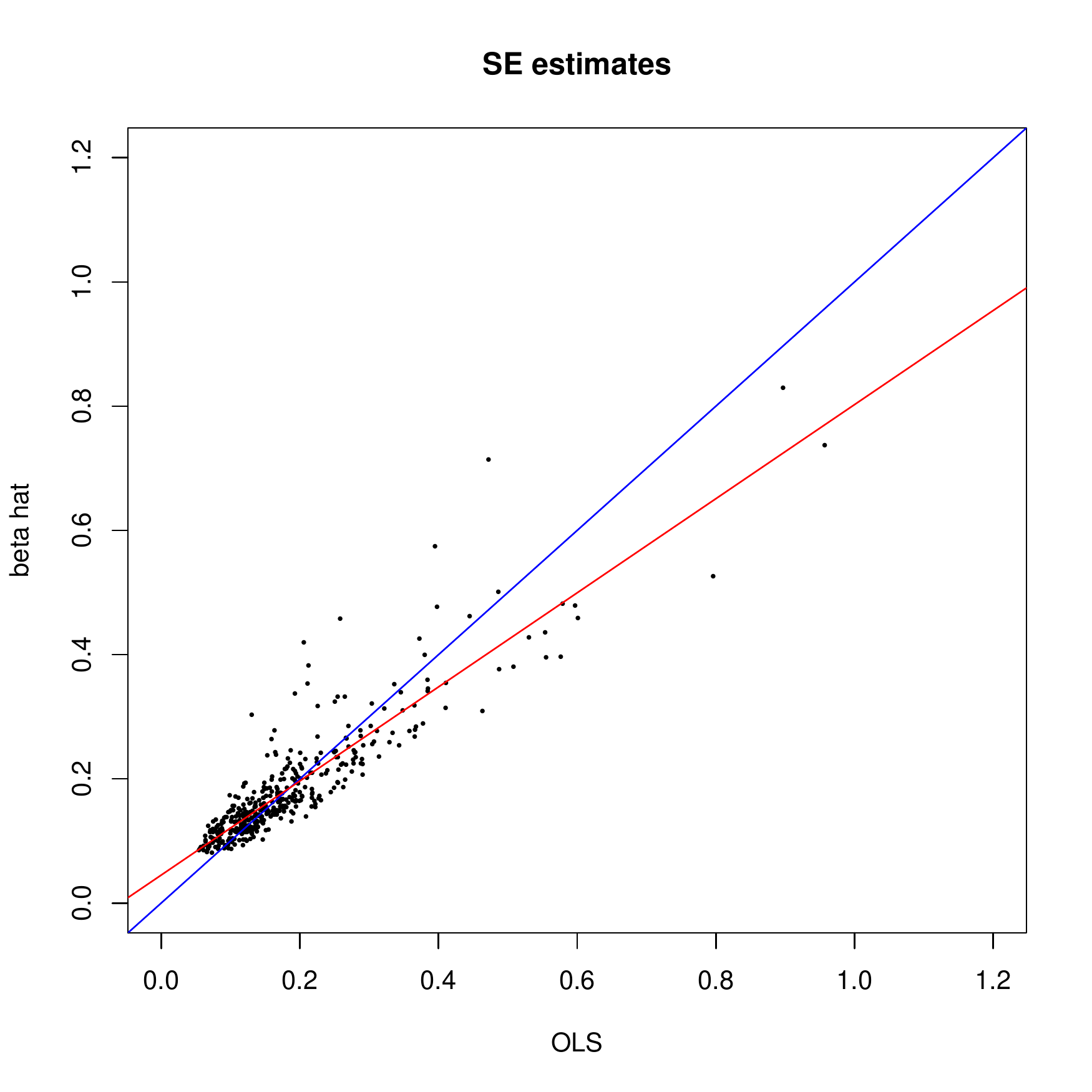}
\caption{The left hand side is a plot of estimated beta for over 400
different stocks based on weekly arithmetic returns, regressing the returns
against the S\&P500 market index. The $y$-axis is $\hat{\protect\beta}$, the 
$x$-axis is $\hat{\protect\beta}_{LS}$. The blue line is a 45 degree line,
going through the origin. The red line is a least squares fitted
straightline from regressing $\hat{\protect\beta}$ on $\hat{\protect\beta}%
_{LS}$. The right hand side is the corresponding estimated standard errors
for $\hat{\protect\beta}$ on $\hat{\protect\beta}_{LS}$.}
\label{fig:SPresults}
\end{figure}

The left hand side of Figure \ref{fig:SPresults} plots $\hat{\beta}$ against 
$\hat{\beta}_{LS}$ in the cross-section. \ The blue line is a 45 degree
line, while a cross-sectional regression of $\hat{\beta}$ against $\hat{\beta%
}_{LS}$ yields an intercept of 0.150 (S.E. of 0.026) and slope of 0.849. The 
$R^{2}\simeq 0.816$. The least squares linear regression line is shown in
the figure by the red line. Overall the picture shows the two sets of
estimates are comparable. \ The $\hat{\beta}$ tend to be slightly pulled up
for low betas and pulled down for high betas, when lined up with the
corresponding $\hat{\beta}_{LS}$. \ \ 

The right hand side of Figure \ref{fig:SPresults} plots $SE(\hat{\beta})$
against $SE(\hat{\beta}_{LS})$ in the cross-section. The blue line is a 45
degree line, while a cross-sectional regression of $SE(\hat{\beta})$ against 
$SE(\hat{\beta}_{LS})$ yields an intercept of 0.045 (S.E. of 0.004) and
slope of 0.757. The $R^{2}\simeq 0.806$. The least squares linear regression
line is shown in the Figure by the red line. When the S.E.s are low the $SE(%
\hat{\beta})$ is materially higher than the $SE(\hat{\beta}_{LS})$. However,
for high S.E.s this is not the case. \ There is more discordance between $SE(%
\hat{\beta})$ and $SE(\hat{\beta}_{LS})$ than $\hat{\beta}$ and $\hat{\beta}%
_{LS}$. In terms of $SE(\widehat{\alpha })$ against $SE(\widehat{\alpha }%
_{LS})$, they have a correlation of about 0.94, so are not plotted here.

Overall, these summary measures suggest that inference based on $\hat{\beta}$
and $SE(\hat{\beta})$ may yield less extreme conclusions than those based on 
$\hat{\beta}_{LS}$ and $SE(\hat{\beta}_{LS})$. \ \ \ \ 
\begin{figure}[tb]
\centering\includegraphics[width=0.48\textwidth]{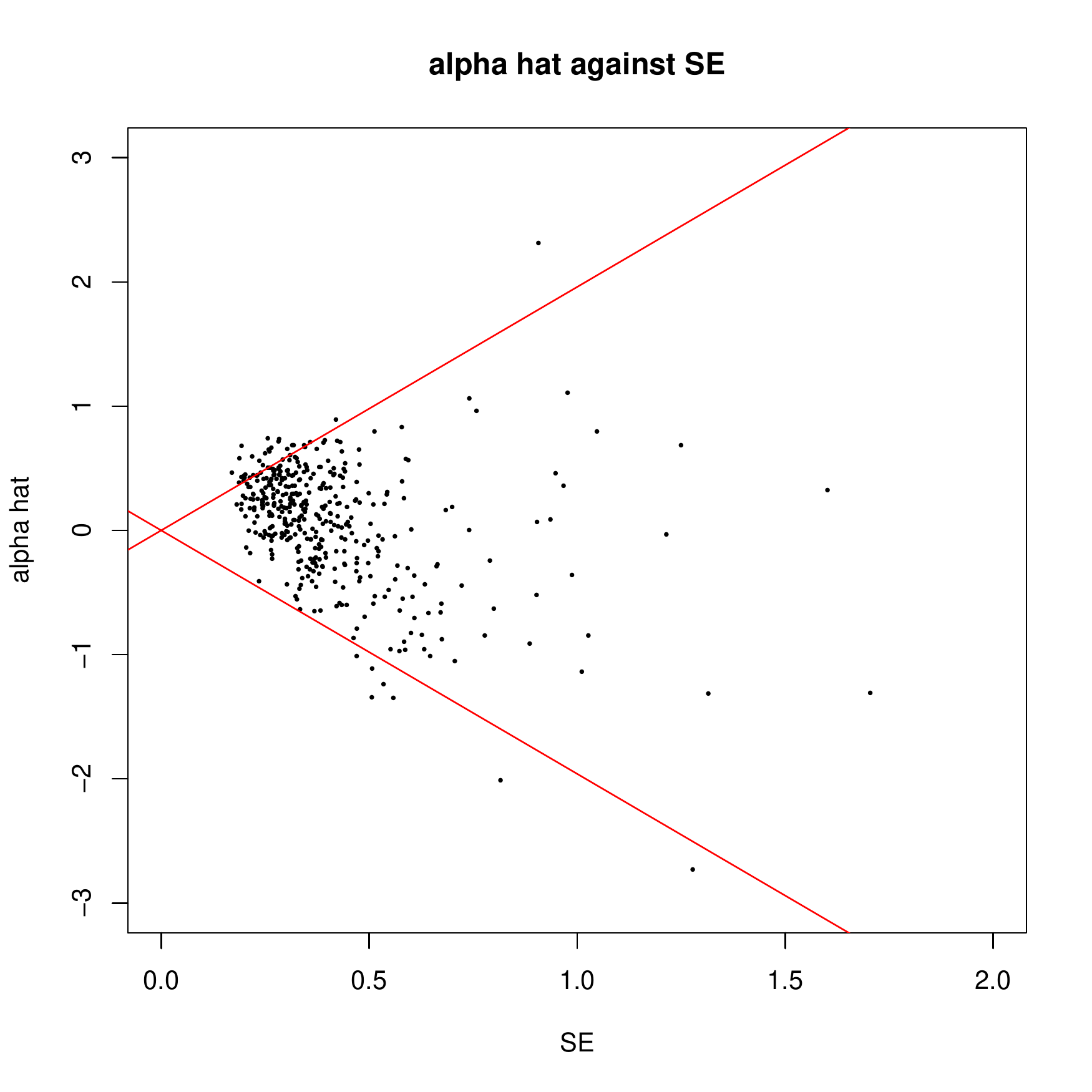}%
\includegraphics[width=0.48\textwidth]{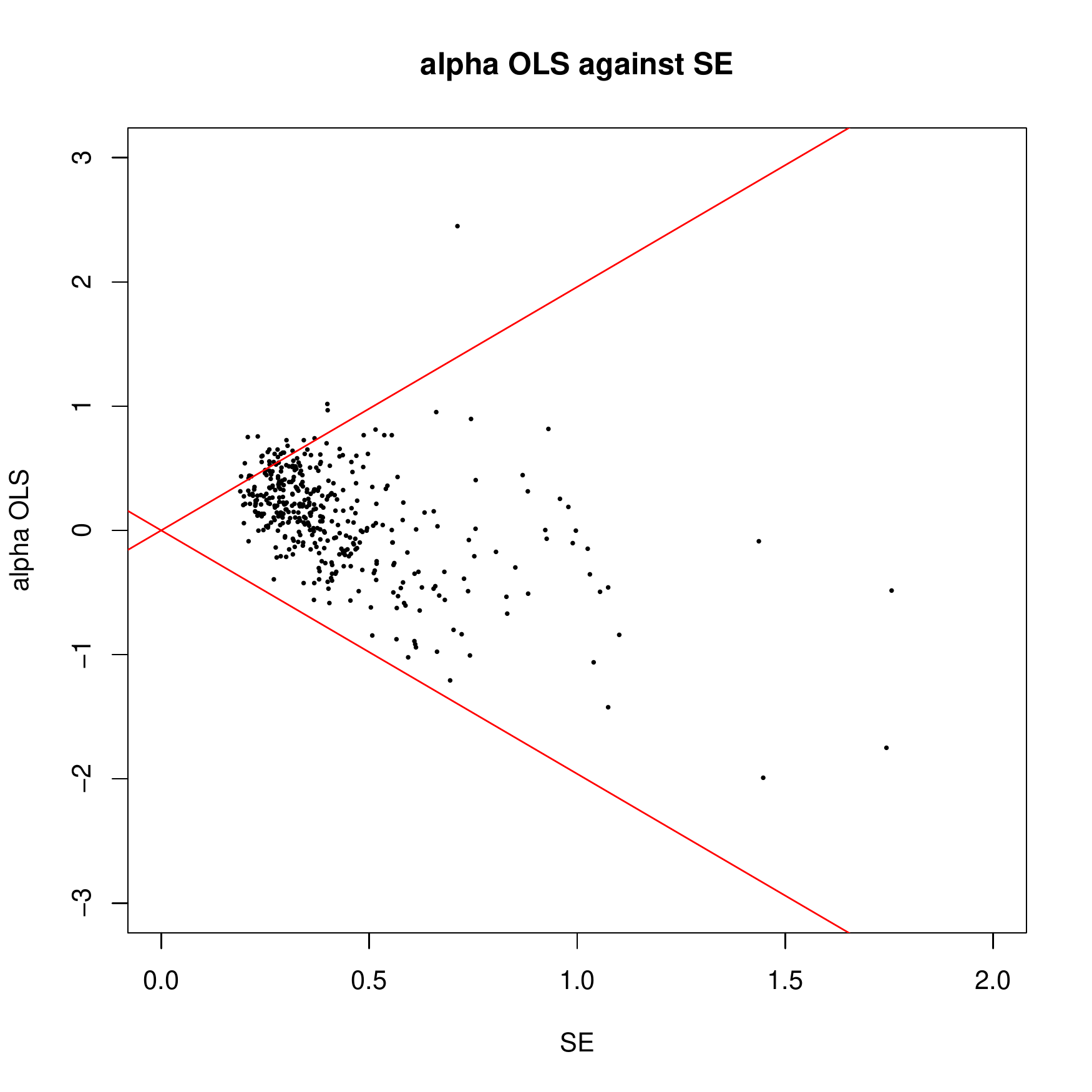}
\caption{The right hand side is the cross-section of least squares estimates 
$\protect\widehat{\protect\alpha }_{LS}$ against the estimated standard
error. \ The left hand side has the same object for $\protect\widehat{%
\protect\alpha }$ and its standard error. \ The red lines have an intercept
of $0$ and slopes of $\pm 1.96$. \ Values above or below the red line are
labelled significant alpha stocks. \ }
\label{fig:alpha}
\end{figure}

Figure \ref{fig:alpha} mimics Figure \ref{fig:summaryFig}, but now for
alpha. \ The differences between $\widehat{\alpha }$,$SE(\widehat{\alpha })$%
\ and $\widehat{\alpha }_{LS}$,$SE(\widehat{\alpha }_{LS})$ are much smaller
than in the beta case. \ Inference based on $\widehat{\alpha }$ and $SE(%
\widehat{\alpha })$ should be very similar to that based on $\widehat{\alpha 
}_{LS}$ and $SE(\widehat{\alpha }_{LS})$.

Figure \ref{fig:summaryFig} shows the critical values for $\hat{\beta}$ and $%
\hat{\beta}_{LS}$ for the high beta null, plotted as the red line (with an
intercept of 1.4 and slope of 1.64). \ Any estimate above the red line
corresponds to a statistically significant high beta. \ The corresponding
results below the green line are statistically significant low beta stocks.\ 

Table \ref{tab:testsummary} provides a summary of the results from the tests
for the nulls for the high and low betas. For both tests, it gives the
number of rejections of the null together with the number of agreements. For
the high beta stocks $\widehat{T}_{LS}$ tests are much more liberal than $%
\widehat{T}_{\widehat{\beta }}$. It is rare for $\widehat{T}_{\widehat{\beta 
}}$ to find a high beta stock which is not found by $\widehat{T}_{LS}$, but
common to see high beta stocks selected by $\widehat{T}_{LS}$ but not $%
\widehat{T}_{\widehat{\beta }}$. \ The results are more scattered for the
test of the low beta stocks. 
\begin{table}[tbp]
\center%
\begin{tabular}{c|cc|c}
& $\widehat{T}_{\widehat{\beta }}$ & $\widehat{T}_{LS}$ & Agree \\ \hline
High beta & 36 & 49 & 32 \\ 
Low beta & 37 & 32 & 23 \\ \hline
\end{tabular}%
\caption{Number of rejections of the null. At a nominal 5\% level, the
procedure would expect around 20 rejections if the null holds just by
chance. The cross-section is over 400 individual stock returns, individually
regressed against the S\&P500 index.}
\label{tab:testsummary}
\end{table}

\begin{figure}[tb]
\centering\includegraphics[width=0.23\textwidth]{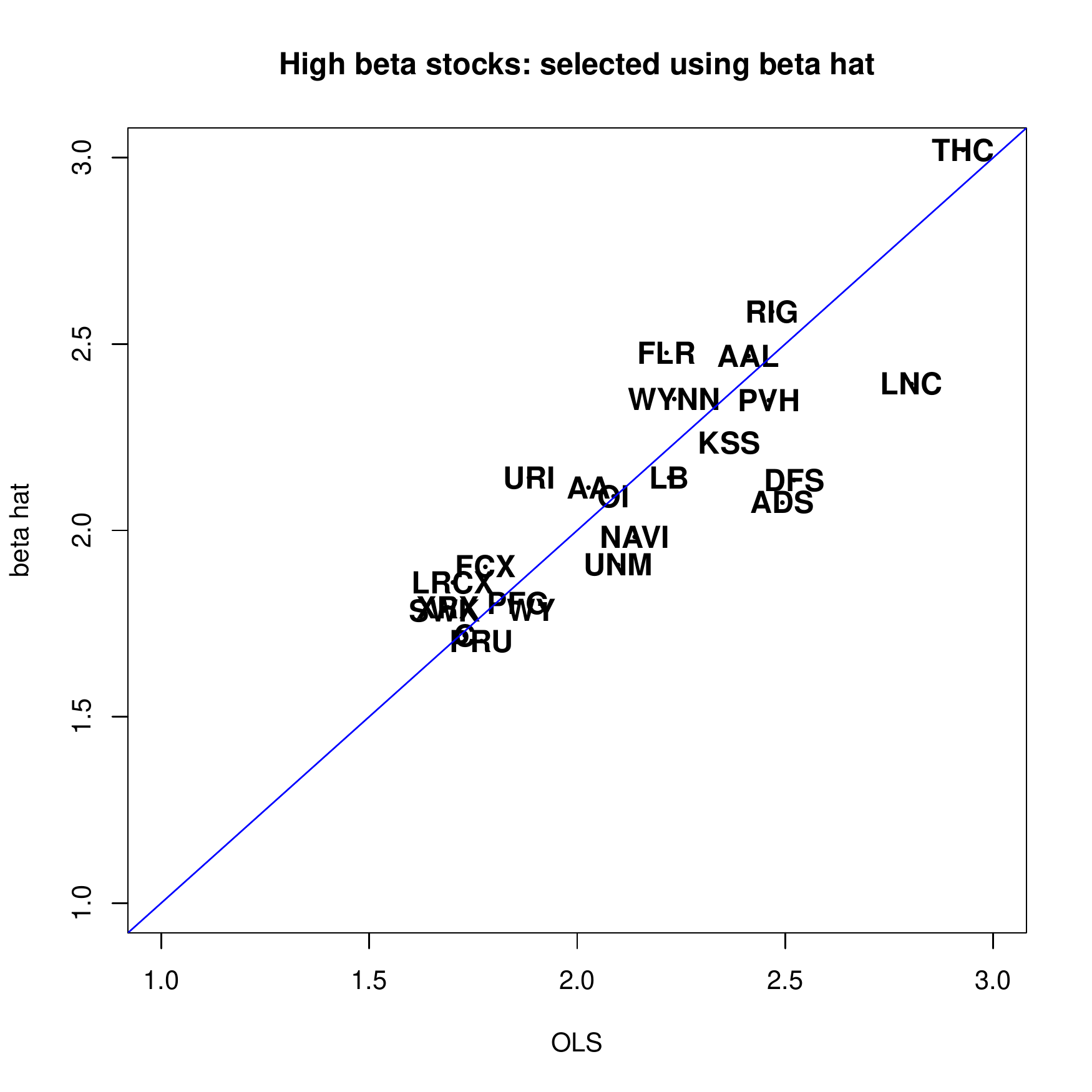} %
\includegraphics[width=0.23\textwidth]{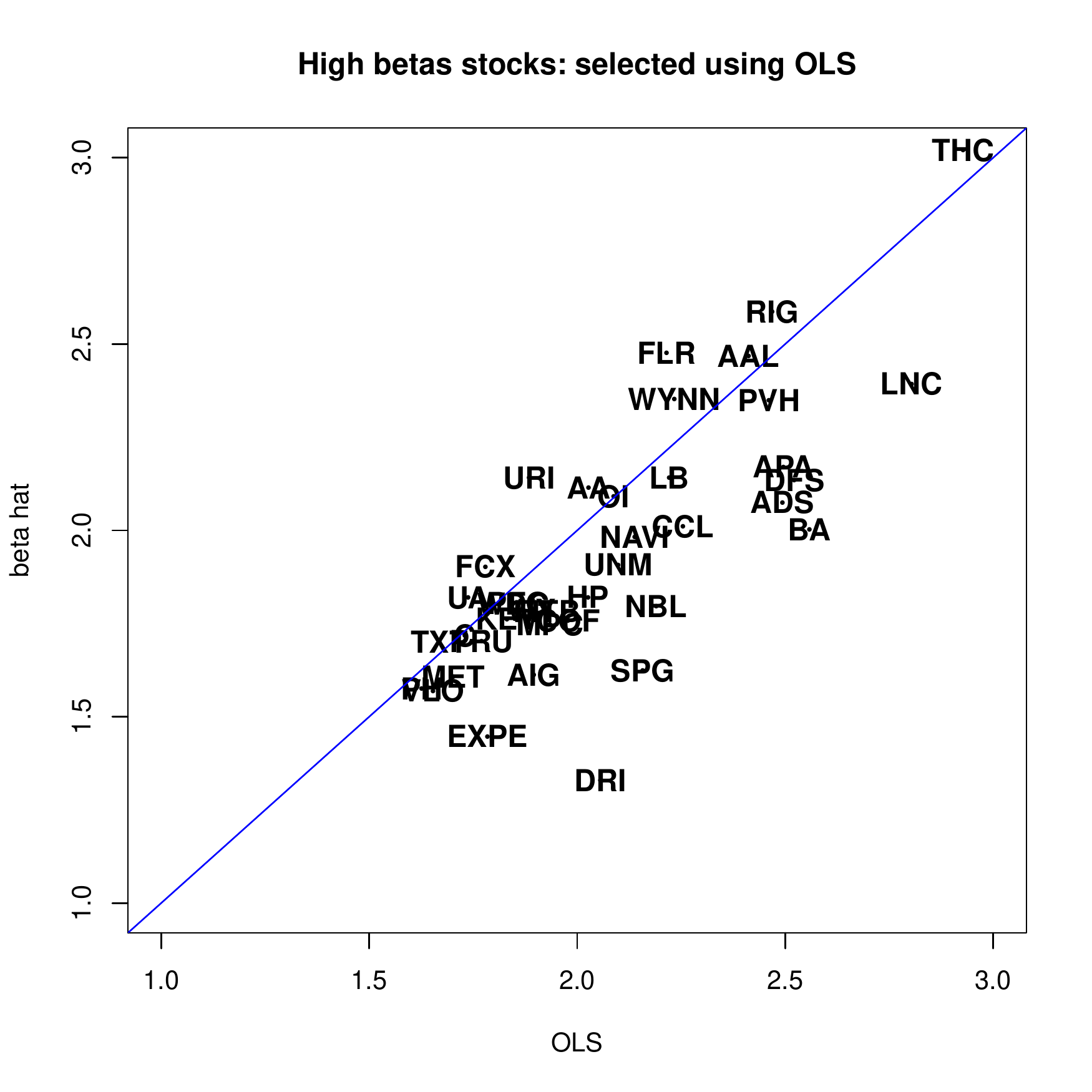}\vrule%
\includegraphics[width=0.23\textwidth]{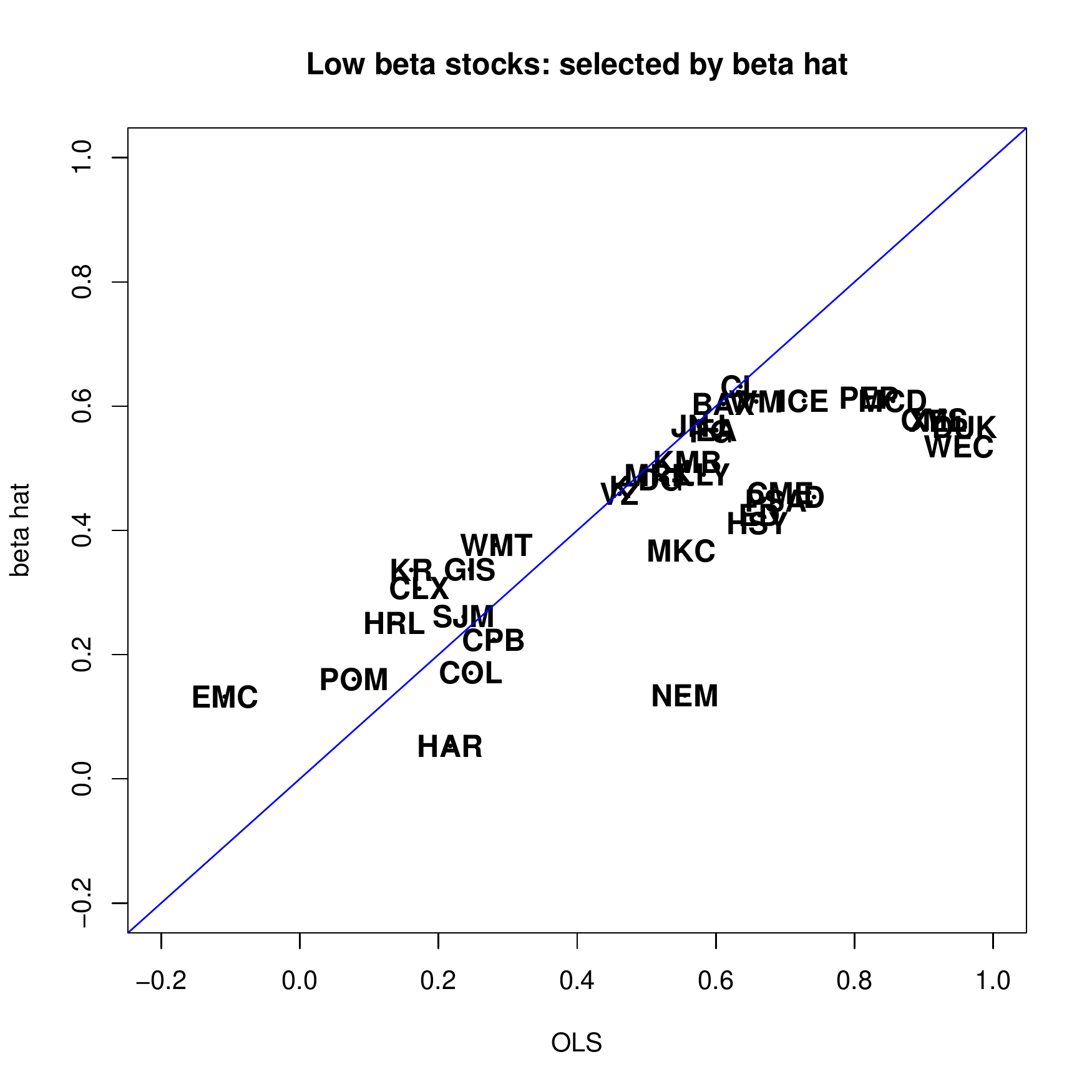} %
\includegraphics[width=0.23\textwidth]{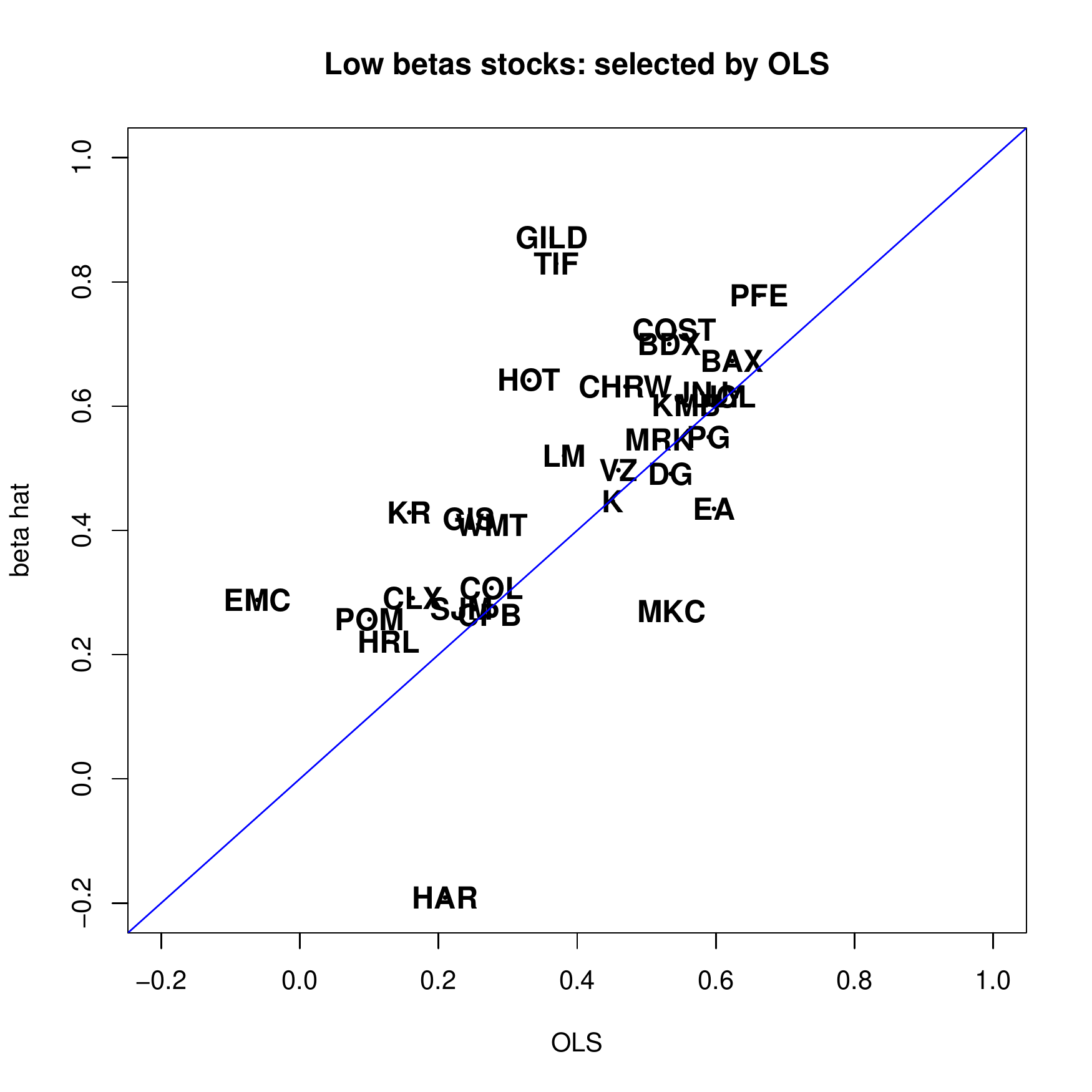}
\caption{ LHS shows selection of large beta stocks, RHS has selection of
small beta stocks. \ The selection in the 1st and 3rd graph is carried out
using $\protect\widehat{T}_{\protect\widehat{\protect\beta }}$, the 2nd and
4th are implemented using $\protect\widehat{T}_{LS}$. The tickers are used
to plot individual stock selections. \ }
\label{fig:SPresultstest}
\end{figure}

This is highlighted by Figure \ref{fig:SPresultstest}, plotting $\hat{\beta}$
against $\hat{\beta}_{LS}$ for selected stocks. The 1st and 3rd selection is
carried out by $\widehat{T}_{\widehat{\beta }}$, the 2nd and 4th by $%
\widehat{T}_{LS}$. \ The left hand side of the Figure shows results for the
selection of high beta stocks. \ The right hand side gives the corresponding
selected low beta stocks. The labels are the individual stock tickers. \ 

For the high beta selections, the $\widehat{T}_{\widehat{\beta }}$ selected
stocks have both estimators $\hat{\beta}$ and $\hat{\beta}_{LS}$ indicating
pretty high betas. When the selection is based on $\widehat{T}_{LS}$ the
results are more variable. \ 

For low beta selection, the story is much more mixed. 

\subsection{Rolling betas}

Empirical researchers often deal with time-varying betas by using moving
block, or rolling, averages --- see \cite{Engle(16)} for an alternative
model based approach and a discussion of the literature. \ In our context a
rolling average approach computes the statistics $\left( \hat{\beta},\hat{%
\beta}_{LS}\right) $ and $(SE(\hat{\beta}),SE(\hat{\beta}_{LS}))$ on the
last 100, say, weeks of data, moving that window through time. \ 

What do the pairs $(\hat{\beta}$,$\hat{\beta}_{LS})$, $(SE(\hat{\beta}),SE(%
\hat{\beta}_{LS}))$ and the two pair of pairs $(\widehat{T}_{\widehat{\beta }%
},\widehat{T}_{LS})$ (corresponding to the high and low beta hypotheses)
look like over the last 15 years for the first stock in our database, ABT,
Abbott Laboratories? Now the data starts on 1 January 2005 and the rolling
window always covers 100 weeks of data. In this entire database there are
358 stocks. \ \ \ 
\begin{figure}[tb]
\centering\includegraphics[width=0.48\textwidth]{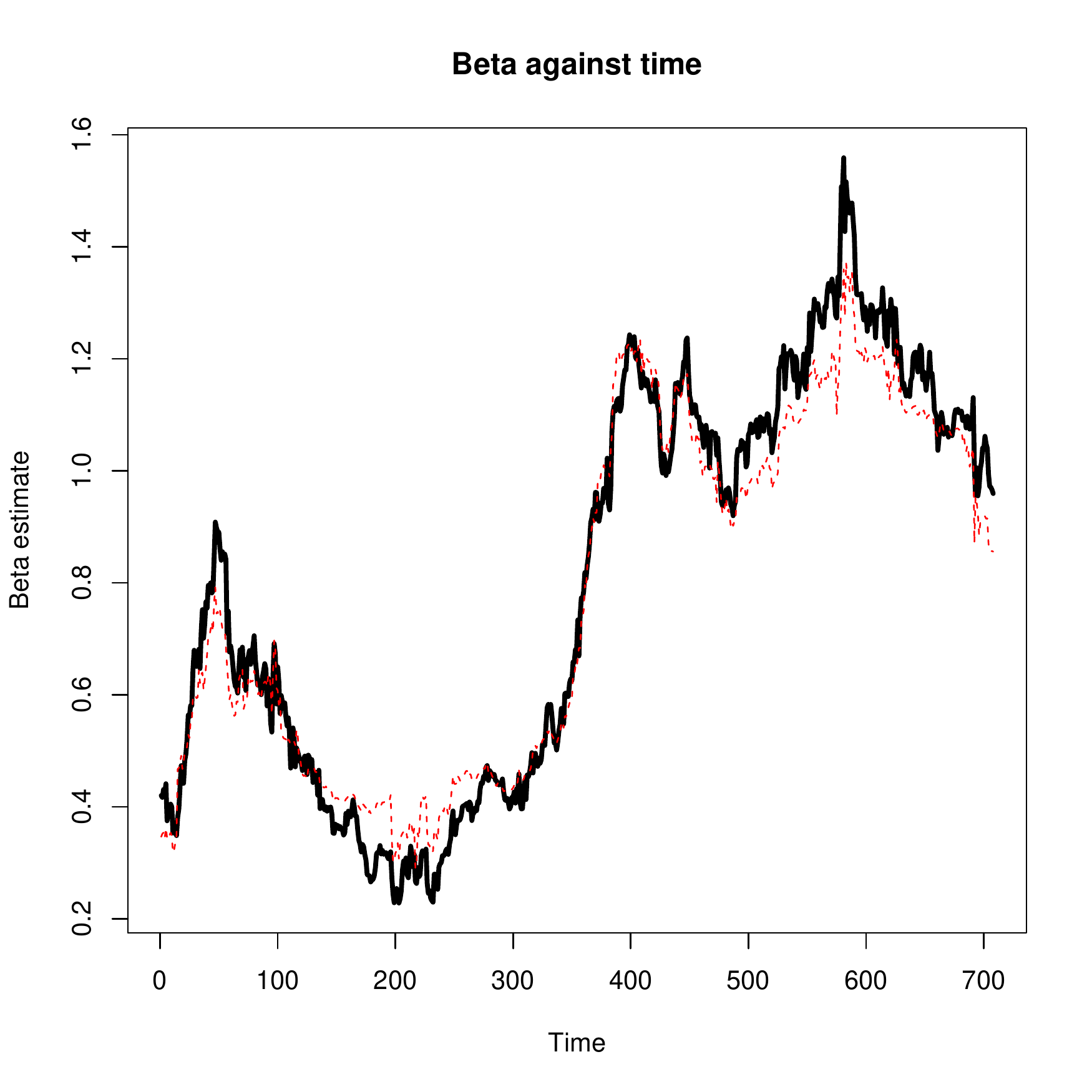}%
\includegraphics[width=0.48\textwidth]{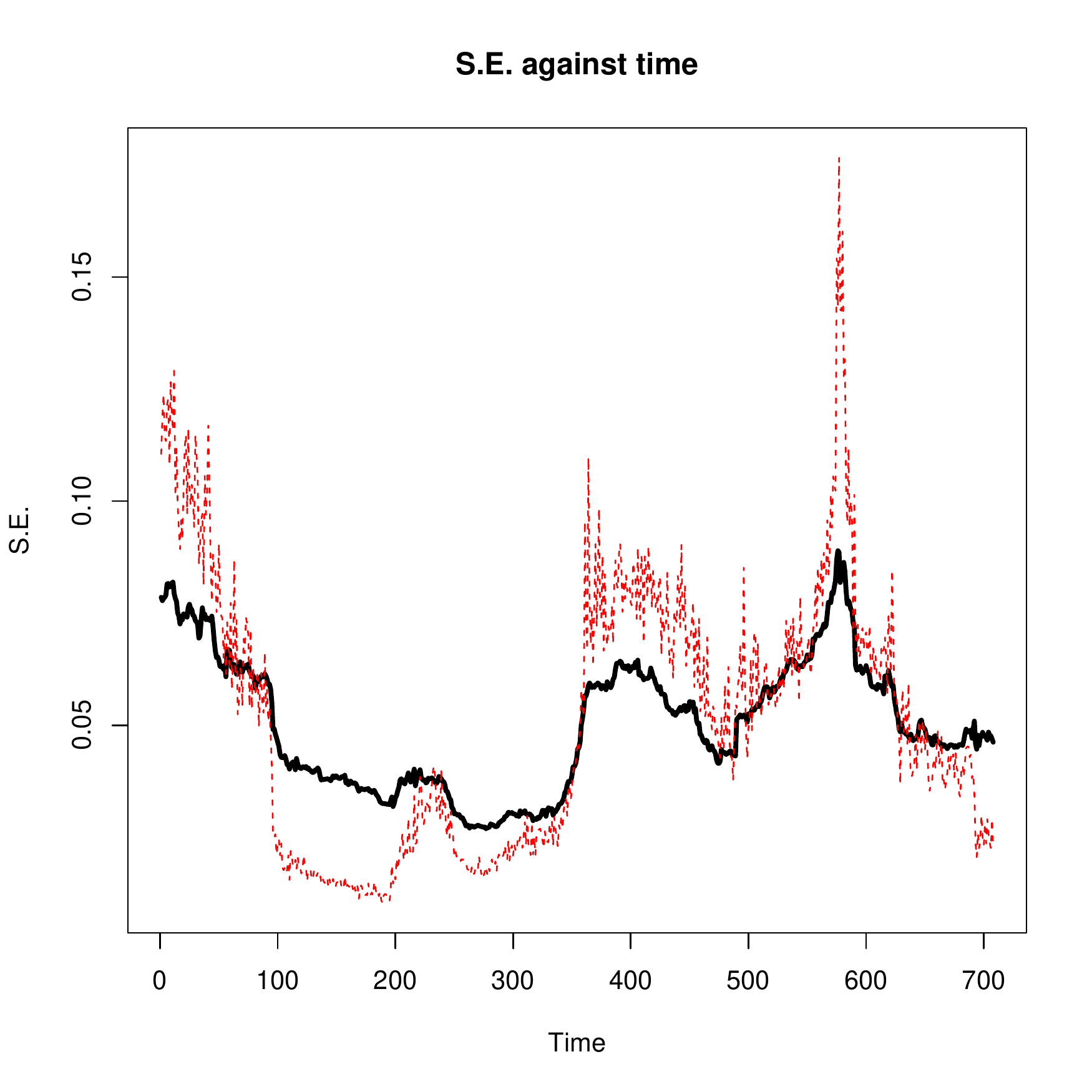}
\caption{LHS shows the pair $(\hat{\protect\beta}$,$\hat{\protect\beta}%
_{LS}) $ through time for ABT, Abbott Laboratories. Black line is $\hat{%
\protect\beta}$, red dotted line is $\hat{\protect\beta}_{LS}$. The RHS
shows the pair $(SE(\hat{\protect\beta}),SE(\hat{\protect\beta}_{LS}))$\
through time. \ Black line is $SE(\hat{\protect\beta})$, red dotted line is $%
SE(\hat{\protect\beta}_{LS})$.}
\label{fig:timeSumm}
\end{figure}

Figure \ref{fig:timeSumm} contains the results for ABT: $\hat{\beta}$ is
drawn using a thick black line; $\hat{\beta}_{LS}$ using a dotted red line.
\ The beta estimates $(\hat{\beta}$,$\hat{\beta}_{LS})$ broadly track one
another. The $\hat{\beta}_{LS}$ coefficient does not go as high as the $\hat{%
\beta}$ during higher beta periods, nor as low during lower beta periods.

I would like you to mostly focus on is the right hand side picture, which
plots the pair $(SE(\hat{\beta}),SE(\hat{\beta}_{LS}))$ through time. \
Although they follow the same general level through time, $SE(\hat{\beta}%
_{LS})$ is very rough, sometimes moving dramatically over a few datapoints,
as individual pairs of data fall in or out of the 100 day window. \ This is
exactly what was feared in the introduction, it is very sensitive to large
moves in the predictors. $SE(\hat{\beta})$ is much smoother, as if it has
been time-series smoother --- but it has not been. \ It drifts down and then
up, the range moving by a factor of 2 in the picture. \ \ 

This result is typical in the cross-section. \ Averaging over all time
periods and across all stocks, the first element of Table \ref%
{tab:variableAns} shows the square root of the average squared 100 times the
daily time series changes in $\widehat{\beta }$. \ This is just below 2.5,
while the corresponding result for $\widehat{\beta }_{LS}$ is around 15\%
higher. \ There are much bigger differences in the roughness of the standard
deviations. \ The average movement of $SE(\widehat{\beta }_{LS})$ is around
70\% higher than the result for $SE(\widehat{\beta })$. 
\begin{table}[tbp]
\center%
\begin{tabular}{cc|cc}
$\sqrt{\mathrm{E}[(100\Delta \widehat{\beta })^{2}]}$ & $\sqrt{\mathrm{E}%
[(100\Delta \widehat{\beta }_{LS})^{2}]}$ & $\sqrt{\mathrm{E}[(100\Delta SE(%
\widehat{\beta }))^{2}]}$ & $\sqrt{\mathrm{E}[(100\Delta SE(\widehat{\beta }%
_{LS}))^{2}]}$ \\ \hline
2.44 & 2.84 & 0.152 & 0.269 \\ \hline
\end{tabular}%
\caption{Measures of the roughness of the path of estimated beta and
corresponding estimated standard errors. The roughness is measured by the
square root of the average squared 100 times daily changes in the estimator.
Here $\Delta $ is a time series difference operator. }
\label{tab:variableAns}
\end{table}

\begin{figure}[tb]
\centering\includegraphics[width=0.48\textwidth]{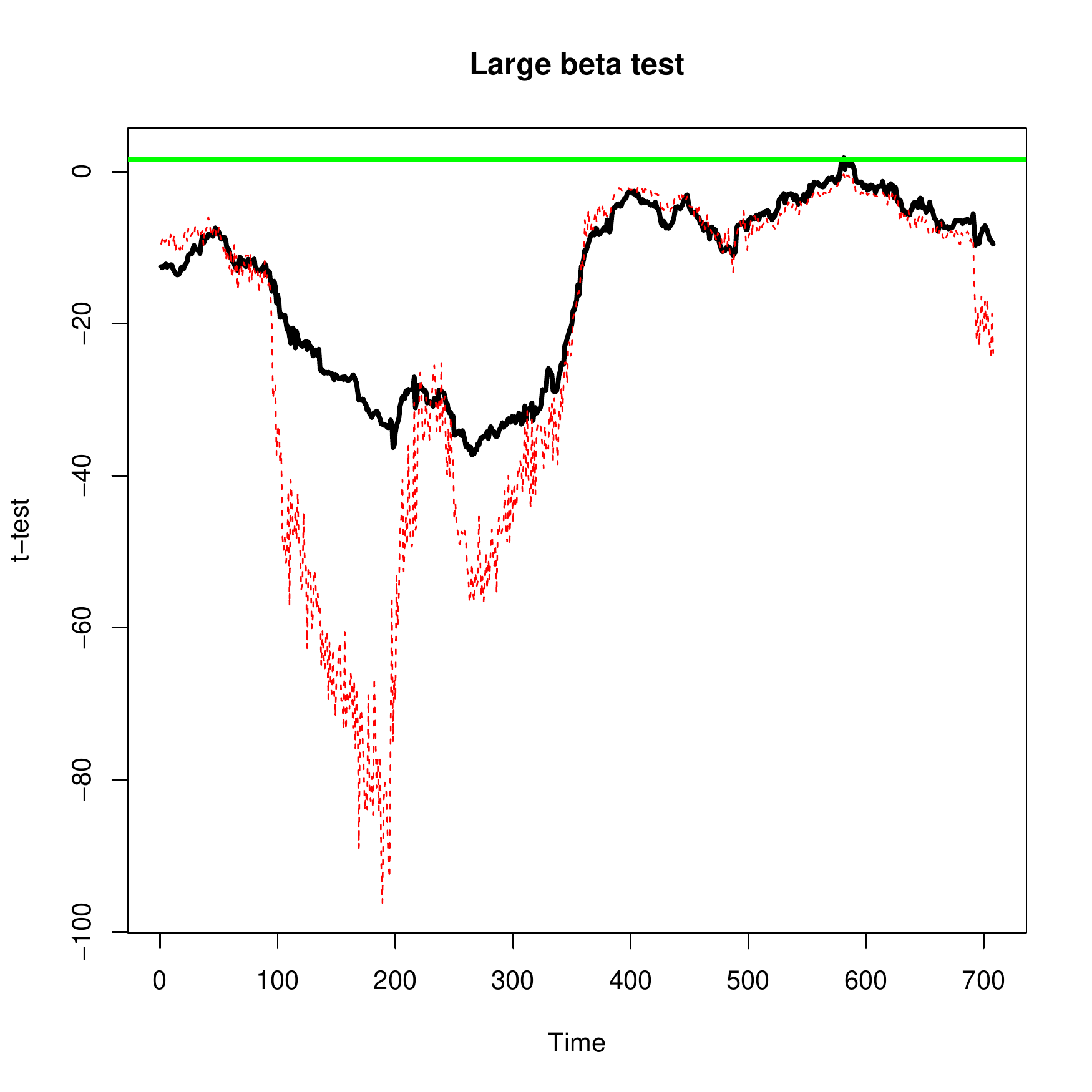} %
\includegraphics[width=0.48\textwidth]{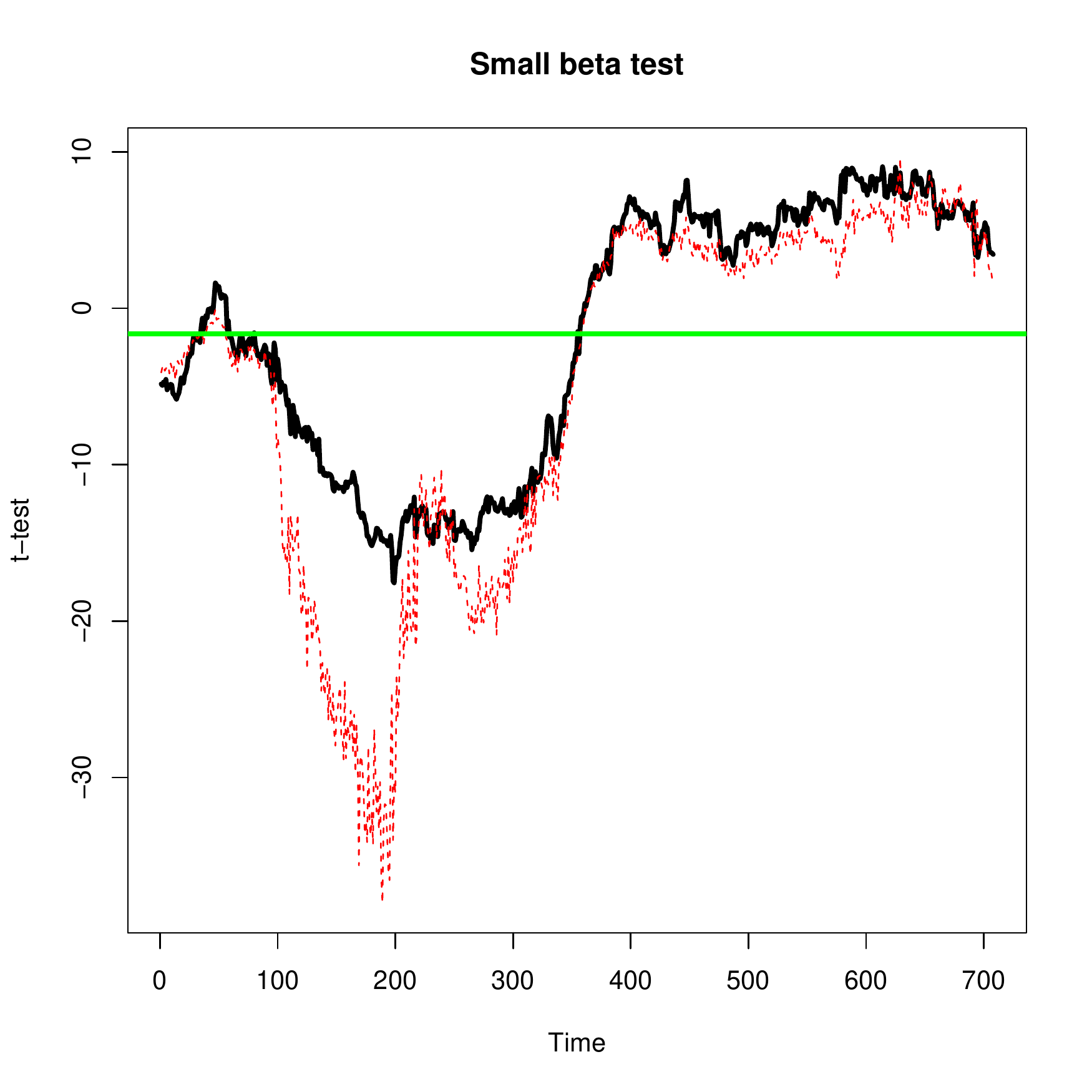}
\caption{ LHS shows $(\protect\widehat{T}_{\protect\widehat{\protect\beta }},%
\protect\widehat{T}_{LS})$ of large beta stocks for ABT, Abbott
Laboratories. \ Black line is $\protect\widehat{T}_{\protect\widehat{\protect%
\beta }}$, red dotted line is $\protect\widehat{T}_{LS}$. \ RHS shows $(%
\protect\widehat{T}_{\protect\widehat{\protect\beta }},\protect\widehat{T}%
_{LS})$ of small beta stocks. \ The green lines are the critical values for
the 5\% nominal level tests. \ \ }
\label{fig:timeTest}
\end{figure}

One implication of this can be seen in Figure \ref{fig:timeTest} for ABT. \
On the left hand side the pair $(\widehat{T}_{\widehat{\beta }},\widehat{T}%
_{LS})$ is plotted against time for the large beta hypothesis test. \ The
test based on least squares $\widehat{T}_{LS}$ is very jagged through time,
moving around week by week dramatically. \ The main driver of these moves
are the jagged standard errors. \ The thick black line, corresponding to $%
\hat{\beta}$ is again smooth. \ The same holds for the small beta test
picture, which is the right hand side picture. \ Here the test rejects the
null and selects this stock as a small beta stock for roughly the same
period, but this is lucky for the evidence is very strong which makes
choosing between using $\widehat{T}_{\widehat{\beta }}$ and $\widehat{T}_{LS}
$\ moot. \ 

The evidence from looking at rolling betas is that the classic $\widehat{%
\beta }_{LS}$ standard errors move around dramatically, easily upended by a
single pair of data. \ More solid inference can be carried out using $\hat{%
\beta}$.

\subsection{Rolling low and high beta portfolios}

Using hypothesis tests based on the rolling estimates $(\hat{\beta}$,$\hat{%
\beta}_{LS})$, and corresponding standard errors $(SE(\hat{\beta}),SE(\hat{%
\beta}_{LS}))$, I built each week an equal weighted high beta and low beta
portfolio. \ The stocks are selected using a hypothesis test, which means
the number of stocks in portfolio is random --- an alternative would be to
put a fixed number of stocks in the portfolio, selecting the stocks with the
lowest $p$-values. \ These portfolios are then used through the next week's
data to form a return over a week. \ This procedure is run through the
entire sample period. \ Table \ref{tab:Portresults} reports out of sample
summary statistics of these two portfolios.

The results for the portfolio selected using test statistics based on $\hat{%
\beta}$ and $\hat{\beta}_{LS}$ are similar. \ 

The high beta portfolios produce what is expected: a higher average return
and a substantially higher standard deviation, compared to the index. \
Their Sharpe ratios are not very different than the index, while their own
betas are high, around 1.75, while the corresponding alphas are
statistically close to 0. \ 

The statistic \textquotedblleft Share\textquotedblright\ is the time series
average of the proportion of the cross-section which is in the portfolio.\
Hence the low beta portfolio contains around 13\% of the stocks, the high
beta portfolios contain around 7\%. \ The \textquotedblleft $|\Delta Share|$%
\textquotedblright\ is the time series average of the difference in
holdings, so if the universe of stocks is 1,000 and the \textquotedblleft $%
|\Delta Share|=0.001$\textquotedblright\ then on average 1 stock moves into
or out of the portfolio each week. \ In our case the universe of stocks is
358, so on average 3 stocks come into or out of the portfolios each week. \ 
\begin{table}[tbp]
\center%
\begin{tabular}{llll|ll|ll}
Portfolio type & E & sd & Sharpe & alpha & beta & Share & $|\Delta Share|$
\\ 
Index & 0.194 & 2.49 & 0.077 & 0 & 1 & 0 & 0 \\ \hline
Low beta $\widehat{\beta }$ & 0.213 & 1.96 & 0.109 & 0.093 {\footnotesize %
(0.041)} & 0.625 & 0.136 & 0.008 \\ 
Low beta $\widehat{\beta }_{LS}$ & 0.222 & 1.88 & 0.118 & 0.111 
{\footnotesize (0.042)} & 0.573 & 0.137 & 0.008 \\ \hline
High beta $\widehat{\beta }$ & 0.361 & 5.05 & 0.071 & 0.018 & 1.77 & 0.076 & 
0.007 \\ 
High beta $\widehat{\beta }_{LS}$ & 0.359 & 5.09 & 0.070 & 0.018 & 1.75 & 
0.076 & 0.006%
\end{tabular}%
\caption{Week by week out of sample portfolio returns. Low and high beta
portfolios are estimated by hypothesis testing. Figures in brackets are
standard errors. }
\label{tab:Portresults}
\end{table}

The low beta portfolios are more interesting economically: their low beta
should deliver a portfolio with low average returns and low standard
deviations, but actually the average returns are higher than for the index,
and this pushes the Sharpe ratio up. \ When these portfolios are regressed
against the index, the betas are around two thirds, while they have positive
alpha (the figure in brackets are standard errors). \ This is certainly not
an unexpected result from the applied finance literature. \ The low beta
premium is reported extensively in the literature, see \cite%
{BakerBradleyWurgler(11)} for a review.

\section{Median predictive regression\label{sect:quantile}}

To finish the substance of this paper, it is useful to note that some of
these points generalize. \ 

Write $Q_{Z}(\tau )$ denotes the $\tau $-th quantile of the generic scalar
random variable $Z$. \ The classic text on quantile regression is \cite%
{Koenker(05)}. \ Here the focus will be on the \textquotedblleft median
predictive regression,\textquotedblright\ the case where $\tau =1/2$, but
the ideas hold for general quantile predictive regressions.\ \ To match the
above treatment of linear predictive regression, the focus will be on the
centered parameterization%
\begin{equation}
Q_{Y_{1}|\mathbf{Z}_{1}=\mathbf{z}_{1}}(1/2)=\mathbf{x}_{1}(\mathbf{\mathbf{%
\psi }})^{\mathrm{T}}\mathbf{\beta ,\quad X}_{1}(\mathbf{\mathbf{\psi }}%
)=\left\{ 1,\left( \mathbf{Z}_{1}\mathbf{-\mathbf{\psi }}\right) ^{\mathtt{T}%
}\right\} ^{\mathtt{T}},\mathbf{\quad \mathbf{\psi =}}\mathrm{E}\mathbf{%
\mathbf{[Z}}_{1}\mathbf{\mathbf{].}}  \label{median prediction}
\end{equation}%
$\mathbf{\beta }$ is the estimand and inference about $\mathbf{\beta }$ is
my goal. \ 

Inference on $\mathbf{\beta }$ is traditionally based on the storied least
absolute deviation estimator 
\begin{eqnarray*}
\widehat{\mathbf{\psi }} &=&\frac{1}{n}\sum_{j=1}^{n}\mathbf{Z}_{j},\quad 
\mathbf{X}_{j}=(1,(\mathbf{Z}_{j}-\widehat{\mathbf{\psi }})^{\mathtt{T}})^{%
\mathtt{T}}, \\
\widehat{\mathbf{\beta }}_{LAD} &=&\underset{\mathbf{b}}{\arg }\underset{}{%
\min }\sum_{j=1}^{n}\left\vert Y_{j}-\mathbf{X}_{j}^{\mathtt{T}}\mathbf{b}%
\right\vert ,
\end{eqnarray*}%
\ whose asymptotic variance is typically estimated by 
\begin{equation*}
\frac{1}{n}S_{1_{U}\mathbf{X},\mathbf{X}}^{-1}S_{\mathbf{X},\mathbf{X}%
}S_{1_{U}\mathbf{X},\mathbf{X}}^{-1},\quad S_{\mathbf{X},\mathbf{X}}=\frac{1%
}{n}\sum_{j=1}^{n}\mathbf{X}_{j}\mathbf{X}_{j}^{\mathtt{T}},\quad S_{1_{U}%
\mathbf{X},\mathbf{X}}=\frac{2}{nh_{n}}\sum_{j=1}^{n}\mathbf{X}_{j}\mathbf{X}%
_{j}^{\mathtt{T}}1_{|\widehat{U}_{j,LAD}<h_{n}|},
\end{equation*}%
where $\widehat{U}_{j,LAD}=Y_{j}-\mathbf{X}_{j}^{\mathtt{T}}\widehat{\mathbf{%
\beta }}_{LAD}$ and the shrinking bandwidth $h_{n}\downarrow 0$ and $%
nh_{n}\rightarrow \infty $ as $n$ increases. \ This approach needs at least
four moments (e.g. assuming the density of $U|\mathbf{X}$ is bounded) of the
predictors to yield asymptotically valid inference. \ Hence LAD based
inference is, in my opinion, also not credible for financial economics. \ It
has the same problem as inference for least squares under
heteroskedasticity. \ This lack of credibility also directly applies, in my
opinion, to the celebrated \cite{KoenkerBassett(78)} check based estimator
for quantile regression (recall it is the LAD estimator in the 0.5-quantile
case). \  \ \ \ 

For median predictive regression I advocate 
\begin{eqnarray*}
\widehat{\mathbf{\psi }} &=&\frac{1}{n}\sum_{j=1}^{n}\mathbf{Z}_{j},\quad 
\mathbf{X}_{j}=(1,(\mathbf{Z}_{j}-\widehat{\mathbf{\psi }})^{\mathtt{T}})^{%
\mathtt{T}}, \\
\widehat{\mathbf{\beta }} &=&\underset{\mathbf{b}}{\arg }\underset{}{\min }%
\sum_{j=1}^{n}S_{j}(\mathbf{b}),\quad S_{j}(\mathbf{b})=\left\Vert \mathbf{X}%
_{j}\right\Vert _{2}^{-1}\left\vert Y_{j}-\mathbf{X}_{j}^{\mathtt{T}}\mathbf{%
b}\right\vert ,
\end{eqnarray*}%
noting $S_{j}(\mathbf{b})$ is convex in $\mathbf{b}$, the $\left\vert S_{j}(%
\mathbf{b})-S_{j}(\mathbf{b}^{\prime })\right\vert \leq \left\Vert \mathbf{b}%
-\mathbf{b}^{\prime }\right\Vert _{2}$ and $S_{j}(\mathbf{b})$ has
subderivative 
\begin{equation*}
\partial S_{j}(\mathbf{b})=-2\mathbf{G}_{j}\left( \frac{1}{2}-1_{Y_{j}<%
\mathbf{X}_{j}^{\mathtt{T}}\mathbf{b}}\right) ,\quad \mathbf{G}%
_{j}=\left\Vert \mathbf{X}_{j}\right\Vert _{2}^{-1}\mathbf{X}_{j}.
\end{equation*}%
As before $\left\Vert \mathbf{G}_{1}\right\Vert _{\infty }\leq 1$ so $%
\left\Vert \partial S_{j}(\mathbf{b})\right\Vert _{\infty }\leq 1$. Hence $%
\widehat{\mathbf{\beta }}$ is robust, having a bounded loss function and a
bounded subderivative (\cite{Mallows(73),Mallows(75)}, \cite%
{KraskerWelsch(82)} and \cite{GiloniSimonoffSengupta(06)})\footnote{%
The approach can be made even more robust by reparameterization the
predictive median regression, replacing the center $\mathrm{E}[X_{1}]$ by $%
\mathbf{\psi =}med(\mathbf{X}_{1})$ the element-by-element median, then
taking $\widetilde{\mathbf{\psi }}$ as the element-by-element sample median
allowing $\mathbf{X}_{j}=(1,(\mathbf{Z}_{j}-\widetilde{\mathbf{\psi }})^{%
\mathtt{T}})^{\mathtt{T}}$. \ This then contributes to $\widetilde{\mathbf{%
\beta }}=\underset{\mathbf{b}}{\arg }\underset{}{\min }\sum_{j=1}^{n}\left%
\Vert \mathbf{X}_{j}\right\Vert _{1}^{-1}\left\vert Y_{j}-\mathbf{X}_{j}^{%
\mathtt{T}}\mathbf{b}\right\vert $, where I have normalized by $\left\Vert 
\mathbf{X}_{j}\right\Vert _{1}$, avoiding any squaring of data. Then $%
\widetilde{\mathbf{\beta }}$ has the same characteristics as before. }. \ 
Inference will be based on 
\begin{equation*}
\frac{1}{n}S_{1_{U}\mathbf{G},\mathbf{G}}^{-1}S_{\mathbf{G},\mathbf{X}%
}S_{1_{U}\mathbf{G},\mathbf{G}}^{-1},\quad S_{\mathbf{G},\mathbf{X}}=\frac{1%
}{n}\sum_{j=1}^{n}\mathbf{G}_{j}\mathbf{X}_{j}^{\mathtt{T}},\quad S_{1_{U}%
\mathbf{G},\mathbf{G}}=\frac{2}{nh_{n}}\sum_{j=1}^{n}\mathbf{G}_{j}\mathbf{G}%
_{j}^{\mathtt{T}}1_{|\widehat{U}_{j,LAD}<h_{n}|},
\end{equation*}%
where $\widehat{U}_{j}=Y_{j}-\mathbf{X}_{j}^{\mathtt{T}}\widehat{\mathbf{%
\beta }}$ and the shrinking bandwidth $h_{n}\downarrow 0$ and $%
nh_{n}\rightarrow \infty $ as $n$ increases. \ 

$\widehat{\mathbf{\beta }}$ is $\widehat{\mathbf{\beta }}_{LAD}$ applied to
the preprocessed data $Y_{j}^{\ast }=$\ $\left\Vert \mathbf{X}%
_{j}\right\Vert _{2}^{-1}Y_{j}$ and $\mathbf{X}_{j}^{\ast }=$ $\left\Vert 
\mathbf{X}_{j}\right\Vert _{2}^{-1}\mathbf{X}_{j}$.\ \ The preprocessing
stabilizes statistical inference, while existing software can be used
without any further changes to find $\widehat{\mathbf{\beta }}$.

\begin{example}
If there is no intercept and a single predictor, then $\left\Vert \mathbf{X}%
_{j}\right\Vert _{2}^{-1}=|Z_{j}-\overline{Z}|$ and 
\begin{eqnarray*}
\left\vert Y_{j}-\left( Z_{j}-\overline{Z}\right) b_{1}\right\vert 
&=&\left\vert \left( Z_{j}-\overline{Z}\right) \left( Y_{j}/\left( Z_{j}-%
\overline{Z}\right) -b_{1}\right) \right\vert =\left\vert sign(Z_{j}-%
\overline{Z})|Z_{j}-\overline{Z}|(Y_{j}/\left( Z_{j}-\overline{Z}\right)
-b_{1})\right\vert  \\
&=&|Z_{j}-\overline{Z}|\left\vert sign(Z_{j}-\overline{Z})\left(
Y_{j}/\left( Z_{j}-\overline{Z}\right) -b_{1}\right) \right\vert =|Z_{j}-%
\overline{Z}|\left\vert Y_{j}/\left( Z_{j}-\overline{Z}\right)
-b_{1}\right\vert 
\end{eqnarray*}%
so $S_{j}(b_{1})=\left\vert Y_{j}/\left( Z_{j}-\overline{Z}\right)
-b_{1}\right\vert $ implying 
\begin{equation*}
\widehat{\beta }_{1}=med\left( Y_{j}/(Z_{j}-\overline{Z})\right) ,
\end{equation*}%
the median slope from $(\overline{Z},0)$ to $(Z_{j},Y_{j})$. This has some
similarities to the Theil--Sen estimator and the repeated median estimator.
\ \ \ \ 
\end{example}

Feasible inference for $\widehat{\mathbf{\beta }}$ will be valid if the
predictors have at least two moments. \ Hence these assumptions are
plausible for data in financial economics. To formalize this, the
Assumptions will again be written in two blocks.

\begin{assumption}

\begin{enumerate}
\item[C1.] Assume $\mathrm{E}|\mathbf{Z}_{1}|<\infty $, write $\mathbf{%
\mathbf{\psi =}}\mathrm{E}[\mathbf{Z}_{1}]$ and define 
\begin{equation*}
\mathbf{X}_{1}(\mathbf{\mathbf{\psi }})^{\mathtt{T}}=\left\{ 1,\left( 
\mathbf{Z}_{1}\mathbf{-\mathbf{\psi }}\right) ^{\mathtt{T}}\right\} ^{%
\mathtt{T}},\quad \text{and\quad }\mathbf{G}_{1}(\mathbf{\mathbf{\psi }}%
)=\left\Vert \mathbf{X}_{1}(\mathbf{\mathbf{\psi }})\right\Vert _{2}^{-1}%
\mathbf{X}_{1}(\mathbf{\mathbf{\psi }}).
\end{equation*}

\item[C2.] There exists a single $\mathbf{\beta }_{0}$ such that 
\begin{equation*}
Q_{Y_{1}|\mathbf{Z}_{1}=\mathbf{z}_{1}}(1/2)=\mathbf{z}_{1}^{\mathtt{T}}%
\mathbf{\beta }_{0},\mathbf{\quad x}_{1}^{\mathtt{T}}=\left\{ 1,\left( 
\mathbf{z}_{1}\mathbf{-\mathbf{\psi }}\right) ^{\mathtt{T}}\right\} ^{%
\mathtt{T}}
\end{equation*}%
so all $\mathbf{z}_{1}$.

\item[C2.] $U_{1}=Y_{1}-\left\{ 1,\left( \mathbf{Z}_{1}\mathbf{-\mathbf{\psi 
}}\right) ^{\mathtt{T}}\right\} ^{\mathtt{T}}\mathbf{\beta }_{0}$ is a
continuous random variable with a conditional density $f_{U_{1}|\mathbf{Z}%
_{1}}$.

\item[D1.] $(\mathbf{Z}_{1},Y_{1}),...,(\mathbf{Z}_{n},Y_{n})$ are i.i.d.;

\item[D2.] $\mathrm{E}[\left\vert \mathbf{Z}_{1}\right\vert f_{U_{1}|\mathbf{%
Z}_{1}}(0)]<\infty $;

\item[D3.] $\mathrm{E}[\mathbf{G}_{1}(\mathbf{\mathbf{\psi }})\mathbf{X}_{1}(%
\mathbf{\mathbf{\psi }})^{\mathtt{T}}f_{U_{1}|\mathbf{Z}_{1}}(0)]$ is
positive definite. \ 
\end{enumerate}
\end{assumption}

\ Theorem \ref{Thm:main prop1} provides the main features of $\widehat{%
\mathbf{\beta }}$.

\begin{theorem}
\label{Thm:main prop1}$S_{j}(\mathbf{b})$ is convex in $\mathbf{b}$ and 
\begin{equation*}
\left\vert S_{j}(\mathbf{b})-S_{j}(\mathbf{b}^{\prime })\right\vert \leq
\left\Vert \mathbf{b}-\mathbf{b}^{\prime }\right\Vert _{2}.
\end{equation*}%
Under Assumptions C1-C3 and D1-D3, as $n$ increases 
\begin{equation*}
\widehat{\mathbf{\beta }}\overset{p}{\rightarrow }\mathbf{\beta }_{0},
\end{equation*}%
further 
\begin{equation*}
\sqrt{n}\left( \widehat{\mathbf{\beta }}-\mathbf{\beta }_{0}\right) \overset{%
d}{\rightarrow }N(0,D_{1}^{-1}M_{1}D_{1}^{-1}),
\end{equation*}%
where, writing $\mathbf{\mathbf{\psi }}_{0}\mathbf{\mathbf{=}}\mathrm{E}[%
\mathbf{Z}_{1}]$, 
\begin{equation*}
M_{1}=\mathrm{E}[\mathbf{G}_{1}(\mathbf{\mathbf{\psi }}_{0})\mathbf{G}_{1}(%
\mathbf{\mathbf{\psi }}_{0})^{\mathtt{T}}],\quad D_{1}=2\mathrm{E}[\mathbf{G}%
_{1}(\mathbf{\mathbf{\psi }}_{0})\mathbf{X}_{1}(\mathbf{\mathbf{\psi }}%
_{0})^{\mathtt{T}}f_{U_{1}|\mathbf{Z}_{1}}(0)].
\end{equation*}
\end{theorem}

Proof: Given in the Appendix.

To carry out inference $M_{1}$ and\ $D_{1}$ need to be estimated well. I
advocate $S_{\mathbf{G},\mathbf{X}}$ and $S_{1_{U}\mathbf{G},\mathbf{G}}$:
this is discussed in Section \ref{sect:est vari med} in the Appendix.

\section{Conclusions\label{sect:conclusion}}

I have advocated estimating linear in parameters predictive regressions
using $\hat{\beta}$ rather than the celebrated least squares estimators $%
\hat{\beta}_{LS}$. \ Why? \ $\hat{\beta}$ has relatively simple standard
errors which are robust to thick tailed predictors. \ This robustness leads,
in practice, to nominal tests which are close to being valid (i.e. correct
size), as well as standard errors which a reasonable smooth through time for
rolling estimators. \ 

There are downsides with $\hat{\beta}$. \ $\hat{\beta}_{LS}$ gains precise
from being super sensitive to unusual predictors. \ $\hat{\beta}$
downweights these low probability but influential predictors. \ Sometimes it
is good to take full advantage of every scrap of available information. \ In
that case it makes sense to use $\hat{\beta}_{LS}$. \ But replication and
testing in that environment will be fragile, sensitive to one or two
datapoints. \ $\hat{\beta}$ is a more conservative estimator, it even works
with data as thick tailed as nearly Cauchy. \ Empirically in finance, the
average standard deviation for $\hat{\beta}$ is close to $\hat{\beta}_{LS}$,
so the practical loss of precision seems modest. \ \ 

The concerns and solutions extend to quantile regression. \ Inference base
on $\hat{\beta}_{KB}$ is not robust to thick tailed predictors, while $\hat{%
\beta}$ is. \ It raises no new issues in terms of computation. \ 

\section{Appendix}

\subsection{Proof of Theorem \protect\ref{thm:unpack}}

Now 
\begin{eqnarray*}
S_{\mathbf{G,X}} &=&\frac{1}{n}\sum_{j=1}^{n}\left\Vert \mathbf{X}%
_{j}\right\Vert _{2}^{-1}\left( 
\begin{array}{cc}
1 & \left( \mathbf{Z}_{j}-\overline{\mathbf{Z}}\right) ^{\mathtt{T}} \\ 
\mathbf{Z}_{j}-\overline{\mathbf{Z}} & \left( \mathbf{Z}_{j}-\overline{%
\mathbf{Z}}\right) \left( \mathbf{Z}_{j}-\overline{\mathbf{Z}}\right) ^{%
\mathtt{T}}%
\end{array}%
\right) =\left( 
\begin{array}{cc}
S_{1} & S_{\mathbf{Z}-\overline{\mathbf{Z}}}^{\mathtt{T}} \\ 
S_{\mathbf{Z}-\overline{\mathbf{Z}}} & S_{\mathbf{Z}-\overline{\mathbf{Z}},%
\mathbf{Z}-\overline{\mathbf{Z}}}%
\end{array}%
\right) \\
&=&S_{1}\left( 
\begin{array}{cc}
1 & \widetilde{S}_{\mathbf{Z}-\overline{\mathbf{Z}}}^{\mathtt{T}} \\ 
\widetilde{S}_{\mathbf{Z}-\overline{\mathbf{Z}}} & \widetilde{S}_{\mathbf{Z}-%
\overline{\mathbf{Z}},\mathbf{Z}-\overline{\mathbf{Z}}}%
\end{array}%
\right) =S_{1}\left( 
\begin{array}{cc}
1 & \left( \widetilde{\mathbf{Z}}-\overline{\mathbf{Z}}\right) ^{\mathtt{T}}
\\ 
\widetilde{\mathbf{Z}}-\overline{\mathbf{Z}} & \widetilde{S}_{\mathbf{ZZ}}-%
\widetilde{\mathbf{Z}}\overline{\mathbf{Z}}^{\mathtt{T}}-\overline{\mathbf{Z}%
}\widetilde{\mathbf{Z}}^{\mathrm{T}}+\overline{\mathbf{Z}}\overline{\mathbf{Z%
}}^{\mathtt{T}}%
\end{array}%
\right) ,
\end{eqnarray*}%
and 
\begin{equation*}
S_{\mathbf{G},Y}=\frac{1}{n}\sum_{j=1}^{n}\left\Vert \mathbf{X}%
_{j}\right\Vert _{2}^{-1}\left( 
\begin{array}{c}
Y_{j} \\ 
\left( \mathbf{Z}_{j}-\overline{\mathbf{Z}}\right) Y_{j}%
\end{array}%
\right) =S_{1}\left( 
\begin{array}{c}
\widetilde{S}_{Y} \\ 
\widetilde{S}_{\mathbf{Z}-\overline{\mathbf{Z}},Y}%
\end{array}%
\right) =S_{1}\left( 
\begin{array}{c}
\widetilde{Y} \\ 
\overline{S}_{\mathbf{Z}Y}-\widetilde{Y}\overline{\mathbf{Z}}%
\end{array}%
\right)
\end{equation*}%
Now 
\begin{equation*}
\left( 
\begin{array}{cc}
1 & \widetilde{S}_{\mathbf{Z}-\overline{\mathbf{Z}}}^{\mathtt{T}} \\ 
\widetilde{S}_{\mathbf{Z}-\overline{\mathbf{Z}}} & \widetilde{S}_{\mathbf{Z}-%
\overline{\mathbf{Z}},\mathbf{Z}-\overline{\mathbf{Z}}}%
\end{array}%
\right) \left( 
\begin{array}{c}
\widehat{\beta }_{0} \\ 
\widehat{\beta }_{1:p}%
\end{array}%
\right) =\left( 
\begin{array}{c}
\widetilde{S}_{Y} \\ 
\widetilde{S}_{\mathbf{Z}-\overline{\mathbf{Z}},Y}%
\end{array}%
\right)
\end{equation*}%
so that 
\begin{eqnarray*}
\widehat{\beta }_{0}+\widetilde{S}_{\mathbf{Z}-\overline{\mathbf{Z}}}^{%
\mathtt{T}}\widehat{\beta }_{1:p} &=&\widetilde{S}_{Y} \\
\widetilde{S}_{\mathbf{Z}-\overline{\mathbf{Z}}}\widehat{\beta }_{0}+%
\widetilde{S}_{\mathbf{Z}-\overline{\mathbf{Z}},\mathbf{Z}-\overline{\mathbf{%
Z}}}\widehat{\beta }_{1:p} &=&\widetilde{S}_{\mathbf{Z}-\overline{\mathbf{Z}}%
,Y}.
\end{eqnarray*}%
Rearranging gives the result for $\widehat{\beta }_{0}$. Note that

\begin{eqnarray*}
\widetilde{S}_{\mathbf{Z}-\overline{\mathbf{Z}},\mathbf{Z}-\overline{\mathbf{%
Z}}}-\widetilde{S}_{\mathbf{Z}-\overline{\mathbf{Z}}}\widetilde{S}_{\mathbf{Z%
}-\overline{\mathbf{Z}}}^{\mathrm{T}} &=&\widetilde{S}_{\mathbf{ZZ}}-%
\widetilde{\mathbf{Z}}\overline{\mathbf{Z}}^{\mathtt{T}}-\overline{\mathbf{Z}%
}\widetilde{\mathbf{Z}}^{\mathrm{T}}+\overline{\mathbf{Z}}\overline{\mathbf{Z%
}}^{\mathtt{T}}-\left( \widetilde{\mathbf{Z}}-\overline{\mathbf{Z}}\right)
\left( \widetilde{\mathbf{Z}}-\overline{\mathbf{Z}}\right) ^{\mathrm{T}} \\
&=&\widetilde{S}_{\mathbf{Z}-\widetilde{\mathbf{Z}},\mathbf{Z}-\widetilde{%
\mathbf{Z}}}.
\end{eqnarray*}%
Then%
\begin{eqnarray*}
\overline{S}_{\mathbf{Z}-\overline{\mathbf{Z}},\mathbf{Z}-\overline{\mathbf{Z%
}}}\widehat{\beta }_{1:p} &=&\overline{S}_{\mathbf{Z}-\overline{\mathbf{Z}}%
,Y}-\overline{S}_{\mathbf{Z}-\overline{\mathbf{Z}}}\widehat{\alpha }=%
\overline{S}_{\mathbf{Z}-\overline{\mathbf{Z}},Y}-\overline{S}_{\mathbf{Z}-%
\overline{\mathbf{Z}}}\overline{S}_{Y}+\overline{S}_{\mathbf{Z}-\overline{%
\mathbf{Z}}}\overline{S}_{\mathbf{Z}-\overline{\mathbf{Z}}}^{\mathtt{T}}%
\widehat{\gamma } \\
\left( \overline{S}_{\mathbf{Z}-\overline{\mathbf{Z}},\mathbf{Z}-\overline{%
\mathbf{Z}}}-\overline{S}_{\mathbf{Z}-\overline{\mathbf{Z}}}\overline{S}_{%
\mathbf{Z}-\overline{\mathbf{Z}}}^{\mathtt{T}}\right) \widehat{\beta }_{1:p}
&=&\overline{S}_{\mathbf{Z}-\overline{\mathbf{Z}},Y}-\overline{S}_{\mathbf{Z}%
-\overline{\mathbf{Z}}}\overline{S}_{Y} \\
\overline{S}_{\mathbf{Z}-\widetilde{\mathbf{Z}},\mathbf{Z}-\widetilde{%
\mathbf{Z}}}\widehat{\beta }_{1:p} &=&\overline{S}_{\mathbf{Z}-\widetilde{%
\mathbf{Z}},Y-\widetilde{Y}},
\end{eqnarray*}%
then using a matrix inverse gives the result.

\subsection{Proof of Theorem \protect\ref{thm:conditional}}

B1 is needed for $\widehat{\mathbf{\beta }}$ to exist. \ Define the
predictive regression errors $U_{j}=Y_{j}-\mathbf{X}_{j}^{\mathtt{T}}\mathbf{%
\beta }$\textbf{,}\ for $j=1,...,n$\textbf{. }Then 
\begin{equation*}
\widehat{\mathbf{\beta }}=S_{\mathbf{G},\mathbf{X}}^{-1}S_{\mathbf{G},Y}=%
\mathbf{\beta +}S_{\mathbf{G},\mathbf{X}}^{-1}\frac{1}{n}\sum_{j=1}^{n}%
\mathbf{G}_{j}U_{j}.
\end{equation*}%
Then under A2 and B2, 
\begin{equation*}
\mathrm{E}[U_{j}|(\mathbf{Z}=\mathbf{z})]=0,
\end{equation*}%
while under B2 and B3 
\begin{equation*}
\mathrm{Var}(U_{j}|(\mathbf{Z}=\mathbf{z}))=\sigma _{j}^{2}<\infty .
\end{equation*}%
Then the conditional unbiasedness and conditional variance follow. \ 

What remains is the Berry-Esseen type result. \ The approach is to use the 
\cite{Bentkus(05)} version. \ It is stated here for convenience.

\begin{theorem}
(\cite{Bentkus(05)}) Suppose $W_{i}$ are zero mean, independent in $R^{d}$,
then 
\begin{equation*}
\sup_{A\in \mathcal{C}^{d}}\left\vert \Pr (\frac{1}{\sqrt{n}}%
\sum_{j=1}^{n}V_{j}\in A)-\Pr (N(0,C^{\ast })\in A)\right\vert \leq c\frac{%
d^{1/4}}{n^{1/2}}\left( \frac{1}{n}\sum_{j=1}^{n}\mathrm{E}[\left\Vert
\left( C^{\ast }\right) ^{-1/2}V_{j}\right\Vert _{2}^{3}]\right) ,
\end{equation*}%
where 
\begin{equation*}
C^{\ast }=n^{-1}\sum_{j=1}^{n}\mathrm{Var}(V_{j}),
\end{equation*}%
where $\mathcal{C}^{d}$ denotes the set of all convex subsets of $R^{p+1}$ \ 
\end{theorem}

Now 
\begin{equation*}
\sqrt{n}(\widehat{\mathbf{\beta }}-\mathbf{\beta })=\sum_{j=1}^{n}V_{j},%
\quad \text{where \quad }V_{j}=\frac{1}{\sqrt{n}}S_{\mathbf{G},\mathbf{X}%
}^{-1}\mathbf{G}_{j}U_{j}.
\end{equation*}%
The $V_{j}$ are conditionally independent with variances 
\begin{equation*}
C_{j}=\mathrm{Var}(V_{j}|(\mathbf{Z}=\mathbf{z}))=\frac{1}{n}\sigma
_{j}^{2}S_{\mathbf{G,X}}^{-1}\mathbf{G}_{j}\mathbf{G}_{j}^{T}S_{\mathbf{G,X}%
}^{-1},
\end{equation*}%
and the corresponding sums 
\begin{equation*}
V^{\ast }=\sum_{j=1}^{n}V_{j},\quad C^{\ast }=\sum_{j=1}^{n}C_{j}=S_{\mathbf{%
G,X}}^{-1}S_{\sigma ^{2}\mathbf{G,G}}S_{\mathbf{G,X}}^{-1}.
\end{equation*}%
$C^{\ast }$ is invertible using B1 and B4. \ Then 
\begin{equation*}
\left( C^{\ast }\right) ^{-1}=S_{\mathbf{G,X}}S_{\sigma ^{2}\mathbf{G,G}%
}^{-1}S_{\mathbf{G,X}},
\end{equation*}%
which is positive definite, so 
\begin{equation*}
\left( C^{\ast }\right) ^{-1/2}=S_{\sigma ^{2}\mathbf{G,G}}^{-1/2}S_{\mathbf{%
G,X}},\quad \text{that is\quad }\left\{ \left( C^{\ast }\right)
^{-1/2}\right\} ^{\mathtt{T}}\left( C^{\ast }\right) ^{-1/2}=\left( C^{\ast
}\right) ^{-1}
\end{equation*}%
Then define the corresponding skewness terms \ \ 
\begin{equation*}
\varsigma _{j}=\mathrm{E}\left\Vert \left( C^{\ast }\right) ^{-1/2}V_{j}|(%
\mathbf{Z}=\mathbf{z})\right\Vert _{2}^{3},\quad \varsigma ^{\ast
}=\sum_{j=1}^{n}\varsigma _{j},
\end{equation*}%
which exists using B5. \ 

Then 
\begin{eqnarray*}
\left( C^{\ast }\right) ^{-1/2}V_{j} &=&S_{\sigma ^{2}\mathbf{G,G}}^{-1/2}S_{%
\mathbf{G,X}}V_{j}=\frac{1}{\sqrt{n}}S_{\sigma ^{2}\mathbf{G,G}}^{-1/2}S_{%
\mathbf{G,X}}S_{\mathbf{G},\mathbf{X}}^{-1}\mathbf{G}_{j}U_{j} \\
&=&\frac{1}{\sqrt{n}}S_{\sigma ^{2}\mathbf{G,G}}^{-1/2}\mathbf{G}_{j}U_{j},
\end{eqnarray*}%
so 
\begin{equation*}
\varsigma _{j}=\mathrm{E}[|U_{j}/\sigma _{j}|^{3}|(\mathbf{Z}=\mathbf{z}%
)]\left\Vert S_{\sigma ^{2}\mathbf{G,G}}^{-1/2}\mathbf{G}_{j}\sigma
_{j}\right\Vert _{2}^{3}.
\end{equation*}%
Then the result follows from \cite{Bentkus(05)}. \ \ \ \ 

\subsection{Proof of Theorem \protect\ref{thm:uncond}}

Stack the moment conditions: 
\begin{equation*}
g(Y_{1},\mathbf{Z}_{1};\mathbf{\theta })=\left( 
\begin{array}{c}
g_{1}(\mathbf{Z}_{1};\mathbf{\psi }) \\ 
g_{2}(Y_{1},\mathbf{Z}_{1};\mathbf{\psi ,\beta })%
\end{array}%
\right) ,\mathbf{\quad }g_{1}(\mathbf{Z}_{1};\mathbf{\psi })=\mathbf{Z}_{1}-%
\mathbf{\psi ,\quad }g_{2}(Y_{1},\mathbf{Z}_{1};\mathbf{\psi ,\beta })=%
\mathbf{G}_{1}(\mathbf{\psi })\left\{ Y_{1}-\mathbf{X}_{1}\mathbf{(\psi )}^{%
\mathtt{T}}\mathbf{\beta }\right\}
\end{equation*}%
where $\mathbf{\theta =(\psi }^{\mathtt{T}}\mathbf{,\beta }^{\mathtt{T}}%
\mathbf{)}^{\mathtt{T}}=\mathbf{(\psi }^{\mathtt{T}}\mathbf{,\alpha ,\gamma }%
^{\mathtt{T}}\mathbf{)}^{\mathtt{T}}$. Crucially $g_{1}$ does not depend
upon $\mathbf{\beta }$. Then 
\begin{equation*}
\mathrm{E}[g(Y_{1},\mathbf{Z}_{1};\mathbf{\theta }_{0})]=0,
\end{equation*}%
while 
\begin{equation*}
\mathrm{Var}[g(Y_{1},\mathbf{Z}_{1};\mathbf{\theta }_{0})]=\left( 
\begin{array}{cc}
\mathrm{Var}(\mathbf{Z}_{1}) & 0 \\ 
0 & \mathrm{Var}(\mathbf{G}_{1}U_{1})%
\end{array}%
\right) ,
\end{equation*}%
where here I write $\mathbf{G}_{1}=\mathbf{G}_{1}(\mathbf{\psi }_{0})$\ and $%
\mathbf{X}_{1}\mathbf{(\psi )=X}_{1}\mathbf{(\psi }_{0}\mathbf{)}$. Of
course 
\begin{equation*}
\mathrm{Var}(\mathbf{G}_{1}U_{1})=\mathrm{E}(\sigma ^{2}\mathbf{G}_{1}%
\mathbf{G}_{1}^{\mathtt{T}}).
\end{equation*}%
The crucial matrix zeros follow by Adam's law applied to 
\begin{equation*}
\mathrm{E}[(\mathbf{Z}_{1}-\mathbf{\psi )G}_{1}U_{1}]=\mathrm{E}[(\mathbf{Z}%
_{1}-\mathbf{\psi )G}_{1}\mathrm{E}[U_{1}|Z_{1}]]=0.
\end{equation*}%
Further, 
\begin{equation*}
\left. \frac{\partial \mathrm{E}[g(Y_{1},\mathbf{Z}_{1};\mathbf{\theta })]}{%
\partial \mathbf{\theta }^{T}}\right\vert _{\mathbf{\theta =\theta }%
_{0}}=\left( 
\begin{array}{cc}
-I_{p} & 0 \\ 
V_{0} & -\mathrm{E}[\mathbf{G}_{1}\mathbf{X}_{1}^{\mathtt{T}}\mathbf{]}%
\end{array}%
\right) .
\end{equation*}%
The term $V_{0}$ is not of central importance, but takes the form 
\begin{eqnarray*}
V_{0} &=&\mathbf{-}\mathrm{E}[\mathbf{G}_{1}(\mathbf{\psi }_{0})\mathbf{%
\beta ^{\mathtt{T}}\frac{\partial \mathbf{X}_{1}\mathbf{(\psi }_{0}\mathbf{)}%
}{\partial \mathbf{\psi }^{\mathtt{T}}}],\quad }\frac{\partial \mathbf{X}_{1}%
\mathbf{(\psi }_{0}\mathbf{)}}{\partial \mathbf{\psi }^{\mathtt{T}}}\mathbf{=%
}\left( 0_{p},-I_{p}\right) \\
&=&\mathbf{-}\mathrm{E}[\mathbf{G}_{1}(\mathbf{\psi })\mathbf{]\beta ^{%
\mathtt{T}}}\left( 0_{p},-I_{p}\right) =\mathrm{E}[\mathbf{G}_{1}(\mathbf{%
\psi })\mathbf{]\gamma ^{\mathtt{T}}.}
\end{eqnarray*}%
Finally, notice A3 means that $\mathrm{E}[\mathbf{G}_{1}\mathbf{X}_{1}^{%
\mathtt{T}}\mathbf{]}$ is symmetric and invertible. \ The result then
follows conventionally. \ 

QED. \ 

\subsection{Proof of Theorem \protect\ref{Thm:main prop1}}

We start with a statement and proof of a preliminary theory.

\begin{theorem}
\label{thm:preliminary}$\left\Vert \mathbf{X}_{1}\right\Vert _{2}>0$, then $%
S_{1}(\mathbf{b})$ is convex in $\mathbf{b}$, writing $\mathbf{\delta }=%
\mathbf{b}-\mathbf{\beta }_{\ast }$, then 
\begin{equation*}
-\left\Vert \mathbf{\delta }\right\Vert _{2}\leq \left\{ S_{1}(\mathbf{b}%
)-S_{1}(\mathbf{\beta }_{\ast })\right\} \leq \left\Vert \mathbf{\delta }%
\right\Vert _{2},
\end{equation*}%
and 
\begin{eqnarray*}
\mathrm{E}\left[ \partial S_{1}(\mathbf{b})|\mathbf{X}_{1}\right]  &=&-2%
\mathbf{G}_{1}\left[ \frac{1}{2}-F_{Y_{1}|\mathbf{X}_{1}}(\mathbf{X}_{1}^{%
\mathrm{T}}\mathbf{b})\right] , \\
\mathrm{Var}\left[ \partial S_{1}(\mathbf{b})|\mathbf{X}_{1}\right]  &=&4%
\mathbf{G}_{1}\mathbf{G}_{1}^{\mathrm{T}}F_{Y_{1}|\mathbf{X}_{1}}(\mathbf{X}%
_{1}^{\mathrm{T}}\mathbf{b})\left\{ 1-F_{Y_{1}|\mathbf{X}_{1}}(\mathbf{X}%
_{1}^{\mathrm{T}}\mathbf{b})\right\} ,
\end{eqnarray*}%
and under C2%
\begin{equation*}
\frac{\partial \mathrm{E}\left[ \partial S_{1}(\mathbf{b})|\mathbf{X}_{1}%
\right] }{\partial \mathbf{b}^{\mathtt{T}}}=2\mathbf{G}_{1}\mathbf{X}_{1}^{%
\mathrm{T}}f_{U_{1}|\mathbf{X}_{1}}(\mathbf{X}_{1}^{\mathtt{T}}\mathbf{%
\delta }).
\end{equation*}%
Under C1 and C2, define $U_{1}=Y_{1}-\mathbf{X}_{1}^{\mathrm{T}}\mathbf{%
\beta }_{0}$ and $\mathbf{\delta }=\mathbf{b}-\mathbf{\beta }_{0}$. \ Then 
\begin{eqnarray*}
\mathrm{E}[S_{1}(\delta )|\mathbf{X}_{1}] &=&2\left\Vert \mathbf{X}%
_{1}\right\Vert _{2}^{-1}\int_{\mathbf{X}_{1}^{\mathtt{T}}\delta }^{0}(s-%
\mathbf{X}_{1}^{\mathtt{T}}\delta )f_{U_{1}|\mathbf{X}_{1}}(s)\mathrm{d}s, \\
\frac{\partial \mathrm{E}[S_{1}(\mathbf{b})|\mathbf{X}_{1}\mathbf{]}}{%
\partial \mathbf{b}} &=&-2\mathbf{G}_{1}\int_{\mathbf{x}_{1}^{\mathtt{T}%
}\delta }^{0}f_{U_{1}|\mathbf{X}_{1}}(s)\mathrm{d}s=\mathrm{E}\left[
\partial S_{1}(\mathbf{b})|\mathbf{X}_{1}\right]  \\
\frac{\partial ^{2}\mathrm{E}[S_{1}(\mathbf{b})|\mathbf{X}_{1}\mathbf{]}}{%
\partial \mathbf{b}\partial \mathbf{b}^{\mathtt{T}}} &=&2\mathbf{G}_{1}%
\mathbf{X}_{1}^{\mathrm{T}}f_{U_{1}|\mathbf{X}_{1}}(\mathbf{X}_{1}^{\mathrm{T%
}}\mathbf{\delta })=D_{1}(\mathbf{X}_{1},\mathbf{\delta }).
\end{eqnarray*}
\end{theorem}

Proof.

Ignore subscripts

\begin{eqnarray*}
\left\vert y+t_{1}\left( x^{\mathtt{T}}\beta _{1}\right) +(1-t_{1})\left( x^{%
\mathtt{T}}\beta _{2}\right) \right\vert &=&\left\vert t_{1}\left( y+x^{%
\mathtt{T}}\beta _{1}\right) +(1-t_{1})\left( y+x^{\mathtt{T}}\beta
_{2}\right) \right\vert \\
&\leq &\left\vert t_{1}\left( y+x^{\mathtt{T}}\beta _{1}\right)
|+|(1-t_{1})\left( y+x^{\mathtt{T}}\beta _{2}\right) \right\vert ,\quad 
\text{triangular inequality} \\
&\leq &t_{1}\left\vert y+x^{\mathtt{T}}\beta _{1}|+(1-t_{1})|y+x^{\mathtt{T}%
}\beta _{2}\right\vert ,
\end{eqnarray*}%
so is convex. \ Scaling by $\left\Vert \mathbf{X}\right\Vert _{2}^{-1}$ has
no impact on convexity.

Define $\mathbf{\delta =b-\beta }_{\ast }$ and $u=y-\mathbf{x}^{\mathtt{T}}%
\mathbf{\beta }_{\ast }$, then 
\begin{equation*}
\left\vert y-\mathbf{x}^{\mathtt{T}}\mathbf{b}\right\vert -\left\vert y-%
\mathbf{x}^{\mathtt{T}}\mathbf{\beta }_{\ast }\right\vert =\left\vert u-%
\mathbf{x}^{\mathtt{T}}\mathbf{\delta }\right\vert -\left\vert u\right\vert
\end{equation*}%
and so by triangular inequality 
\begin{equation*}
\left\vert u-\mathbf{x}^{\mathtt{T}}\mathbf{\delta }\right\vert -\left\vert
u\right\vert \leq \sum_{i=1}^{p+1}\left\vert x_{i}\delta _{i}\right\vert .
\end{equation*}%
By Cauchy-Schwartz 
\begin{equation*}
\sum_{i=1}^{p+1}\left\vert x_{i}\delta _{i}\right\vert \leq \sqrt{%
\sum_{i=1}^{p+1}x_{i}^{2}}\sqrt{\sum_{i=1}^{p+1}\delta _{i}^{2}}=\left\Vert 
\mathbf{x}\right\Vert _{2}\left\Vert \mathbf{\delta }\right\Vert _{2}.
\end{equation*}%
Applying the same argument the other way, we have 
\begin{equation*}
-\left\Vert \mathbf{x}\right\Vert _{2}\left\Vert \mathbf{\delta }\right\Vert
_{2}\leq \left\vert u-\mathbf{x}^{\mathtt{T}}\mathbf{\delta }\right\vert
-\left\vert u\right\vert \leq \left\Vert \mathbf{x}\right\Vert
_{2}\left\Vert \mathbf{\delta }\right\Vert _{2},
\end{equation*}%
so 
\begin{equation*}
-\left\Vert \mathbf{\delta }\right\Vert _{2}\leq \left\Vert \mathbf{x}%
\right\Vert _{2}^{-1}\left( \left\vert y-\mathbf{x}^{\mathtt{T}}\mathbf{b}%
\right\vert -\left\vert y-\mathbf{x}^{\mathtt{T}}\mathbf{\beta }_{\ast
}\right\vert \right) \leq \left\Vert \mathbf{\delta }\right\Vert _{2}.
\end{equation*}%
This bounding implies all moments of $S(\delta )$ must exist. \ 

The subderivative is straightforward, as is the stated conditional variance
as%
\begin{equation*}
\mathrm{E}\left( \partial S_{1}(\mathbf{\beta }_{0})|\mathbf{X}_{1}=\mathbf{x%
}_{1}\right) =0,
\end{equation*}%
and the squared sign function is $1$. \ Now let $\mathbf{\beta }_{\ast }=%
\mathbf{\beta }_{0}$, then 
\begin{eqnarray*}
\mathrm{E}[S(\delta )|\mathbf{X} &=&\mathbf{x}] \\
&=&\left\Vert \mathbf{x}\right\Vert _{2}^{-1}\mathrm{E}[(U-\mathbf{x}^{%
\mathtt{T}}\delta )sign(U-\mathbf{x}^{\mathtt{T}}\delta )-sign(U)|\mathbf{X}=%
\mathbf{x}] \\
&=&\left\Vert \mathbf{x}\right\Vert _{2}^{-1}\mathrm{E}[(U-\mathbf{x}^{%
\mathtt{T}}\delta )\{sign(U-\mathbf{x}^{\mathtt{T}}\delta )-sign(U)\}|%
\mathbf{X}=\mathbf{x}],\quad \text{by }\mathrm{E}[sign(U)|\mathbf{X}=\mathbf{%
x}]=0
\end{eqnarray*}%
Now, for $a>0$, 
\begin{equation*}
sign(u-a)-sign(u)=\left\{ 
\begin{array}{ccc}
-2 &  & 0<U<a \\ 
0 &  & elsewhere%
\end{array}%
\right. 
\end{equation*}%
and, for $a<0$,%
\begin{equation*}
sign(u-a)-sign(u)=\left\{ 
\begin{array}{ccc}
2 &  & 0>U>a \\ 
0 &  & elsewhere%
\end{array}%
\right. 
\end{equation*}%
Hence 
\begin{eqnarray*}
\mathrm{E}[S(\delta )|\mathbf{X} &=&\mathbf{x}]=2\left\Vert \mathbf{x}%
\right\Vert _{2}^{-1}\left\{ 1_{\mathbf{x}^{\mathtt{T}}\delta <0}\int_{%
\mathbf{x}^{\mathtt{T}}\delta }^{0}(s-\mathbf{x}^{\mathtt{T}}\delta )f_{U|%
\mathbf{X}=\mathbf{x}}(s)\mathrm{d}s-1_{\mathbf{x}^{\mathtt{T}}\delta
>0}\int_{0}^{\mathbf{x}^{\mathtt{T}}\delta }(s-\mathbf{x}^{\mathtt{T}}\delta
)f_{U|\mathbf{X}=\mathbf{x}}(s)\mathrm{d}s\right\}  \\
&=&2\left\Vert \mathbf{x}\right\Vert _{2}^{-1}\int_{\mathbf{x}^{\mathtt{T}%
}\delta }^{0}(s-\mathbf{x}^{\mathtt{T}}\delta )f_{U|\mathbf{X}=\mathbf{x}}(s)%
\mathrm{d}s,
\end{eqnarray*}%
using the convention $\int_{0}^{a}f\mathrm{d}s=-\int_{a}^{0}f\mathrm{d}s$. \
Then so assuming range of $X$ does not depend upon $\delta $, using
Leibnitz's rule 
\begin{eqnarray*}
\frac{\partial \mathrm{E}[S_{1}(\delta )|\mathbf{X}_{1}]}{\partial \delta }
&=&-2\mathbf{G}_{1}\int_{\mathbf{X}_{1}^{\mathtt{T}}\delta }^{0}f_{U|\mathbf{%
X}_{1}}(s)\mathrm{d}s \\
\frac{\partial ^{2}\mathrm{E}[S_{1}(\delta )|\mathbf{X}_{1}]}{\partial
\delta \partial \delta ^{\mathtt{T}}} &=&2\mathbf{G}_{1}\mathbf{X}_{1}^{%
\mathrm{T}}f_{U_{1}|\mathbf{X}_{1}}(\mathbf{X}_{1}^{\mathtt{T}}\delta ).
\end{eqnarray*}%
The results on the moments of $\partial S_{1}(\delta )|\mathbf{X}_{1}$ are
straightforward.

QED.

We now move to the proof of the main parts of the theorem. \ 

Proof of Theorem \ref{thm:preliminary}, shows that 
\begin{equation*}
-\left\Vert \delta \right\Vert _{1}\leq S_{1}(\mathbf{\beta })\leq
\left\Vert \delta \right\Vert _{1}
\end{equation*}%
so the mean of $S_{1}(\mathbf{\beta })$ exists. Theorem \ref{thm:preliminary}
says $S_{1}(\mathbf{\beta })$ is convex and thus so is $S_{n}^{\ast }(%
\mathbf{\beta })$, as indeed is $\mathrm{E}[S_{1}(\mathbf{\beta })]$. Under
D1, the strong law of large numbers (\ref{sLLN}) implies 
\begin{equation}
S_{n}^{\ast }(\mathbf{\beta })=\frac{1}{n}\sum_{j=1}^{n}S_{j}(\mathbf{\beta }%
)\overset{p}{\rightarrow }\mathrm{E}[S_{1}(\mathbf{\beta })].  \label{sLLN}
\end{equation}%
\ As $S_{n}^{\ast }(\mathbf{\beta })$ is convex, pointwise convergence
implies uniform convergence. \ 

All that remains is to check that $\mathbf{\beta }_{0}$ is the global
minimizer of the convex $\mathrm{E}[S_{1}(\mathbf{\beta })]$? We know this
from the derivative and second derivative of $\mathrm{E}[S_{1}(\mathbf{\beta 
})]$\ with respect to $\mathbf{\beta }$ evaluated at $\mathbf{\beta }_{0}$
so long as $\mathrm{E}[D_{1}(\mathbf{X}_{1})]$ exists. \ But Assumption Q4
is enough to guarantee this. D3 is enough for this to $\mathrm{E}[S_{1}(%
\mathbf{\beta })]$ to be guaranteed to be a unique minimum at $\mathbf{\beta 
}=\mathbf{\beta }_{0}$.

Now turn to the CLT. \ 

Recall under C1 and D2, $M_{1}$ exists. \ Further, under C1,C2 and D2 $D_{1}$
exists. \ Additionally under D3, it is also positive definite. The sole
challenge here is that $S_{n}^{\ast }(\mathbf{\beta })$ is not
differentiable at $\beta $ for which $y_{j}=\mathbf{x}_{j}^{T}\mathbf{\beta }
$ and $\partial S_{1}(\mathbf{\beta })$ has derivatives which are either 0
or not defined.

We use a stochastic equicontinuity argument. A review is provided by \cite%
{Andrews(94)}. \ Now write 
\begin{equation*}
\partial \overline{S}(\mathbf{\beta })=\frac{1}{n}\sum_{j=1}^{n}\partial
S_{j}(\mathbf{\beta }),
\end{equation*}%
which is the sums of i.i.d. terms, while recall $\left\Vert \partial S_{j}(%
\mathbf{\beta })\right\Vert _{\infty }$ is bounded above by 1. \ Further, in
the special case $\mathbf{\beta =\beta }_{0}$, $\mathrm{E}[\partial S_{1}(%
\mathbf{\beta }_{0})]=0$ and $\mathrm{Var}[\partial S_{1}(\mathbf{\beta }%
_{0})]=M_{1}$.

By mean value Theorem, there exists a $\widetilde{\mathbf{\beta }}$ such
that $\left\Vert \widetilde{\mathbf{\beta }}-\mathbf{\beta }_{0}\right\Vert
_{2}\leq $\ $\left\Vert \widehat{\mathbf{\beta }}-\mathbf{\beta }%
_{0}\right\Vert _{2}$\ 

\begin{eqnarray*}
0 &=&\left. \mathrm{E}[\partial \overline{S}(\mathbf{\beta })]\right\vert _{%
\mathbf{\beta =\beta }_{0}}=\left. \mathrm{E}[\partial \overline{S}(\mathbf{%
\beta })]\right\vert _{\mathbf{\beta =}\widehat{\mathbf{\beta }}}-\left. 
\frac{\partial \mathrm{E}[\partial \overline{S}(\mathbf{\beta })]}{\partial 
\mathbf{\beta }^{T}}\right\vert _{\mathbf{\beta =}\widehat{\mathbf{\beta }}}(%
\widehat{\mathbf{\beta }}-\mathbf{\beta }_{0}). \\
&\simeq &\left. \mathrm{E}[\partial \overline{S}(\mathbf{\beta }%
)]\right\vert _{\mathbf{\beta =}\widehat{\mathbf{\beta }}}-nD_{1}(\widehat{%
\mathbf{\beta }}-\mathbf{\beta }_{0}).
\end{eqnarray*}%
Now $\widehat{\mathbf{\beta }}\overset{p}{\rightarrow }\mathbf{\beta }_{0}$
so%
\begin{equation*}
\left. \frac{\partial \mathrm{E}[\partial \overline{S}(\mathbf{\beta })]}{%
\partial \mathbf{\beta }^{T}}\right\vert _{\mathbf{\beta =}\widehat{\mathbf{%
\beta }}}=D_{1}+o_{p}(1),\quad \text{recalling\quad }D_{1}=\left. \frac{%
\partial \mathrm{E}[\partial \overline{S}(\mathbf{\beta })]}{\partial 
\mathbf{\beta }^{T}}\right\vert _{\mathbf{\beta =\beta }_{0}}.
\end{equation*}%
Hence 
\begin{equation*}
\sqrt{n}(\widehat{\mathbf{\beta }}-\mathbf{\beta }_{0})\simeq D_{1}^{-1}%
\sqrt{n}\left. \mathrm{E}[\partial \overline{S}(\mathbf{\beta })]\right\vert
_{\mathbf{\beta =}\widehat{\mathbf{\beta }}}.
\end{equation*}%
The challange is the limit law of 
\begin{equation*}
\left. \mathrm{E}[\partial \overline{S}(\mathbf{\beta })]\right\vert _{%
\mathbf{\beta =}\widehat{\mathbf{\beta }}}.
\end{equation*}

As $\partial S_{j}(\mathbf{\beta })$ is always bounded above and below by 1,
this setup is contained in type I class of \cite{Andrews(94)}, so \ the law
of $\left. \sqrt{n}\mathrm{E}[\partial \overline{S}(\mathbf{\beta }%
)]\right\vert _{\mathbf{\beta =}\widehat{\mathbf{\beta }}}$ is the same as
the law of $\sqrt{n}\partial \overline{S}(\mathbf{\beta }_{0})$. But that
law follows from Lindeberg-Levy CLT 
\begin{equation*}
\sqrt{n}\partial \overline{S}(\mathbf{\beta }_{0})\overset{d}{\rightarrow }%
N(0,M_{1}).
\end{equation*}%
The rest is straightforward. \ 

QED.

\subsection{Estimating the asymptotic variance\label{sect:est vari med}}

Focus on the two stage method of moment estimator. \ Now 
\begin{eqnarray*}
\widehat{\mathbf{\psi }} &=&\overline{\mathbf{Z}}\overset{p}{\rightarrow }%
\mathbf{\psi }_{0}\mathbf{,} \\
\widehat{M}_{1}(\mathbf{\psi }) &=&\frac{1}{n}\sum_{j=1}^{n}\mathbf{G}_{j}(%
\mathbf{\psi })\mathbf{G}_{j}(\mathbf{\psi })^{\mathtt{T}}\overset{p}{%
\rightarrow }\mathrm{E}[\mathbf{G}_{1}(\mathbf{\mathbf{\psi }})\mathbf{G}%
_{1}(\mathbf{\mathbf{\psi }})^{\mathtt{T}}],
\end{eqnarray*}%
by, respectively, the strong law of large numbers and the uniform strong law
of large numbers as $\left\Vert \mathbf{G}_{1}(\mathbf{\psi })\right\Vert
_{\infty }\leq 1$ for all $\mathbf{\psi }$. So 
\begin{equation*}
\widehat{M}_{1}=\frac{1}{n}\sum_{j=1}^{n}\mathbf{G}_{j}\mathbf{G}_{j}^{%
\mathtt{T}}\overset{p}{\rightarrow }\mathrm{E}[\mathbf{G}_{1}(\mathbf{%
\mathbf{\psi }}_{0})\mathbf{G}_{1}(\mathbf{\mathbf{\psi }}_{0})^{\mathtt{T}%
}].
\end{equation*}

The harder term is and, for a bandwidth $h_{n}>0$, the%
\begin{equation*}
\int \mathbf{G}_{1}\mathbf{X}_{1}^{\mathtt{T}}f(U_{1}=0|\mathbf{Z}_{1})f(%
\mathbf{Z}_{1})\mathrm{d}\mathbf{Z}_{1}=\int \mathbf{G}_{1}\mathbf{X}_{1}^{%
\mathtt{T}}f(u=0,\mathbf{Z}_{1})\mathrm{d}\mathbf{Z}_{1}
\end{equation*}%
Suppose we had some data $\left( u_{1},\mathbf{z}_{1}\right) ,...,\left(
u_{n},\mathbf{z}_{n}\right) $ then estimate the joint density by the kernel%
\begin{equation*}
\frac{1}{n}\sum_{j=1}^{n}k_{h}(-u_{j})k_{h}(z-\mathbf{z}_{j}),
\end{equation*}%
where $k_{h}$ are zero mean kernel densities, and plug it into 
\begin{eqnarray*}
&&\int \mathbf{G}_{1}\mathbf{X}_{1}^{\mathtt{T}}\left\{ \frac{1}{n}%
\sum_{j=1}^{n}k_{h_{n}}(-U_{j})k_{h_{n}}(\mathbf{Z}_{1}-\mathbf{z}%
_{j})\right\} \mathrm{d}\mathbf{Z}_{1} \\
&=&\frac{1}{n}\sum_{j=1}^{n}k_{h}(-U_{j})\int \mathbf{G}_{1}\mathbf{X}_{1}^{%
\mathtt{T}}k_{h_{n}}(\mathbf{Z}_{1}-\mathbf{z}_{i})\mathrm{d}\mathbf{Z}_{1}
\\
&=&\frac{1}{n}\sum_{j=1}^{n}k_{h_{n}}(-U_{j})\mathbf{G}_{j}\mathbf{X}_{j}^{%
\mathtt{T}}.
\end{eqnarray*}%
This suggests the use a rectangular kernel 
\begin{equation*}
\widehat{D}_{1}=\frac{2}{nh_{n}}\sum_{j=1}^{n}\mathbf{G}_{j}\mathbf{X}_{j}^{%
\mathtt{T}}1_{|\widehat{U}_{j}<h_{n}|},
\end{equation*}%
which needs $h_{n}\downarrow 0$ and $nh_{n}\rightarrow \infty $ as $n$
increases so long as 
\begin{equation*}
\mathrm{Var}(\mathbf{X}_{1,i}f_{U_{1}|\mathbf{X}_{1}}(0))<\infty ,\quad
i=1,...,p+1.
\end{equation*}%
Here $\widehat{U}_{j}=Y_{j}-\mathbf{X}_{j}^{\mathtt{T}}\widehat{\mathbf{%
\beta }}$. \ 

\baselineskip=14pt

\bibliographystyle{chicago}
\bibliography{neil}

\baselineskip=20pt

\end{document}